\documentclass{jfm}

\usepackage{graphicx}
\usepackage{epsfig,psfrag}
\usepackage{color}
\usepackage{psfrag}
\usepackage{multirow}

\usepackage{mathrsfs}
\usepackage[all]{xy}
\usepackage{subfig}
\usepackage{syntonly}

\usepackage{amsmath}
\usepackage{eucal}
\usepackage{multirow}
\usepackage{array}

\ifCUPmtlplainloaded \else
  \checkfont{eurm10}
  \iffontfound
    \IfFileExists{upmath.sty}
      {\typeout{^^JFound AMS Euler Roman fonts on the system,
                   using the 'upmath' package.^^J}%
       \usepackage{upmath}}
      {\typeout{^^JFound AMS Euler Roman fonts on the system, but you
                   dont seem to have the}%
       \typeout{'upmath' package installed. JFM.cls can take advantage
                 of these fonts,^^Jif you use 'upmath' package.^^J}%
      }
  \else
  \fi
\fi

\ifCUPmtlplainloaded \else
  \checkfont{msam10}
  \iffontfound
    \IfFileExists{amssymb.sty}
      {\typeout{^^JFound AMS Symbol fonts on the system, using the
                'amssymb' package.^^J}%
       \usepackage{amssymb}%
         \let\leq=\leqslant
         \let\geq=\geqslant
      }{}
  \fi
\fi

\ifCUPmtlplainloaded \else
  \IfFileExists{amsbsy.sty}
    {\typeout{^^JFound the 'amsbsy' package on the system, using it.^^J}%
     \usepackage{amsbsy}}
    {\providecommand\boldsymbol[1]{\mbox{\boldmath $##1$}}}
\fi

\providecommand\upartial{\partial}

\providecommand\bnabla{\boldsymbol{\nabla}}
\providecommand\bcdot{\boldsymbol{\cdot}}

\newcommand\ex{\boldsymbol{\hat{x}}}
\newcommand\ey{\boldsymbol{\hat{y}}}
\newcommand\ez{\boldsymbol{\hat{z}}}

\newcommand\ek{\boldsymbol{k}}

\newcommand\bs{\boldsymbol}

\newcommand\lf{\left }
\newcommand\rg{\right }

\newcommand\op{\mathrm}

\newcommand\Rey{\mbox{\textit{Re}}}  

\newsavebox{\astrutbox}
\sbox{\astrutbox}{\rule[-5pt]{0pt}{20pt}}

\newcommand\bq{\mbox{\boldmath $q$}}
\newcommand\bu{\mbox{\boldmath $u$}}
\newcommand\bx{\mbox{\boldmath $x$}}

\newcommand\bOmega{{\bf \Omega}}
\newcommand\z{{\bf \hat{z}}}
\newcommand\btimes{\mbox{\bf $\times$}}
\newcommand\beq{\begin{equation}}
\newcommand\eeq{\end{equation}}
\newcommand\beqa{\begin{eqnarray}}
\newcommand\eeqa{\end{eqnarray}}
\newcommand{\half}{\mbox{$\frac12$}}

\newcommand{\R}{{\cal R}}

\newcommand{\K}{{\cal K}}
\newcommand{\T}{\mathsfi {T}} 

\title[2D Kolmogorov Turbulence]{Simple invariant solutions embedded in 
2D Kolmogorov turbulence}
\author[Gary J. Chandler and Rich R. Kerswell]
{G\ls A\ls R\ls Y\ns J.\ns C\ls H\ls A\ls N\ls D\ls L\ls E\ls R$^1$ \and 
R\ls I\ls C\ls H\ns R.\ns K\ls E\ls R\ls S\ls W\ls E\ls L\ls L$^1$}

\affiliation{$^1$School of Mathematics, University of Bristol,\\
University walk, Bristol, UK}

\pubyear{xxxx}
\volume{xxx}
\pagerange{xxx--xxx}
\date{?? and in revised form ??}

\begin{document}
\maketitle

\begin{abstract}
We consider long simulations of 2D Kolmogorov turbulence body-forced
by $\sin4y \ex$ on the torus $(x,y) \in [0,2\pi]^2$ with the purpose
of extracting simple invariant sets or `exact recurrent flows'
embedded in this turbulence. Each recurrent flow represents a
sustained closed cycle of dynamical processes which underpins the
turbulence. These are used to reconstruct the turbulence statistics in
the spirit of Periodic Orbit Theory derived for certain types of low
dimensional chaos. The approach is found to be reasonably successful
at a low value of the forcing where the flow is close to but not fully
in its asymptotic (strongly) turbulent regime. Here, a total of 50
recurrent flows are found with the majority buried in the part of
phase space most populated by the turbulence giving rise to a good
reproduction of the energy and dissipation probability density
functions.  However, at higher forcing amplitudes now in the
asymptotic turbulent regime, the generated turbulence data set proves
insufficiently long to yield enough recurrent flows to make viable
predictions. Despite this, the general approach seems promising
providing enough simulation data is available since it is open to
extensive automation and naturally generates dynamically important
exact solutions for the flow.

\end{abstract}

\section{Introduction}
Ideas from dynamical systems have recently provided fresh insight into
transitional and weak turbulent flows where the system size is smaller
than the spatial correlation length. Viewing such flows as a
trajectory through a phase space littered with invariant (`exact')
solutions and their stable and unstable manifolds has proved a
fruitful way of understanding such flows (Eckhardt et al. 2002,
Kerswell 2005, Eckhardt et al. 2007, Gibson et al. 2008, Cvitanovic \&
Gibson 2010, Kawahara et al. 2012). It is therefore natural to ask
whether any ideas attempting to rationalise chaos may have something
to say about developed turbulence. This is not to presuppose the two
phenomena are simply related - that they are not has surely been
appreciated for over 30 years - but merely an approach found useful in
one may provide some insight into the other.  One promising line of
thinking in low-dimensional, hyperbolic dynamical systems stands out
as a possibility - Periodic Orbit Theory.

%
%
The study of periodic orbits as a tool to understand chaos has been a
longstanding theme in dynamical systems dating back to Poincar\'{e}'s
original work on the three body problem in the 1880s (Poincar\'{e}
1892; Ruelle 1978, Eckmann \& Ruelle 1985, MacKay \& Miess 1987) The
fact that chaotic solutions can fleetingly, but also recurringly,
resemble different periodic flows over time has always suggested that
the statistics of the former may be expressible as a weighted sum of
properties of the latter. However, this has generally remained a vague
hope except for a special subclass of dynamical system where Periodic
Orbit Theory has formalised this link (Auerbach et al. 1987,
Cvitanovic 1988, Artuso et al. 1990a and for a recent review, Lan
2010). For these systems - very low dimensional, fully hyperbolic
invariant sets in which periodic orbits are dense (`axiom A'
attractors) - there have been some notable successes (e.g. Artuso et
al. 1990b, Cvitanovic 1992 and later papers in the same journal issue,
see also the evolving webbook Cvitanovic et al 2005). Here the
invariant measure across the attractor can be expressed in terms of
the periodic orbits which are dense within it so that ergodic averages
can be determined from suitably weighted sums across the periodic
orbits. Central to applying the approach is identifying a symbolic
dynamics which can catalogue and order the infinite periodic orbits
present in a chaotic attractor to give convergent expressions.

%
%
Extending Periodic Orbit Theory to higher dimensional dynamical
systems - most notably spatiotemporal systems - would obviously be
highly desirable but represents a very considerable challenge.
However, there are encouraging signs that something approaching this
could be possible in fluid turbulence. The fact that a turbulent flow
fleetingly yet recurringly resembles a series of smoother coherent
structures or spatiotemporal patterns is a familiar observation
perhaps first recorded by Leonardo da Vinci in his famous drawings and
made mathematically by Hopf (Hopf 1948: see Robinson 1991, Holmes et
al. 1996 and Panton 1997 for overviews of subsequent work). Hopf's
vision of turbulence was of a flow exploring a repertoire of distinct
spatiotemporal patterns where the implication was that these patterns
were simple invariant solutions of the governing equations
(e.g. equilibria, periodic orbits, tori etc - hereafter also referred
to as {\em recurrent flows}). This viewpoint then advocates a
dynamical systems approach even for fully turbulent flows.  However,
an attempt to build a prediction of turbulence statistics from the
recurrent flows present is fraught with difficulties. Not only is
there the daunting problem of initially identifying enough of them in
such high dimensional systems (typically $10^4$-$10^5$ degrees of
freedom) to make such a prediction seem feasible, but there is the
problem of understanding how each should be weighted in any
expansion. Finally, in the very likely eventuality that there is no
symbolic dynamics for turbulence, it is impossible to know if
important recurrent flows have been missed thereby compromising any
prediction.

%
%

The situation although very difficult, promises much and is not
without hope.  Efforts to extend the ideas of Periodic Orbit Theory to
higher dimensional systems have focussed on 1-space and 1-time partial
differential equations, most notably the 1-dimensional
Kuramoto-Sivashinsky system (Christiansen et al. 1997, Zoldi \&
Greenside 1998, Lan \& Cvitanovic 2008, Cvitanovic et al. 2010) and
the complex Ginzburg-Landau equation (Lopez et al. 2005). The emphasis
in this work has mostly been to establish the feasibility of
extracting recurrent flows directly from the `turbulent' dynamics
although some tentative predictions were made (Christiansen et
al. 1997, Lopez et al. 2005). The first attempt to extract a recurrent
flow from 3 dimensional Navier-Stokes turbulence was made in a
landmark calculation by Kawahara \& Kida in 2001. In this work they
managed to find one periodic orbit embedded in the turbulent attractor
in a 15,422 degree-of-freedom (d.o.f.) simulation of small box plane
Couette flow.  This immediately raised the `bar' of what had been
thought possible and interestingly, they found that this {\em one}
orbit was a very good proxy for their turbulence statistics. Van Veen
et al. (2006) drew a similar conclusion albeit after discarding all
but one of the few orbits they found when studying highly symmetric 3D
body-forced box turbulence. Subsequent work in plane Couette flow by
Viswanath (2007) essentially confirmed the existence of Kawahara \&
Kida's (2001) periodic orbit (using 180,670 d.o.f.), found another and
identified 4 new {\em relative} periodic orbits (see also Lopez et
al. 2005). These are periodic orbits where the flow repeats in time
but drifts spatially in directions where the system has a continuous
translational symmetry. Cvitanovic \& Gibson (2010) report (using
61,506 d.o.f.) having identified 40 periodic solutions, 15 relative
periodic solutions with streamwise shifts and one relative periodic
orbit with a small spanwise shift in low Reynolds number and small box
plane Couette flow.

%
%

The state of the field is then that recurrent flows can be found in 3
dimensional Navier-Stokes turbulence calculations requiring up to
O($10^5$) d.o.f.  (weak turbulence at low Reynolds numbers) but
understanding how many can be found in a reasonable (tolerable) time and
then identifying how dynamically important they are, remain
outstanding issues. As a result, making useful predictions with any
confidence using the set of recurrent motions found seems some way
off. With this background, our objective here is to make some
contribution to this effort by mounting a systematic investigation of
the issues in the simpler context of 2-dimensional Navier-Stokes
turbulence. 

It's worth emphasizing that even if a `turbulence' version
of Periodic Orbit Theory ultimately proves beyond our grasp, the
procedure of identifying different recurrent flows buried within a
turbulent solution has considerable value in its own right. This is
because each recurrent flow can be thought of as a sustainable
dynamical process which helps underpin the turbulent state. Since they
are `closed' (recur exactly), their spatial and temporal structure
can be dissected to reveal the fundamental physics involved. Just such
an approach has helped uncover the `self-sustaining process' ( Waleffe
1997) - streamwise vortices generate streaks which are unstable to
streamwise-dependent flows which subsequently invigorate the streamwise
vortices - in wall-bounded shear flows following the discovery of a
quasi-cycle in highly constrained plane Couette flow by Hamilton, Kim
\& Waleffe (1995).  Beautifully, this quasi-cycle turned out to
indicate the presence of families of exact (unstable) travelling wave
solutions to the Navier-Stokes equations (Waleffe 1997), the existence
of which have revolutionized our thinking in transitional and weakly
turbulent shear flows (see the reviews by Kerswell 2005, Eckhardt et
al. 2007 and Kawahara et al. 2012).

%
%

The specific framework investigated here is 2-dimensional Kolmogorov flow
on a $[0,2\pi]^2$ torus (efficiently simulated using spectral
methods) where the flow is forced monochromatically and steadily at a
large length scale. This flow has been extensively studied since
Kolmogorov introduced the model in 1959 (Arnold \& Meshalkin 1960) as
a simple example of linear instability which could be studied
analytically (Meshalkin \& Sinai 1961). The flow has many possible
variations: torus aspect ratio (e.g. Marchioro 1986, Okamoto \& Shoji
1993, Sarris et al. 2007), forcing wavelength (She 1988, Platt et
al. 1991, Armbruster et al. 1996), forcing form (e.g. Gotoh \& Yamada
1986, Kim \& Okamoto 2003), and 3-dimensionalisation (e.g. Borue \&
Orszag 1996, Shebalin \& Woodruff 1997, Sarris et al. 2007). It has
been experimentally realised using magnetohydrodynamic forcing
(e.g. Bondarenko et al 1979, Obukhov 1983, Sommeria 1986) and latterly
in soap films (e.g. Burgess et al. 1999). With an additional Coriolis
term, Kolmogorov flow can also be used as a barotropic ocean model on
the $\beta$-plane (e.g. Kazantsev 1998, 2001 and Tsang \& Young 2008).

The work by Kazantsev (1998, 2001) is particularly relevant for this
study as this made the first attempt to apply Periodic Orbit Theory in
a 211 d.o.f. discretization of a 2D Kolmogorov-like flow (differences
include the addition of non-periodic boundary conditions, rotation and
bottom friction).  The work is most notable for his use of a
minimisation procedure to identify periodic orbits (59 found) as well
as a good survey of relevant atmospheric literature.
More recent work by Fazendeiro et al. (2010) (see also Boghosian et
al. 2011) has started to study triply-periodic body-forced turbulence
using Lattice-Boltzmann computations. Their focus was on developing
another variational approach for identifying periodic orbits based
upon the idea of Lan \& Cvitanovic (2004) and they describe
convergence evidence for 2 periodic orbits. The approach starts with a
closed orbit that does not satisfy the Navier-Stokes equations and
uses a variational method to  adjust the orbit until it
does. This requires manipulating the whole orbit at once and
requires massive computations which are facilitated by the inherent
parallelism of the Lattice-Boltzmann approach.
In contrast, the approach adopted here is to start with a flow
trajectory which {\em does} satisfy the Navier-Stokes equations but
is not closed and to adjust the start of the the trajectory until it
does. This boils down to a Newton-Raphson root search in very high
dimensions and iterative methods have to be employed to make things
feasible. We adopt a {\em Newton-GMRES-Hookstep} procedure developed
by Viswanath (2007,2009) and subsequently used with success by
Cvitanovic \& Gibson (2010) (see Duguet et al. 2008 for a slight
variation which replaces the `Hook step' with the `Double Dogleg'
step; Dennis \& Schnabel 1996).

The plan of the paper is as follows. Section 2 describes 2D Kolmogorov
flow in detail, discusses its symmetries (\S2.1) and makes connections
with some previous direct numerical simulations (DNS) (\S2.2).  Key
flow measures to be used subsequently are listed in \S2.3.  Section 3
describes the methodology used starting with the time-stepping code in
\S3.1, how initial guesses for recurrent flows are identified in
\S3.2, and then the Newton-GMRES-Hookstep algorithm in \S3.3 (this
draws its inspiration from Viswanath (2009)).  \S3.4 discusses how the
algorithms were tested. Section 4 describes the results, first giving
a flow orientation in \S4.1, then reporting on how recurrent flows
were actually extracted, before giving details of the recurrent flows
found in \S 4.3. Section 5 describes an attempt to reproduce
properties of 2D Kolmogorov turbulence before section 6 discusses the
results and the outlook for future work.

%
\section{Formulation}
%

\noindent
The incompressible
Navier--Stokes equations with what is called `Kolmogorov forcing' is
\beqa
\frac{\partial \bu^*}{\partial t^*}
+ \bu^* \bcdot \bnabla^* \bu^*+\frac{1}{\rho} \bnabla^* p^* 
&=& \nu \nabla^{*2} \bu^* + \chi \sin \lf(2 \pi ny^*/L_y \rg) \ex, \\
\bnabla^* \bcdot \bu^*=0
\eeqa
where $\rho$ is the density, $\nu$ the kinematic viscosity, $n$ an
integer describing the scale of the (monochromatic) Kolmogorov forcing
and $\chi$ is the forcing amplitude per unit mass of fluid over a
doubly-periodic domain $[0,L_x] \times [0,L_y]$ (and in this
section only $*$ indicates a dimensional quantity). The system is
non-dimensionalised by the lengthscale $L_y/2 \pi$ and timescale
$\sqrt{L_y /2 \pi \chi}$ so that the equations become
\beqa
\frac{\partial \bu}{\partial t} + \bu \bcdot \bnabla \bu+\bnabla p &=&
\frac{1}{Re} \nabla^2 \bu + \sin \lf( n y \rg) \ex, \\ \bnabla \bcdot
\bu=0 
\eeqa 
where the Reynolds number is 
\beq 
\Rey :=\frac{\sqrt{\chi}}{\nu}\left( \frac{L_y}{2 \pi} \right)^{3/2}
\eeq
to be solved over the domain $[0,2\pi/\alpha] \times
[0,2\pi]$ ($\alpha:=L_y/L_x$).  Given the doubly-periodic boundary
conditions, dealing with the cross-plane vorticity equation is more
natural and reduces simply to the scalar equation
\beq 
\frac{\partial \omega}{\partial t}=\z \bcdot \bnabla
\btimes (\bu \btimes \omega \z)+ \frac{1}{Re} \nabla^2 \omega - n\cos \lf( n y \rg)  
\label{governing}
\eeq 
where $\omega \ez :=\bnabla \btimes \bu$.
(The form of the nonlinearity on the RHS is convenient for computation
but can be further reduced to simply $-\bu \bcdot \bnabla \omega$ as
the vortex stretching term $\boldmath{\omega}.\bnabla \bu=0$ is, of
course, absent in 2D.)  Dealing with this equation is analogous to
working with the streamfunction $\bu=\bnabla \btimes \psi(x,y)\ez$
since spatially-constant velocity and vorticity fields are not present
so $\psi=\nabla^{-2}\omega$.

\subsection{Symmetries}\label{symmetries}

There is a shift-\&-reflect symmetry \beq {\cal S}:[u,v](x,y)
\rightarrow [-u,v](-x,y+\displaystyle{\pi \over n}).  \eeq which
shifts half a wavelength of the forcing function in $y$ and reflects
in $x$. Since there are $n$ wavelengths in the domain, this
transformation forms a cyclic group of order $2n-1$. There is also a
rotation-through-$\pi$ symmetry \beq {\cal R}:[u,v](x,y) \rightarrow
[-u,-v](-x,-y) \eeq and the continous group of translations \beq {\cal
  T}_l:[u,v](x,y) \rightarrow [u,v](x+l,y) \qquad {\rm for} \quad 0
\leq l < {2\pi \over \alpha}.  \eeq The focus here is (unusually) not
to take advantage of these, that is, the flow is allowed to fully
explore phase space.

\subsection{Past literature}

Of all the previous work on 2D Kolmogorov flow, Platt et al. (1991)
seem to have carried out the most detailed study with $n=4$ over the
non-dimensional domain $[0,2\pi] \times [0, 2\pi]$. The same choices
$n=4$ and $\alpha=1$ were therefore made throughout the calculations
reported here. With this, $\Rey = 8 \sqrt{\Rey^{Platt}}$ and so the
critical Reynolds number for linear instability is $\Rey_{c} = 8
\sqrt[4]{2}$ ($Re_c^{Platt}=\sqrt{2}$). Platt et al. (1991) looked at
the flow regime $\Rey/\Rey_c \leq 3.54$ over a $32 \times 32$ spatial
grid so that $9.51 \leq \Rey \leq 33.6$. Here we consider a $256
\times 256$ grid and look at $9.5\leq \Rey \leq 100$ ($\Rey/\Rey_c
\leq 10.5$). Unfortunately, we were only able to confirm the detailed
dynamics reported by Platt et al. (1991) if we reduced our resolution
down to theirs.

The next closest study was She's (1988) which took $n=8$, a $64 \times
64$ grid and examined $\sqrt{2} \leq \Rey_{She} \leq 30$ ($26.9 \leq
\Rey \leq 123.9$ as $\Rey = 8^{3/2} \sqrt{\Rey_{She}}$) which
corresponds to $\Rey/\Rey_c \leq 4.6$. More recently, Sarris et
al. (2007) considered 3D Kolmogorov flow over a variety of box aspect
ratios considering $65 \leq Re^{Sarris} \leq 180$ including the
3-dimensionalisation of the flow considered here (then
$Re=8Re^{Sarris}$). Typically they use 128 mesh points per wavelength
of the forcing. At the time of writing, the world record for resolution when
simulating doubly-periodic body-forced turbulence seems to be
$32,768^2$ (Boffetta \& Musacchio 2010).

%
%
\subsection{Key measures of the flow}

Key measures of the flow which will aid the subsequent discussion are
as  follows ($\bu=u \ex +v \ey$): the mean flow,
\beq
\overline{u}(y)  := \langle \bu \cdot \ex \rangle_{t,x},
\eeq
(initial conditions are such that $\langle \bu(\bx,0) \cdot \ey
\rangle_{x}=0$ so that  $\langle \bu(\bx,t) \cdot \ey
\rangle_{x}=0$ for all time ); the bulk mean
square of the fluctuations around the mean,
\beq
\hat{u}_{rms}^2(t) := \langle (u-\overline{u})^2 \rangle_{V},\qquad
\hat{v}_{rms}^2(t) := \langle v^2 \rangle_{V},
\eeq
and root mean square of the fluctuations as a function of $y$,
\beq
u_{rms}(y) := \sqrt{ \langle (u-\overline{u})^2 \rangle_{t,x}},\qquad
v_{rms}(y) := \sqrt{ \langle v^2 \rangle_{t,x}}\,;
\label{fluctuation}
\eeq
the total kinetic energy and the kinetic energy of the fluctuation field
\beq
E(t)       :=  \half \langle \bu^2 \rangle_{V},\qquad
E_t(t)     :=  \half \langle (\bu-\langle \bu\rangle_{t,x} )^2\rangle_{V};
\eeq
the total dissipation rate and the instantaneous power input
\beq
D(t)       :=  \frac{1}{Re}\langle |\bnabla \bu|^2 \rangle_{V},\qquad
I(t)       := \langle \, u \sin(ny)\, \rangle_V, 
\eeq
with finally the laminar state, bulk laminar kinetic energy and
bulk dissipation rate
\beq
\bu_{lam}:=\frac{Re}{n^2} \sin ny, \ex, \quad
\quad E_{lam} :=  \frac{Re^2}{4n^4},\qquad
\quad D_{lam} :=  \frac{Re}{2n^2},
\label{base}
\eeq
where the various averagings are defined as 
\beqa
\langle \quad \rangle_V &:=& \frac{\alpha}{4 \pi^2} \int^{2
  \pi/\alpha}_0 \int^{2 \pi}_0 \quad dx dy, \nonumber \\
\langle \quad \rangle_x &:=&
\frac{\alpha}{2 \pi} \int^{2\pi/\alpha}_0 \quad dx, \nonumber \\ 
\langle \quad \rangle_t &:=&
\lim_{T \rightarrow \infty} \frac{1}{T} \int^{T}_0 \quad dt. \nonumber
\eeqa

\section{Methodology}

\subsection{Time Stepping Code}

A 2D fully de-aliased pseudospectral code was used as developed in
Bartello \& Warn (1996). The original leapfrog+filter approach was
replaced by the Crank-Nicolson method for the viscous terms and Heun's
method (Euler predictor method) for the nonlinear and forcing terms so
that only 1 state vector was required to accurately restart the code.
This together with a constant time step size (except for the last
step) means that the discretised flow is a dynamical system which
closely matches the Navier-Stokes flow.  Specifically, if
$\Omega(\ek)=\texttt{fft}(\omega(\bx))$ is the Fourier transform of
$\omega$ with $\ek=(k_x,k_y)$, the vorticity equation
(\ref{governing}) in spectral space is
\begin{gather}
\frac{\upartial \Omega }{\upartial t} 
=   - G \Omega  
+ f(\Omega) \\
{\rm with} \quad G(k_x,k_y) := \frac{k_x^2 + k_y^2}{ \Rey } = \frac{|\bs{k}|^2}{ \Rey }
\\
{\rm and} \quad f(\Omega) :=  -\op{i} \lf( k_x \, \texttt{fft}[ u  \omega ] 
+ k_y \, \texttt{fft}[ v  \omega ]\rg)   
- \frac{n}{2} \delta_{k_x} \delta_{(|k_y|-n)}.
\end{gather}
Here $\delta_i$ is the Kronecker delta function and takes the value $1$
when $i=0$ and $0$ otherwise.
A time step is performed by solving:
\begin{gather}
\frac{\widetilde{\Omega}^{i+1} - \Omega^{i}}{\Delta t} = 
- \frac{G}{2} \lf( \widetilde{\Omega}^{i+1} + \Omega^{i} \rg)
+  f(\Omega^{i})
\end{gather}
followed by solving:
\begin{gather}
\frac{\Omega^{i+1} - \Omega^{i}}{\Delta t} = 
- \underbrace{ \frac{G}{2} \lf( \Omega^{i+1} + \Omega^{i} \rg)
}_{\text{C-N}}
+ \underbrace{ \frac{1}{2} \lf( f(\widetilde{\Omega}^{i+1}) 
+ f(\Omega^{i}) \rg) }_{\text{Heun}}
\end{gather}
where the superscript is a time step index. With de-aliasing, a
resolution of $N_x \times N_y$ corresponds in practice to the
vorticity representation
\beq 
\omega(x,y,t)=\sum^{N_x/3-1}_{j=-N_x/3} \sum^{N_y/3-1}_{l=-N_y/3}
\Omega_{jl}(t) e^{i(\alpha j x+l y)} 
\label{omega}
\eeq 
where $\ek=(\alpha j,l)$, $\Omega_{00}=0$ and a mask is employed so
that $\Omega_{jl}=0$ for wavenumbers outside a specified domain
$\Sigma$. Calculations reported here have $\alpha=1$, $n=4$ and
$N_x=N_y$ so $\Sigma:=\{(j,l): j^2+l^2 \leq (N_x/3)^2\}$ is used. The
number of active (real) degrees of freedom is therefore $\approx \pi
N_x^2/9$ which is $\approx 22,800$ (or exactly $22,428$) for the
$N_x=256$ used here ($0\leq j \leq 85$ \& $-85 \leq l \leq 85$ since
$\Omega(-j,-l)=\Omega^*(j,l)$).

\subsection{Near Recurrences}

\begin{figure}
\centerline{\includegraphics[scale=0.75]{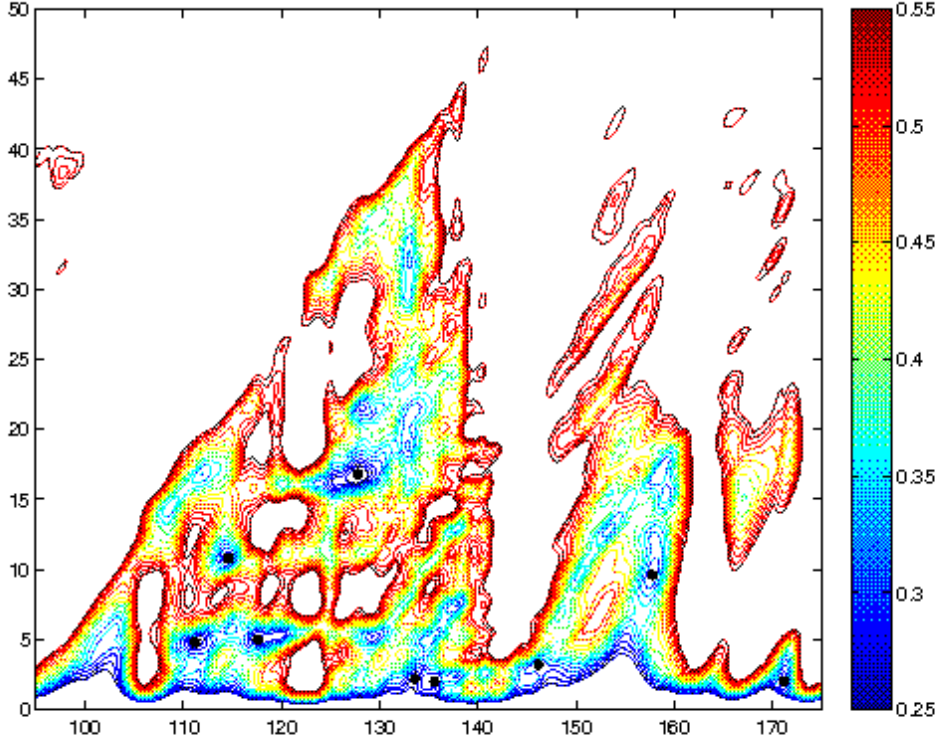}}
\caption{(Low resolution version for the arXiv) $R(t,T)$ at $Re=40$ contoured over $t
  \in [95,175]$ ($x$-axis) and $T \in [thres,50]$ ($y$-axis). The 9
  black dots are the guesses identified by the code ($R<R_{thres}$)
  over this time interval. All except one (the last dot at $t \approx
  171$) subsequently converged to exact solutions - the 4 dots for
  $t<130$ to a periodic orbit with period 5.3807 ($P1$ in Table 2) and
  the next 4 dots with $t \in [130,160]$ to a TW ($T1$ in Table 2)
  with phase speed $c=0.0198$ ($R$ values above 0.55 are not
  drawn/coloured for clarity.}
\label{R_40}
\end{figure}

The key idea pursued here is to extract recurrent flows directly from
the turbulent DNS data with the implication that they are clearly
dynamically important.  With this in mind, the time stepping code was
run for $10^5$ time units starting from random initial conditions and
`near recurrences' of the flow field searched for. These near
recurrences were defined as episodes where
%
%
%
\beq
\omega(x+s,y+\half \pi m,t+T)=\omega(x,y,t)
\label{recurrence}
\eeq
`approximately' holds for some choice of the continuous shift $0 \leq
s < 2 \pi$, the discrete shift $m \in {0,1,2,\dots,n-1}$ and $T>0$
over $0 \leq x,y < 2\pi$.  Periodic orbits correspond to $s=m=0$ and
some period $T>0$, travelling waves (TWs) to $m=0$ and $s=cT$ with $T>0$
free where $c$ is the phase speed, equilibria have $s=m=0$ and $T$
free and relative periodic orbits have one or both of $s$ and $m$
not equal to zero with period $T>0$. (The possible existence of
relative periodic orbits - permitted by the inclusion of 2 free
parameters ($s$ and $m$) - is, of course, a reflection of the discrete
and continuous translational symmetries of the system). The key is
understanding what `approximately' means, that is, how close should
(\ref{recurrence}) be to holding for it to signify the presence of a
recurrent flow structure nearby. The only way to answer this seems to
be to do computations and experiment.  The search for near
recurrences was done most efficiently by calculating every, say t=0.1
or 0.2 steps, the normalised difference between states in wavenumber
space suitably minimised over continuous shifts in $x$ and discrete
shifts in $y$ as follows
\beq 
R(t,T):=\min_{0\leq s < 2\pi} \min_{m \in {0,1,2,..., n-1}} 
\frac{\sum_j \sum_l | \Omega_{jl}(t)e^{i \alpha j
  s+2iml \pi/n}-\Omega_{jl}(t-T)|^2}
{\sum_j \sum_l | \Omega_{jl}(t)|^2}
\eeq
where $\sum_j \sum_l |\Omega_{jl}|^2=\alpha/(4 \pi)^2\int^{2\pi/x}_0
\int^{2 \pi}_0 \omega^2 dxdy$, $0< T_{thres} \approx 0.5 < T <
100$. Since $R(t,0)=0$ and $dR(t,0)/dT>0$, the offset $T_{thres}$ is
defined adaptively as the first time at which $dR(t,T_{thres})/dT
<0$. Figure \ref{R_40} is a typical example of how $R(t,T)$ looks as a
function of $t$ and $T$ during a recurrent episode.  The 9 black dots
are the guesses identified by the code ($R<R_{thres}=0.3$) over this
time interval. All except one (the last dot at $t \approx 171$)
subsequently converged to an exactly recurrent solution (the 4 dots
for $t<130$ to a periodic orbit ($P1$ in Table 2) with period 5.3807
and the next 4 dots with $t \in [130,160]$ to a TW ($T1$ in Table 2)
with phase speed $c=0.0198$). The threshold $R_{thres}$ was chosen
judiciously to give enough good quality guesses.

%
%
%

%
\subsection{UPO extraction method: Newton-GMRES-Hookstep}

Once a near-recurrence has been found by the above stated criterion,
we then attempted to find if an exact recurrent flow was lurking
nearby in phase space. This required a high-dimensional root finding
algorithm acting on a state vector which completely specifies the
velocity field
\beq
\bx=\left[\begin{array}{c} \bOmega\\ s\\ T \end{array} \right]
\eeq
and contains information about the potential recurrence ($\bOmega$ is
a vector containing the scalars $\Omega_{jl}$ arranged in some
fashion). The shift $s$ is included since it can be adjusted
continuously whereas the discrete shift $m$ cannot and therefore is
pre-set.  To set up the Newton-Raphson algorithm (and we follow the
excellent description in Viswanath 2009), it is convenient to define
the infinitesimal generators $\T_x$ and $\T_y$ of translations in $x$
and $y$ respectively
\beqa
\T_x\, \omega(x,y,t) &\rightarrow& 
\frac{\partial \omega}{\partial x}=
\sum^{N_y/3-1}_{-N_y/3} \sum^{N_x/3-1}_{-N_x/3}
i \alpha j \Omega_{jl}(t) e^{i(\alpha j x+l y)},  \nonumber \\
\T_y\, \omega(x,y,t) &\rightarrow& 
\frac{\partial \omega}{\partial y}=
\sum^{N_y/3-1}_{-N_y/3} \sum^{N_x/3-1}_{-N_x/3}
i l \Omega_{jl}(t) e^{i(\alpha j x+l y)}  
\nonumber 
\eeqa
as they act in spectral space
\beq
\T_x \, \bOmega \rightarrow  {\bf \Omega_x}  \quad {\rm and} \quad 
\T_y \, \bOmega  \rightarrow  {\bf \Omega_y} 
\eeq
where each element $\Omega_{jl}$ of $\bOmega$ is mapped to $i \alpha
j\Omega_{jl}$ in ${\bf \Omega_x}$ and $il\Omega_{jl}$ in ${\bf
  \Omega_y}$. Then, in
spectral space the recurrence condition (\ref{recurrence}) becomes
\beq 
{\bf F}(\bOmega_0,s,T;m):=\exp(s\T_x+\half \pi m \T_y)
{\bf \hat{\Omega}}(\bOmega_0,T)-\bOmega_0={\bf 0} 
\eeq
where $\bOmega_0=\bOmega(t)$ and ${\bf \hat{\Omega}}=\bOmega(t+T)$.
If $\bx_0=(\bOmega_0,s_0,T_0)^T$ is an initial guess for a solution,
then a better (next) guess
$\bx_0+\delta \bx_0=(\bOmega_0+\delta \bOmega, s_0+\delta s,
T_0+\delta T)^T$ is given by 
\beq
\frac{\partial {\bf F}}{\partial \bOmega_0} {\bf \delta} \bOmega+
\frac{\partial {\bf F}}{\partial s} \delta s +
\frac{\partial {\bf F}}{\partial T} \delta T = -{\bf F}(\bOmega_0,s_0,T_0;m)
\eeq
These are  $\dim(\bOmega)$ equations for $\dim(\bOmega)+2$
unknowns.  The extra two equations come from removing the degeneracy
associated with these translational symmetries (the system is
invariant under $(x,t) \rightarrow (x+s,t+T)$). This can be done by
imposing that ${\bf \delta} \bOmega$, has no component which shifts
the solution infinitesimally in the $x$-direction or the $t-$direction
(i.e. just redefines the time origin of the flow). The Newton-Raphson
problem is then to solve
\beq
\left[\begin{array}{c|c|c}
\ddots \hspace*{2cm} & \vspace{0.0cm} \vdots & \vspace{0.0cm} \vdots\\
\hspace*{1cm} {\displaystyle \frac{\partial {\bf \hat{\Omega}_s}}{\partial \bOmega_0}}-{\bf I}   
\hspace*{1cm} & \T_x {\bf \hat{\Omega}_s} &
{\displaystyle \frac{\partial {\bf 
 \hat{\Omega}_s}}{\partial T}}\\
\hspace*{2cm}\ddots& \vdots & \vdots \\\hline
\cdots \quad (\T_x \bOmega_0)^T \quad \cdots & 0 & 0\\\hline
\cdots \quad {\displaystyle \frac{\partial \bOmega_0}{\partial t}^T} \quad\cdots & 0 & 0
\end{array}
\right]
\left[
\begin{array}{c}
\vdots\\
\\
\vspace{0.25cm}\delta \bOmega\\
\vdots\\
\\
\delta s\\
\\
\\
\delta T
\end{array}
\right]
=-\left[ \begin{array}{c}
\vdots\\
\\
\vspace{0.25cm}{\bf F}(   {\bf \Omega}_0,s_0,T_0;m)\\
\vdots\\
\\
0\\
\\
\\
0
\end{array} \right]
\label{NR}
\eeq
where ${\bf \hat{\Omega}_s}:=\exp(s\T_x+\half \pi m \T_y) {\bf
  \hat{\Omega}}$ is the `back-shifted' final state and ${\bf I}$ is
the $\dim(\bOmega)\times \dim(\bOmega)$ identity matrix. This is now
in the standard form ${\bf A}{\boldsymbol{\delta} {\bf x}} ={\bf b}$
with only the Jacobian matrix $\partial {\bf \hat{\Omega}_s}/\partial
\bOmega$ not straightforward to evaluate ($\partial {\bf
  \hat{\Omega}_s}/\partial T$ and $\partial \bOmega_0/\partial t$ are
found by substituting ${\bf \hat{\Omega}_s}$ or $\bOmega_0$ into the
Navier-Stokes equations). 

Typically, the size of the matrix ${\bf A}$ is too large to store
explicitly let alone attempt to solve ${\bf A}{\boldsymbol{\delta}
  {\bf x}} ={\bf b}$ directly. As a result, the only way to proceed is
iteratively and GMRES (Saad \& Schultz 1986) is convenient (see the
excellent description by Trefethen \& Bau 1997). Here only the effect
of ${\bf A}$ on an arbitrary vector is needed. The effect of
the troublesome Jacobian can be handled easily by a
forward difference approach since
\beq
\frac{\partial {\bf \hat{\Omega}_s}}{\partial \bOmega_0} {\bf y} 
\approx
\frac{{\bf \hat{\Omega}_s}(\bOmega_0+\epsilon {\bf y})
-{\bf \hat{\Omega}_s}(\bOmega_0)}{\epsilon}
\label{action}
\eeq
where $\epsilon$ is chosen such that $|| \epsilon {\bf y}||=10^{-7}
||\bOmega_0||$ which balances truncation error with round-off error
using double precision arithmetic and $||\cdot ||$ is the Euclidean
norm (using a more-physically orientated norm is clearly an
interesting direction awaiting exploration). 

Straight Newton-GMRES is typically not good enough as guesses are
usually not in the region where linearisation holds sufficiently well
and divergence to infinity is commonplace. Instead it proves useful to
modify the approach to incorporate a trust region.  Following
Viswanath (2007, 2009), we use the `hook-step' method (Dennis \&
Schnabel 1996 \S 6.4.1) which can be easily built on top of the GMRES
process. Exactly how the approach is implemented can vary and we adopt
what looks to be slightly different algorithm to Viswanath (2007,
2009) in which GMRES is used first to derive an approximate solution
to ${\bf A}{\boldsymbol{\delta} {\bf x}} ={\bf b}$ before this
`solution' ${\boldsymbol{\delta}{\bf x}}$ is moved into the trust
region. The advantage of this is there is a clear convergence
criterion that can be imposed to terminate the initial GMRES
algorithm. Before stating this, it's worth first briefly describing
the GMRES algorithm itself which is based upon a simple idea. The
GMRES algorithm for solving ${\bf A}{\boldsymbol{\delta} {\bf x}}
={\bf b}$ at iteration $n$ approximates ${\boldsymbol{\delta}}{\bf x}$
by the vector ${\boldsymbol {\delta}} {\bf x}_n$ in the Krylov space
${\cal K}_n:=\langle \, {\bf b}, {\bf A b}, {\bf A^2 b},\ldots, {\bf
  A^{n-1}b} \,\rangle$ that minimises the norm of the residual
\beq
\|{\bf A}{\boldsymbol{\delta} {\bf x}_n} -{\bf b}\|
\label{GMRES}
\eeq
(Trefethen \& Bau 1997).
For numerical stability, an orthonormal basis for ${\cal K}_n$ is
constructed using a Gram-Schmidt-style iteration as follows
\beq
\bq_1=\frac{{\bf b}}{\|{\bf b}\|}, \qquad \tilde{\bq}_{i+1}={\bf A}
\bq_i-\sum_{j=1}^{i}\bq_j \frac{\bq_j^T{\bf
    A}\bq_i}{\|\bq_j\|^2} \qquad  \bq_{i+1}=\tilde{\bq}_{i+1}/\|\tilde{\bq}_{i+1}\|
\qquad i \in {1\ldots,n-1}
\eeq
so that if ${\bf Q}_n$ is the matrix with columns
$\bq_1,\bq_2,\ldots,\bq_n$, then ${\bf A Q}_n={\bf Q}_{n+1} {\bf
  H}_{n+1,n}$ where ${\bf H}_{n+1,n}$ is the upper $(n+1) \times n$
left section of an upper Hessenberg matrix generated by the basis
orthonormalisation (Trefethen \& Bau 1997, p252). With this basis the
solution ${\boldsymbol{\delta}{\bf x}_n}={\bf Q}_n {\bf y}_n$ (where
${\bf y}_n$ is an $n$-vector) minimises $\|{\bf Q}_{n+1} {\bf
  H}_{n+1,n} {\bf y}_n-{\bf b} \|$ ($N$ equations and $n$ unknowns) or
equivalently $\|{\bf H}_{n+1,n} {\bf y}-\|{\bf b}\|{\bf \hat{e}_1}\|$
($n+1$ equations and $n$ unknowns) since the only non-zero entry in
${\bf Q}_{n+1}^T{\bf b}$ is the first entry. This can be accomplished
by a singular value decomposition (SVD) of ${\bf H}_{n+1,n}$ into
${\bf U}_{n+1}{\bf D}{\bf V}_n^T$ (where ${\bf U}_{n+1}$ and ${\bf
  V}_{n}$ are orthonormal matrices and ${\bf D}$ is an $(n+1) \times
n$ diagonal matrix with a zeroed bottom row) through straightforwardly
solving the first $n$ equations ${\bf D} {\bf z}_n ={\bf
  p}_{n+1}:=||{\bf b}||{\bf U}_{n+1}^T {\bf \hat{e}_1}$ followed by
${\bf V}^T {\bf y}_n={\bf z}_n$ and then ${\boldsymbol{\delta} {\bf
    x}_n}={\bf Q}_n {\bf y}_n$. The modulus of the remaining
unbalanced component $p_{n+1}(n+1)$ then gives the minimum value or
residual.  The iterations are continued until
\beq
\frac{\| {\bf A} {\boldsymbol{\delta} {\bf x}_n}-{\bf b}\|}{\|{\bf b}\|} = 
\frac{\| {\bf D} {\bf z}_n-{\bf p}_{n+1}\|}{\| {\bf p}_{n+1}\|}=
\frac{|p_{n+1}(n+1)|}{\|{\bf p}_{n+1}\|}\leq tol
\label{tol}
\eeq
where $tol$ is a small number typically chosen in the range
$10^{-4}$ to $10^{-2}$ (the majority of the computations reported here
were obtained using a value of $10^{-3}$). If $\|{\bf F}({\bf
  x}+{\boldsymbol{\delta} {\bf x}_n})\|$ is not smaller than $\|{\bf
  F}({\bf x})\|$ or more specifically not well predicted by the
linearisation around ${\bf x}$ i.e. $\|{\bf F}({\bf x})+{\bf
  A}{\boldsymbol{\delta} {\bf x}_n}\|$, then the approximate solution
of the linearised problem is transformed back to a smaller trust
region where the linearised problem {\em is} valid.  This is done by
adding the constraint $\|{\boldsymbol{\delta} {\bf x}_n}\| \leq
\Delta$ or equivalently $\|{\bf y}_n\| \leq \Delta$ to the GMRES
minimisation (\ref{GMRES}): this is the hook step. The beauty of this
adjustment is that it is a very natural modification of the GMRES
approximate solution since the (innermost) problem for ${\bf z_n}$ is
then
\beq
{\rm min} \|{\bf D} {\bf z}_n-{\bf p}_{n+1}\| \qquad {\rm s.t.} 
\quad \|{\bf y}_n\|=\|{\bf z_n}\| \leq \Delta
\eeq
Constructing the Lagrangian
\beq
{\cal L}:=({\bf D}{\bf z}_n-{\bf p}_{n+1})^2+\mu({\bf z}_n^2+\beta^2-\Delta^2)
\eeq
where $\mu$ is a Lagrange multiplier imposing the trust region
constraint, leads to the minimisation equations
\beqa
2d(i)(d(i)z_n(i)-p_{n+1}(i))+2\mu z_n(i) &=&0 \\
2 \mu \beta &=& 0\\
{\bf z}_n^2+\beta^2-\Delta^2 &=& 0
\eeqa
where $d_i$ is the $i$th diagonal element of ${\bf D}$.  The solution
to this is 
\beq
z_n(i):=\frac{p_{n+1}(i)d_i}{d_i^2+\mu} \qquad 1 \leq i\leq n
\eeq
with either $\mu=0$ as $\|{\bf z}_n\| < \Delta$ (the original GMRES
solution) or $\mu \neq 0$ chosen so that $\|{\bf z}_n\|=\Delta$ (in
practice $\mu$ is just increased until $\|{\bf z}_n\| < \Delta$).  An
acceptable solution ${\boldsymbol{\delta} {\bf x}_n}$ is signalled by
\beq
{\bf F}({\bf x}+{\boldsymbol {\delta}} {\bf x}_n) \leq 
{\bf F}({\bf x})+c {\bf A} {\boldsymbol {\delta}} {\bf x}_n
\eeq
where some value of $c \in (0,0.5)$ is chosen (eqn 6.4.14, Dennis \&
Schnabel 1996: we took the least demanding value of $c=0$). If this
does not hold, $\Delta$ is decreased and the hook step repeated until
it is. Depending on how easily this improvement condition is met, the
trust region may be relaxed (e.g. if linearisation holds well - ${\bf
  F}({\bf x}+{\boldsymbol {\delta}} {\bf x}_n) \approx {\bf F}({\bf
  x})+{\bf A} {\boldsymbol {\delta}} {\bf x}_n$) or not for subsequent
Newton steps. 

This algorithm can be readily extended to perform
solution branch continuation: see appendix A. Furthermore, since we
know how to calculate the action of the Jacobian on any vector (see
(\ref{action}), the linear stability of an exactly recurrent flow can
also be readily found using the Arnoldi technique (e.g. using ARPACK
to extract extremal eigenvalues).\\

\subsection{Testing}

The modified (Crank-Nicholson+Heun) time-stepping code was thoroughly
validated against the well-tested Leapfrog+filter code developed by
Bartello \& Warn (1996). The Newton-GMRES-Hookstep algorithm developed
on top of this was tested by attempting to converge onto a known
periodic orbit. This orbit was originally found by tracing
bifurcations up from the basic state. For $n=4$, the 1D basic state
becomes linearly unstable at $Re=9.9669$ for disturbances
$2\pi$-periodic in $x$ giving rise to a steady 2D state which is
${\cal R}$-symmetric. This state loses stability to a stable periodic
orbit within the ${\cal R}$-symmetric subspace for $30<Re<31$
before this orbit becomes unstable at $Re \gtrsim 32$ through a torus
bifurcation. The periodic orbit at $Re=31$ was easily found by
time-stepping within the ${\cal R}$-symmetric subspace yet is unstable
to ${\cal R}$-asymmetric disturbances in the full unrestricted space.
Having such an orbit to experiment with was invaluable for building up
confidence in the code and some feel for how the tolerances of the
algorithm should be set (e.g. $tol$ in (\ref{tol})).

%
\section{Results}
%

\begin{figure}
\centerline{\includegraphics[scale=0.6]{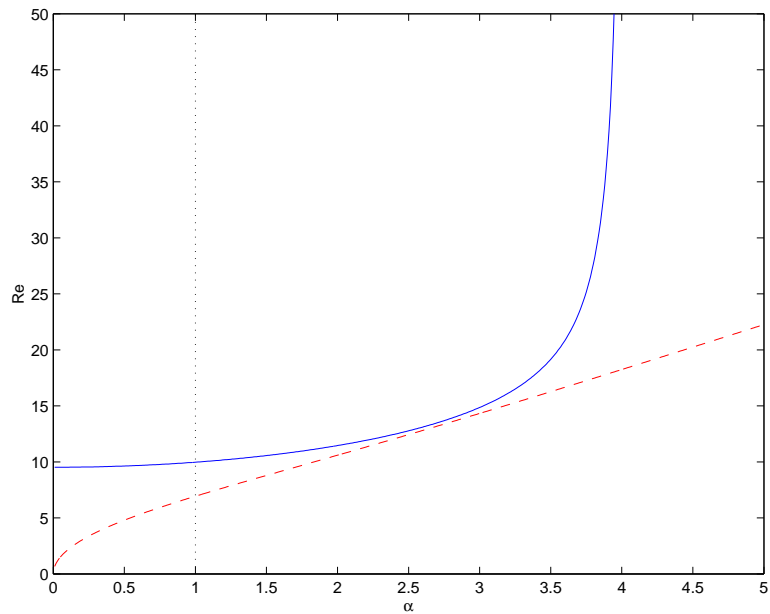}}
\caption{The energy stability Reynolds number $Re_E$ (red dashed line)
  and the linear instability Reynolds number $Re_{lin}$ (blue solid
  line) for $\sin 4y \ex$ forcing over the torus $[0,2\pi/\alpha]
  \times [0,2\pi]$. Note that $Re_{lin} \rightarrow \infty$ as $\alpha
  \rightarrow 4$ so the domain is squeezed down to $[0,\pi/2]\times
              [0, 2\pi]$. The dotted line at $\alpha=1$ is the present
              case ($Re_E=6.8297$ and $Re_{lin}=9.9669$). $Re_{E}
              \rightarrow 0$ and $Re_{lin} \rightarrow 8\sqrt[4]{2}
              \approx 9.5137$ as $\alpha \rightarrow 0$.}
\label{en_lin}
\end{figure}

%
%
\begin{figure}
\begin{center}\setlength{\unitlength}{1cm}
\begin{picture}(14,11)
\put(0,0){\epsfig{figure=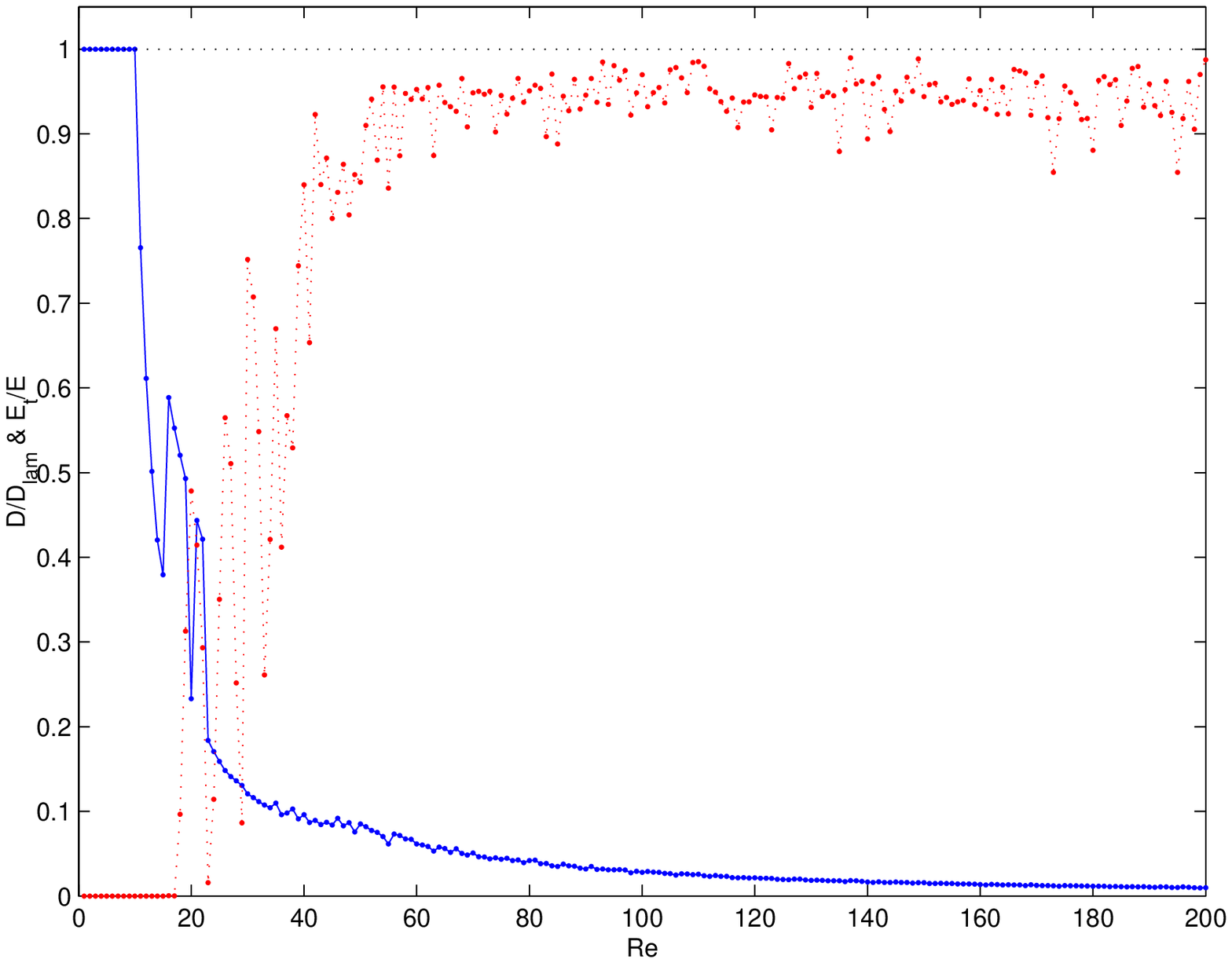,width=13cm,height=10cm,clip=true}}
\put(4.5,1.5){\epsfig{figure=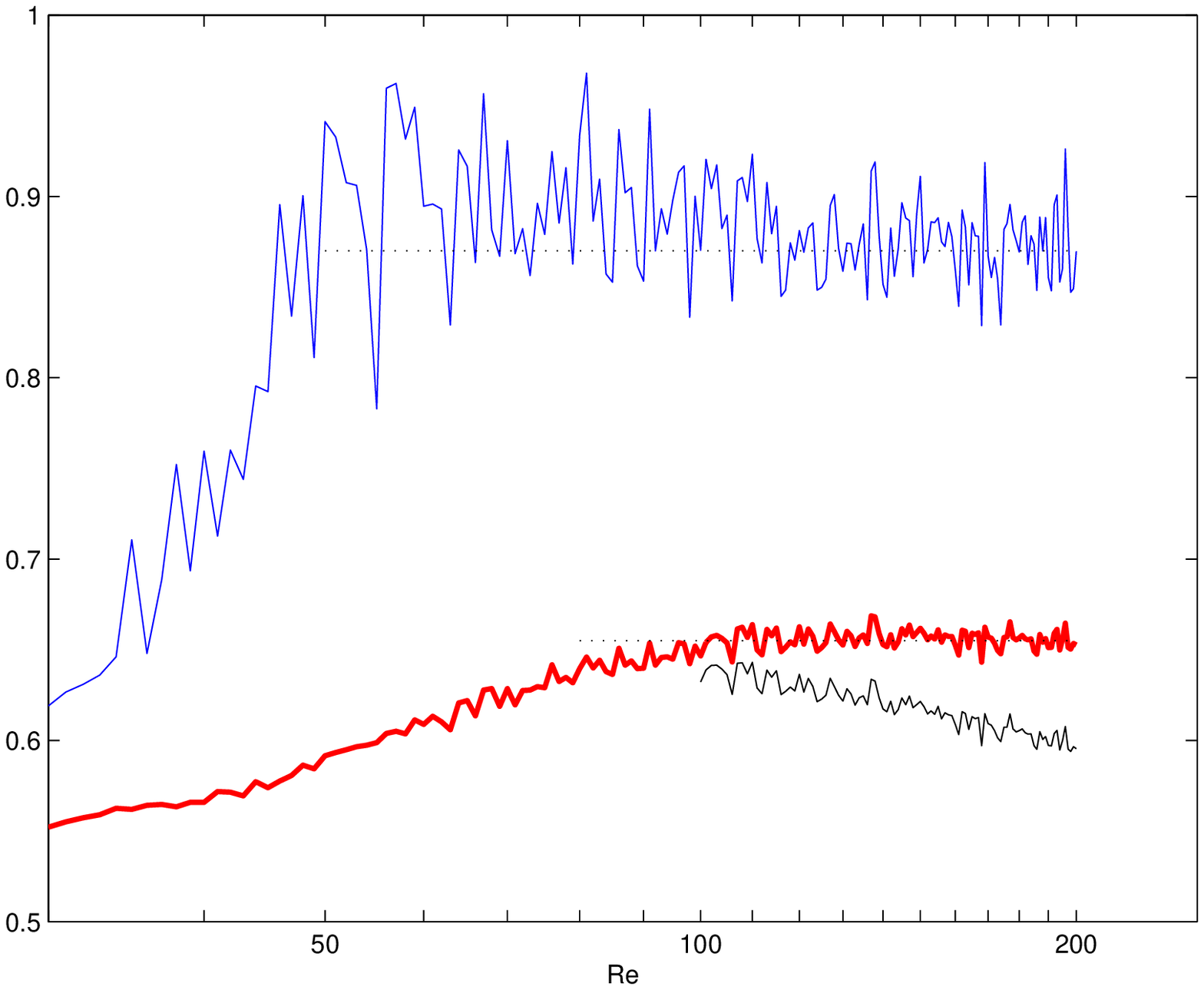,width=8cm,height=6cm,clip=true}}
\end{picture}
\end{center}
\caption{The normalised dissipation $D/D_{lam}$ (blue solid line with
  dots) and fractional fluctuation kinetic energy $E_t/E$ (red dotted
  line connecting dots) to illustrate how the flow changes with
  $Re$. The first bifurcation at $Re_{lin}=9.9669$ is steady so
  $D/D_{lam}$ decreases below 1 but $E_t/E$ remains 0 until
  time-dependent flows appear at around $Re=15$.  Curves were
  generated by using a random initial condition, running for 500 time
  units and then calculating averages over the next 1000 time
  units. This produces a unique average except over the interval $15
  \lesssim Re \lesssim 23$ where there is some metastability as
  shown. Inset: $Re^{1/2}D$ (upper blue line) and $3Re^{-2/5}E^{1/2}$
  (lower thick red line) against $Re$ on a log scale showing the
  apparent start of the asymptotic regime $D \sim Re^{-1/2}$ and
  $U:=\sqrt{2E}\sim Re^{2/5}$. The downward sloping lowest line shows
  $U/Re^{1/2}$ for comparison.}
\label{D_Re}
\end{figure}

%
%
\begin{figure}
\centerline{\includegraphics[scale=0.75]{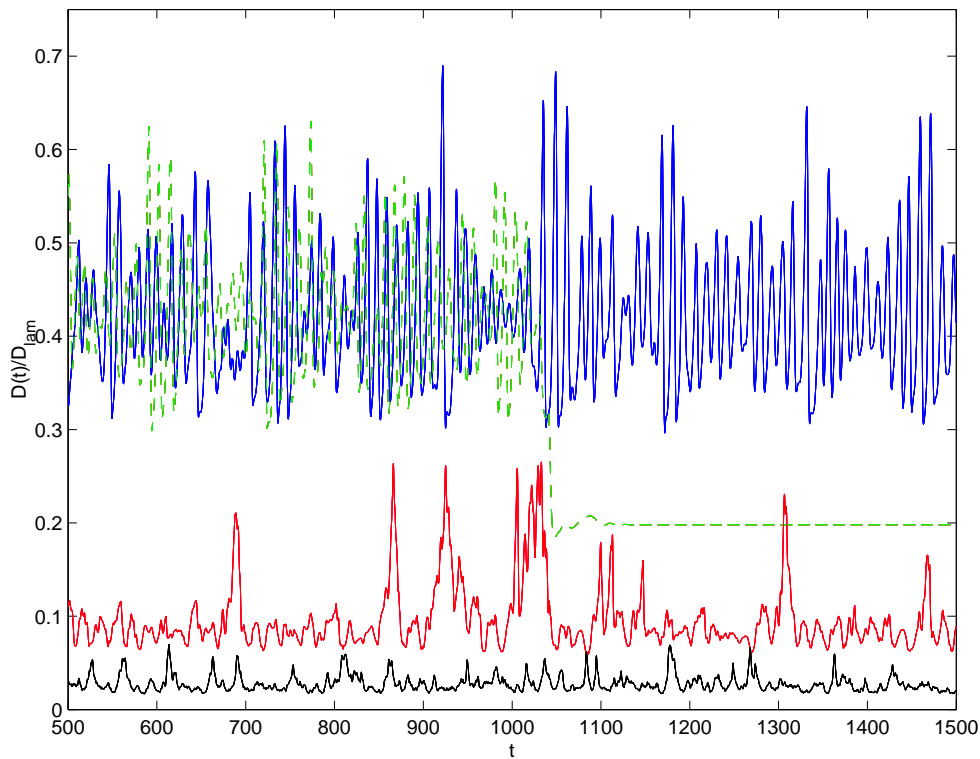}}
\caption{ The normalised dissipation $D(t)/D_{lam}$ vs $t \in
  [500,1500]$ for Re=$22$ (two different initial conditions: top solid
  blue \& green dashed  lines), $Re=40$ (second lowest red
  line) and $Re=100$ (lowest black line). $Re=22$ shows chaotic saddle
  behaviour with one trajectory seemingly dropping out back randomly
  to the travelling wave T1 (stable at $Re=22$).}
\label{D_t}
\end{figure}

%
%
\begin{figure}
\begin{center}\setlength{\unitlength}{1cm}
\begin{picture}(14,9)
\put(0,4.5){\epsfig{figure=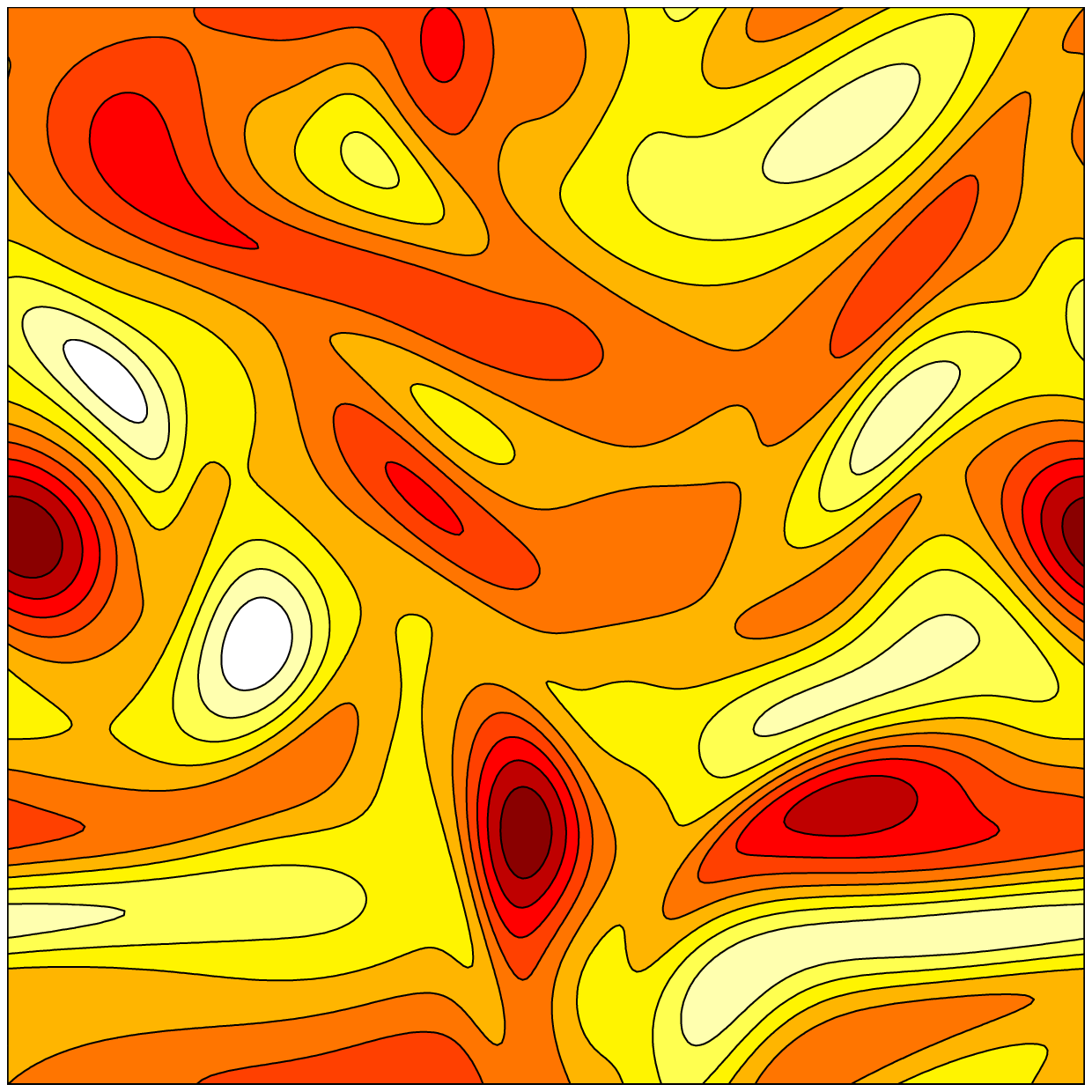,width=4.25cm,height=4.25cm,clip=true}}
\put(4.5,4.5){\epsfig{figure=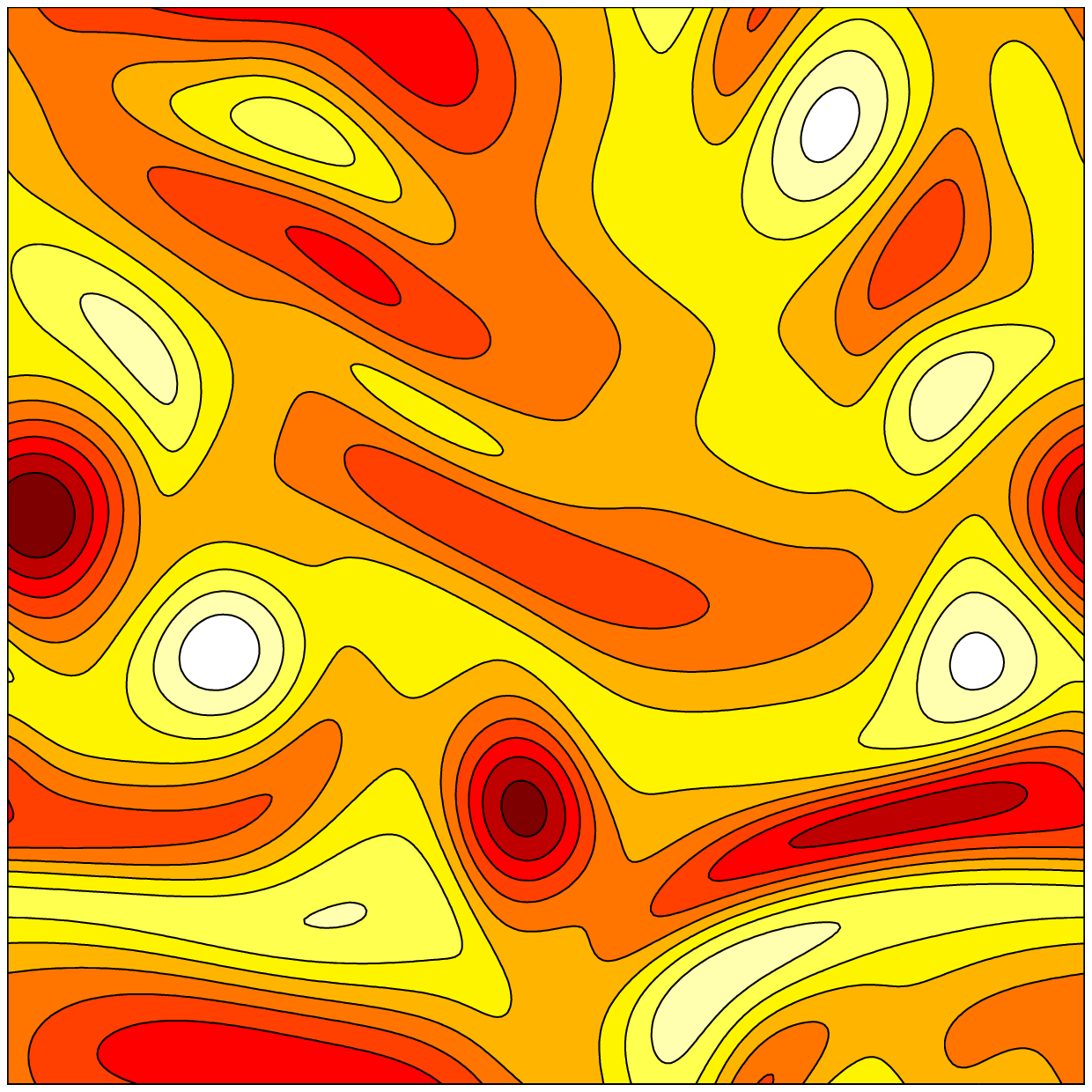,width=4.25cm,height=4.25cm,clip=true}}
\put(9,4.5){\epsfig{figure=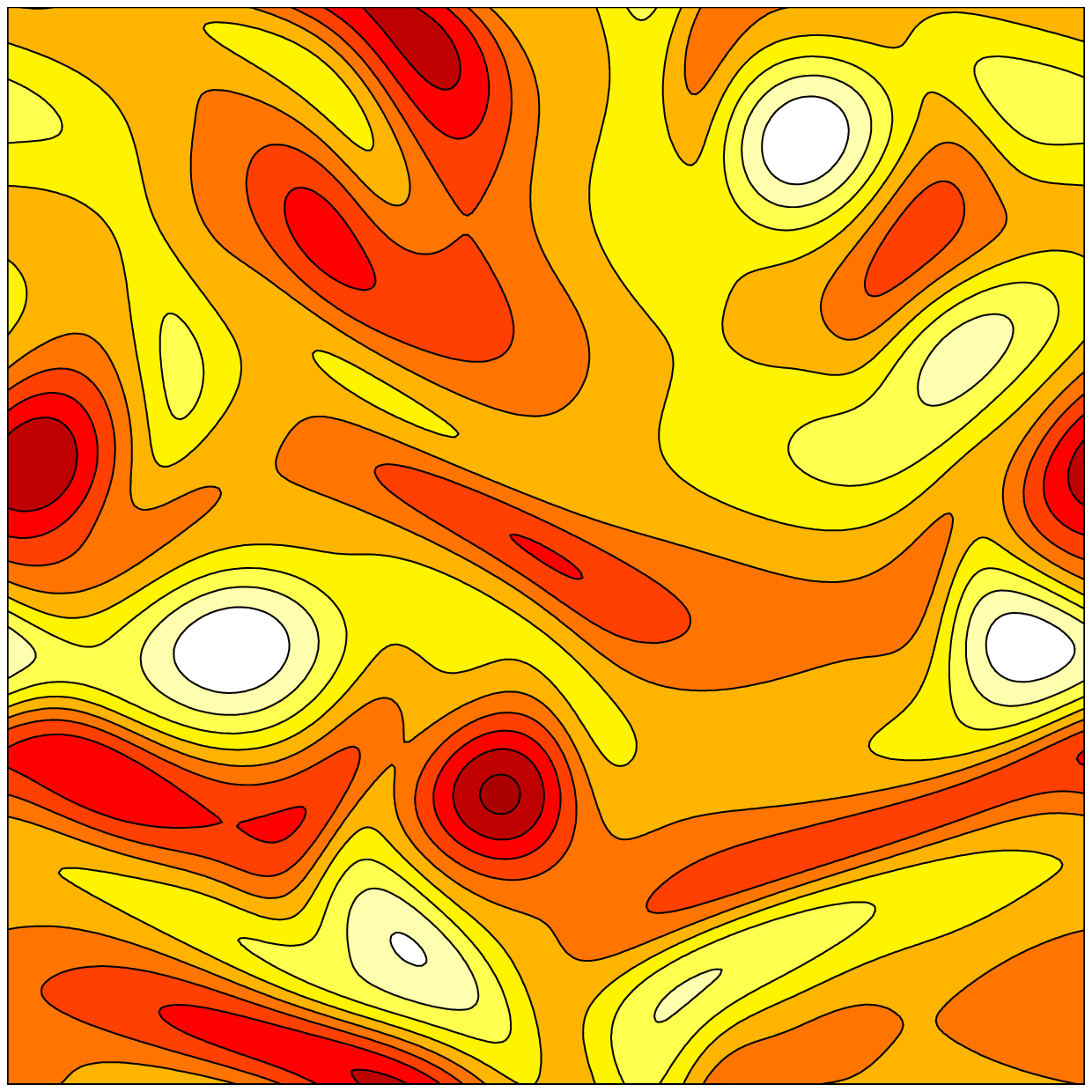,width=4.25cm,height=4.25cm,clip=true}}
\put(0,0){\epsfig{figure=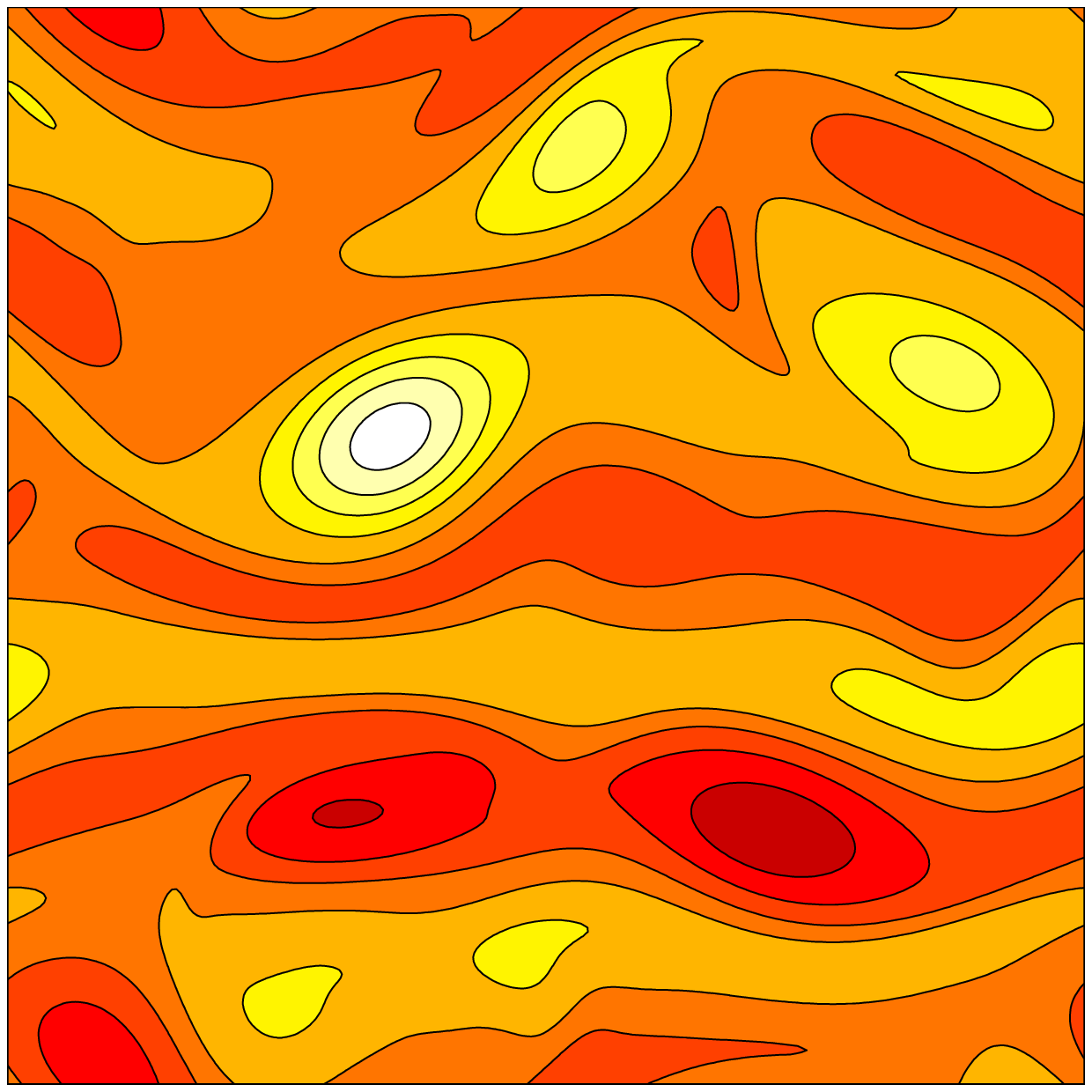,width=4.25cm,height=4.25cm,clip=true}}
\put(4.5,0){\epsfig{figure=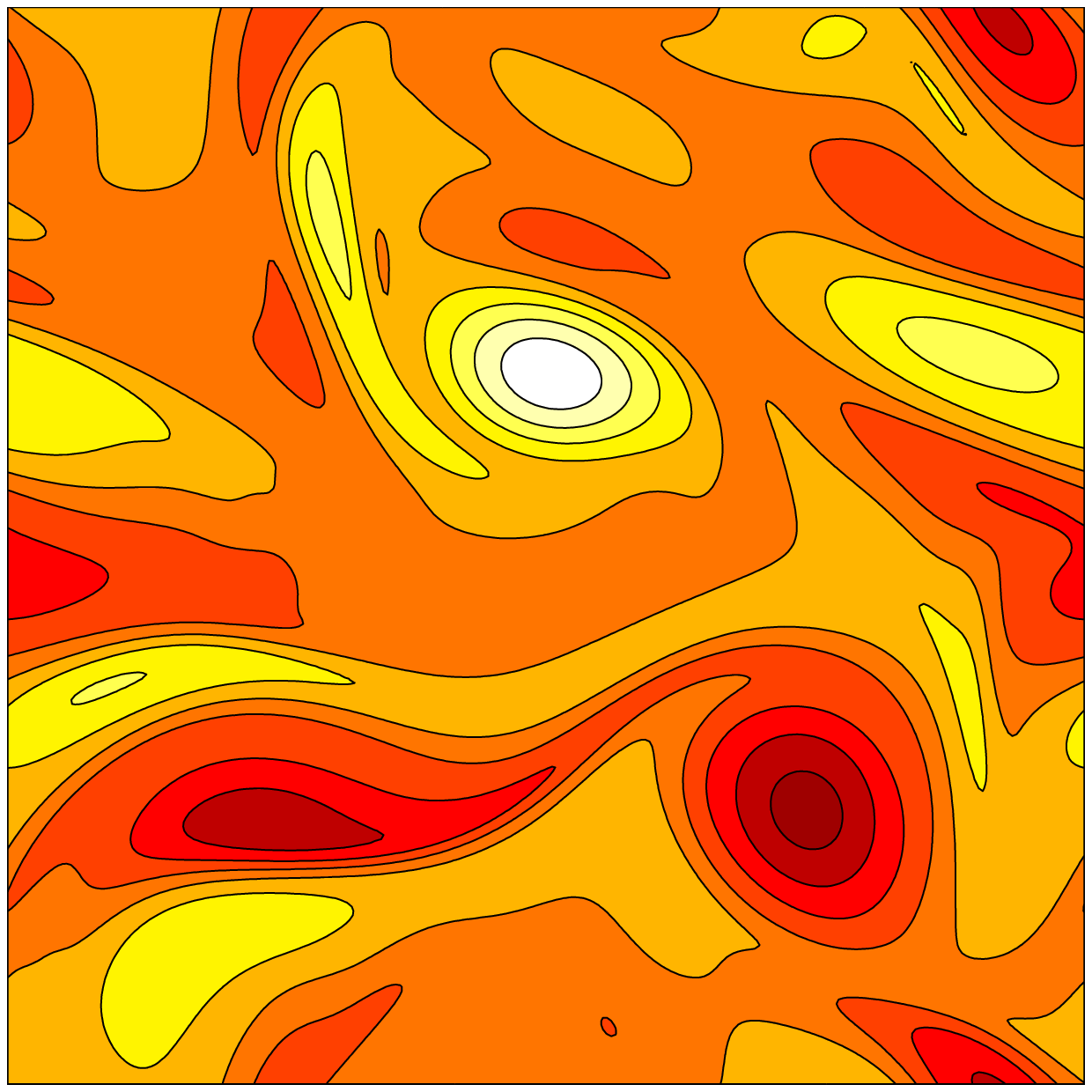,width=4.25cm,height=4.25cm,clip=true}}
\put(9,0){\epsfig{figure=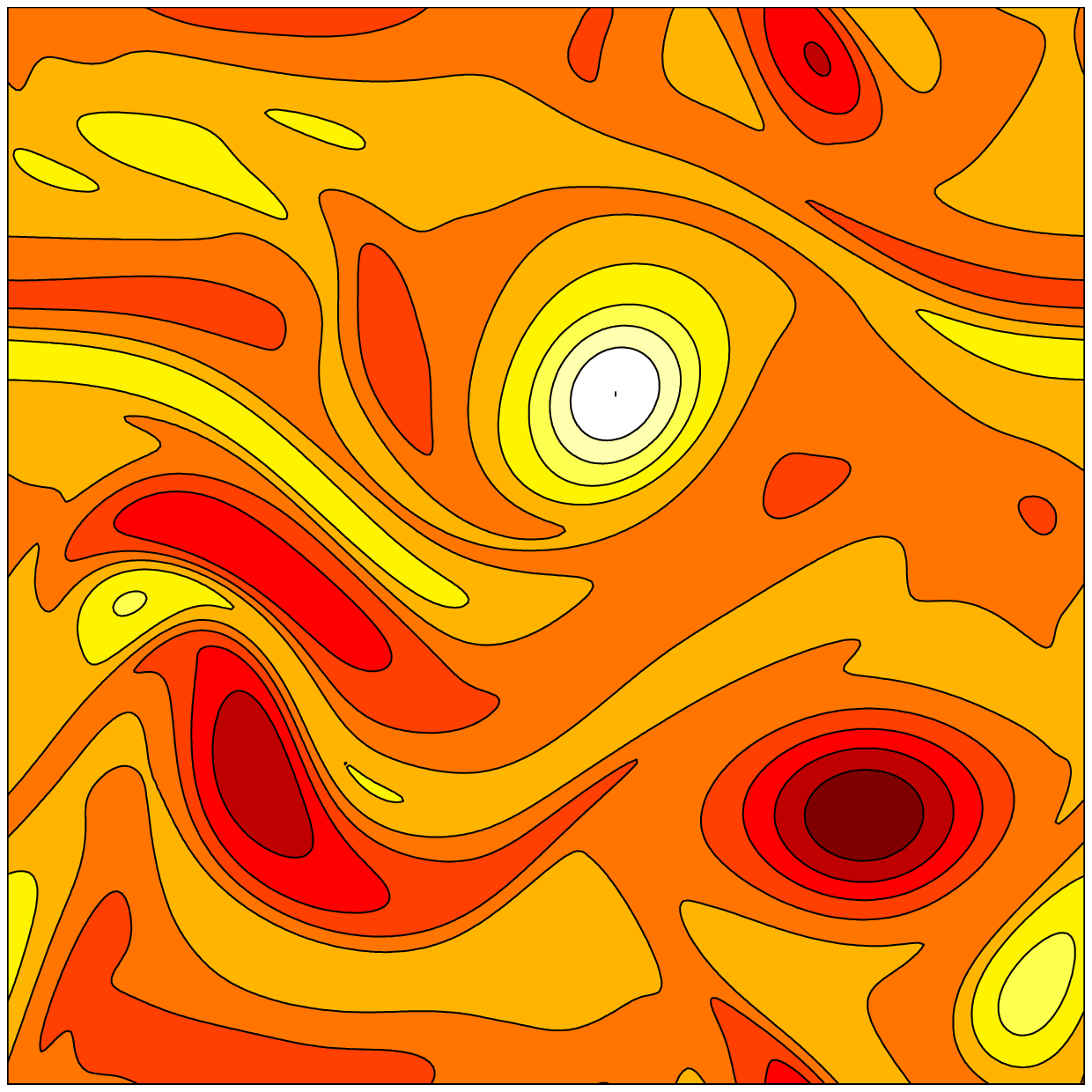,width=4.25cm,height=4.25cm,clip=true}}
\end{picture}
\end{center}
\caption{A time sequence of vorticity plots at $Re=40$
  (upper) and $Re=100$ (lower) over the $2\pi \times 2\pi$
  domain. Time difference between plots is 1 time unit and 10 contours
  are drawn between the maximum (white/light) and minimum values
  (red/dark) of the perturbation vorticity $-10.25 \leq \omega \leq
  8.6$ at $Re=40$ and $-18.4 \leq \omega \leq 20.3$ at $Re=100$.  }
\label{vorticity}
\end{figure}

\begin{figure}
\centerline{\includegraphics[scale=0.5]{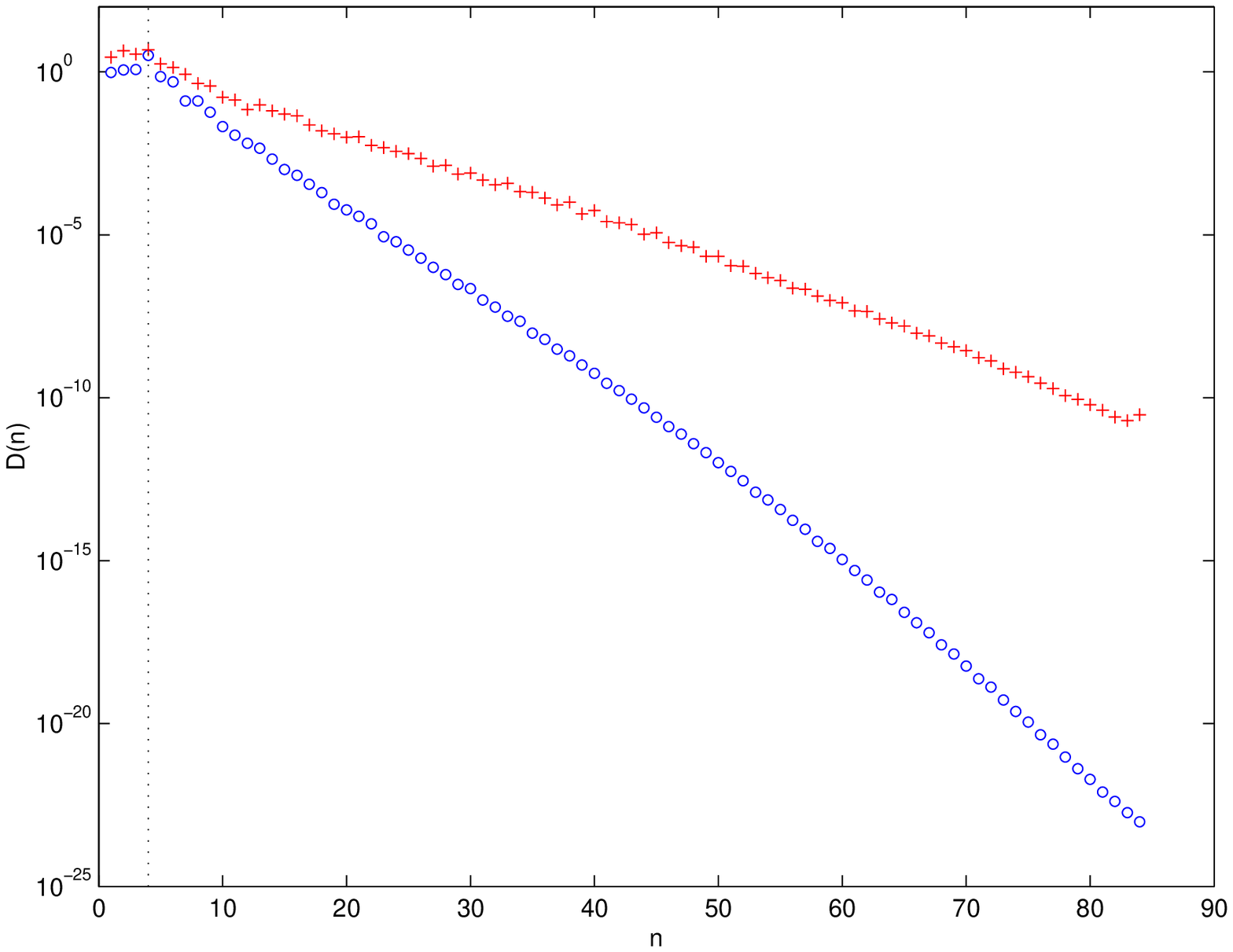}}
\caption{The enstrophy spectrum for the rightmost snapshots in figure
  \ref{vorticity} at $Re=40$ (blue circles) and $Re=100$ (red crosses)
  $D(n):=\sum_{n-\half \leq \sqrt{j^2+l^2} < n+\half} |\Omega_{jl}|^2$
  for $n=1,2...85$: recall definition (\ref{omega} with $\alpha=1$)
  (e.g. there are 264 and 516 wavenumbers included for $n=40$ and $84$
  respectively). The energy spectrum $E(n) \approx D(n)/n^2$ has a
  steeper drop off.  The dotted line indicates the wavenumber of
  energy injection. }
\label{spectral}
\end{figure}

\subsection{Flow Orientation}

2D Kolmogorov flow is linearly unstable at a comparatively low $Re$
which depends strongly on the imposed periodicity in the forcing
direction: see figure \ref{en_lin}. For the domain studied here
($\alpha=1$), disturbances to the base flow (\ref{base}) fail to decay
monotonically at $Re_E=6.8297$ and then start to grow exponentially at
$Re_{lin}=9.9669$. Figure \ref{D_Re} shows that this initial
bifurcation is to a steady flow ($D/D_{lam}<1$ and $E_t/E=0$) until
$Re \approx 15$ whereupon time dependence appears. For $15 \lesssim Re
\lesssim 23$, some metastability is noticed which is illustrated in
figure \ref{D_t} at $Re=22$ for two different initial conditions. One
leads to a chaotic-looking dissipation signal across the time interval
$[500,1500]$ whereas the other drops out of this chaotic state at just
over t=1000 to converge on a stable travelling wave solution (later
named T1). Beyond $Re \approx 23$, the chaotic state presumably
becomes an attractor or the probability of dropping out of this state
becomes so small that it is not picked up over the time windows
studied ($10^3$ units here and $10^5$ later). Finally, an asymptotic
regime is approached for $Re \gtrsim 50$. The preliminary calculations
performed here for $100<Re<200$ tentatively support the asymptotic
scaling laws $D \sim Re^{-1/2}$ and $U:=\sqrt{2E} \sim Re^{2/5}$
although the noisy data clearly warrants much longer time averaging to
confirm this.

Given this general flow behaviour, we chose to concentrate on
analysing the flow at $Re=40$ (approaching the asymptotic regime), and
three values, $Re=60, 80$ and $100$, which get deeper into the
asymptotic regime. Figures \ref{D_t} and \ref{vorticity} give an idea
of the temporal and spatial scales in the flow at the two extremes,
$Re=40$ and $Re=100$, of our study. Both indicate a hierarchy of
temporal and spatial scales (which broaden with $Re$) indicative
of 2D turbulence. Figure \ref{spectral} confirms that the flows
studied for $Re \leq 100$ are well-resolved: there is 10 orders of
drop off in the enstrophy spectrum in the most demanding case
($Re=100$) for the $256^2$ resolution used throughout this work.


\subsection{Finding recurrent structures}

A standard hunt for recurrent flows involved integrating the flow from
random initial conditions for a period of $10^5$ time units.
Initially, $R_{thres}$ was set at 0.15 for three runs at $Re=40$
(labelled $a$,$b$ and $c$ in Table 1) which produced only 9, 7 and 13
guesses respectively. Relaxing $R_{thres}$ to 0.3 (run $d$), however,
produced 885. This threshold value proved adequate at $Re=60$ (nearly
300 near-recurrences detected over runs $e$,$f$ and $g$) but had to be
further relaxed to $0.35$ at $Re=80$ and $0.4$ for $Re=100$: see Table
1. Unfortunately, it was noticed after these (Series A) runs had been
completed and the guesses tested for convergence that only $s=m=0$
shifts had been searched over. So the
runs were repeated (Series B runs $o$, $p$, $q$ and $r$) searching
specifically for recurrences which selected either $s \neq 0$ and/or
$m \neq 0$ to minimise $R$. This was done to indicate the frequency of
observing strictly periodic near-recurrences and relative periodic
near-recurrences.

The initial trawl for near-recurrences took a few weeks (each case run
on a Xeon X5670 processor) with the DNS code slowed considerably by
the need to search for near-recurrences every $0.1$ or $0.2$ units in
time (which is anything from 20 to 100 numerical time steps). The more
time-consuming activity, however, was attempting to converge the
near-recurrent guesses to exact solutions. Adopting fairly
conservative limits for the Newton-GMRES-Hookstep procedure - maximum
period considered  was 100, maximum number of Newton, GMRES and Hook steps
were 75, 500 and 50 respectively - typically lead to run times of a
couple of months for each of the $Re=60,80$ and $100$ runs. The data
for $Re=40$ had to be subdivided 12 ways to make the process
manageable. These numbers make it clear why a very efficient DNS code was
important for this work.

Table 1 also indicates the conversion rate of near-recurrences guesses
to exactly recurrent solutions. There is considerable duplication of
such solutions so that a  much smaller set of distinct recurrent
structures is obtained.

%
%
\begin{table}
\begin{center}
\begin{tabular}{@{}cccrrrr@{}}
  &  Run  & $R_{thres}$& 
\hspace*{0.5cm}dt &
\hspace*{0.1cm} duration&
\hspace*{0.1cm} \# of guesses &
\qquad \# of convergences\\ 
  & & & & & & \\\hline
  & & & & & & \\
\underline{Series A} &  $(s=m=0)$  & & & & & \\ 
  & & & & & & \\
$Re=40$& a & 0.15 & 0.005 &$10^{5}$&9  &5\\
       & b & 0.15 & 0.005 &$10^{5}$&7  &3\\
       & c & 0.15 & 0.005 &$10^{5}$&13  &5\\
       & d & 0.30 & 0.005 &$10^{5}$&885&553\\
       &   &      &       &       &   &   \\
$Re=60$& e & 0.30 & 0.003 &$10^{5}$&102&64\\
       & f & 0.30 & 0.003 &$10^{5}$&104&67\\
       & g & 0.30 & 0.003 &$10^{5}$&78&58\\
       &   &      &       &       &   &  \\
$Re=80$& h & 0.35 & 0.0025&$10^{5}$&53&31\\
       & i & 0.35 & 0.0025&$10^{5}$&60&37\\
       & j & 0.35 & 0.0025&$10^{5}$&41&25\\
       &   &      &       &       &  &  \\
$Re=100$&l & 0.4  & 0.002&$10^{5}$&75&34\\
       & m & 0.4  & 0.002&$10^{5}$&91&33\\
       &n  & 0.4  & 0.002&$10^{5}$&93&42\\
       &   &      &       &       &  &  \\
\underline{Series B} & $(s \neq 0$ or $ m \neq 0)$   &      &       &       &  &  \\
       &   &      &       &       &  &  \\
$Re=40$& o & 0.30 & 0.005 &$10^{5}$&1223&540\\
       &   &      &       &       &  &  \\
$Re=60$& p & 0.30 & 0.003 &$3 \times 10^{5}$&163&7\\
       &   &      &       &       &  &  \\
$Re=80$& q & 0.35 & 0.0025&$3 \times 10^{5}$&66&15\\
       &   &      &       &       &  &  \\
$Re=100$&r & 0.4  & 0.002&$3 \times 10^{5}$&84&12\\
        &   &      &       &       &  &  \\
\end{tabular}
\end{center}
\caption{DNS data used to extract equilibria, travelling waves,
  periodic orbits and relative periodic orbits at $Re=40,60,80$ and
  $100$. The value of $R_{thres}$ cannot be set too ambitiously. Run d
  yielded all the solutions thrown up by runs a,b and c combined.}
\end{table}

%
%
%
%
\begin{table}
\begin{center}
\begin{tabular}{@{}crccccrrc@{}}
& UPO & \hspace*{0.25cm}frequency\hspace*{0.25cm} 
&\hspace*{0.125cm} $c$\hspace*{0.125cm}& \hspace*{0.25cm} $T$\hspace*{0.25cm} 
&\qquad $-s$& \qquad $-m$ 
&\hspace*{0.25cm} $N$ & \qquad $\sum_{j=1}^{N} \Re
  e(\lambda_j)$\,(\,max$\, \Re e(\lambda_j$)\,) \\
       &     &            &        &        &       &   &    &              \\ 
$Re=40$& E1  & $\geq$ 261 &        &        &       & 0 &  9 & 1.296(0.249) \\ 
       & T1  &        127 & 0.0198 &        &       & 0 &  4 & 0.142(0.068) \\ 
       & T2  &          1 & 0.0096 &        &       & 0 &  4 & 1.227(0.454) \\ 
       & P1  &        143 &        &  5.380 &       & 0 &  7 & 0.570(0.191) \\ 
       & P2  &          6 &        &  2.830 &       & 0 &  5 & 0.742(0.223) \\ 
       & P3  &          2 &        &  2.917 &       & 0 &  7 & 0.992(0.236) \\
       & R1  &          1 &        & 56.677 & 0.092 & 0 &  3 & 0.156(0.077) \\
       & R2  &          1 &        & 25.401 & 0.199 & 0 &  5 & 0.254(0.123) \\
       & R3  &          1 &        & 54.280 & 0.200 & 0 &  3 & 0.195(0.108) \\
       & R4  &          1 &        &  6.720 & 0.106 & 0 &  8 & 0.870(0.343) \\
       & R5  &          1 &        & 23.780 & 0.022 & 0 &  4 & 0.376(0.156) \\
       & R6  &          4 &        & 20.808 & 0.060 & 0 &  3 & 0.258(0.172) \\
       & R18 &          1 &        & 37.233 & 0.270 & 0 &  5 & 0.242(0.165) \\
       &     &            &        &        &       &   &    &              \\
       & R19 &            &        & 12.207 & 0.243 & 0 &  2 & 0.141(0.070) \\
       & R20 &            &        & 16.586 & 5.827 & 1 &  4 & 0.289(0.103) \\
       & R21 &            &        & 17.470 & 5.765 & 3 &  5 & 0.348(0.143) \\
       & R22 &            &        & 19.723 & 0.222 & 0 &  4 & 0.297(0.172) \\
       & R23 &            &        & 19.762 & 0.513 & 0 &  4 & 0.302(0.127) \\
       & R24 &            &        & 19.779 & 6.035 & 0 &  5 & 0.292(0.202) \\
       & R25 &            &        & 20.201 & 5.898 & 3 &  6 & 0.380(0.138) \\
       & R26 &            &        & 20.385 & 1.334 & 2 &  7 & 0.714(0.270) \\
       & R27 &            &        & 20.632 & 5.871 & 3 &  4 & 0.365(0.127) \\
       & R28 &            &        & 20.885 & 5.987 & 1 &  4 & 0.360(0.121) \\
       & R29 &            &        & 20.909 & 0.306 & 1 &  5 & 0.380(0.124) \\
       & R30 &            &        & 21.310 & 5.694 & 0 &  5 & 0.330(0.100) \\
       & R31 &            &        & 21.725 & 5.799 & 0 &  3 & 0.319(0.133) \\
       & R32 &            &        & 22.560 & 0.006 & 1 &  4 & 0.283(0.096) \\
       & R33 &            &        & 22.617 & 5.660 & 0 &  5 & 0.478(0.156) \\
       & R34 &            &        & 23.157 & 0.265 & 0 &  3 & 0.260(0.113) \\
       & R35 &            &        & 23.417 & 5.936 & 3 &  4 & 0.489(0.183) \\
       & R36 &            &        & 24.465 & 6.010 & 3 &  4 & 0.358(0.191) \\
       & R37 &            &        & 25.870 & 0.182 & 0 &  3 & 0.272(0.122) \\
       & R38 &            &        & 25.934 & 0.227 & 0 &  4 & 0.263(0.125) \\
       & R39 &            &        & 27.138 & 6.248 & 0 &  4 & 0.391(0.107) \\
       & R40 &            &        & 28.817 & 5.971 & 0 &  5 & 0.238(0.116) \\
       & R41 &            &        & 32.541 & 0.349 & 1 &  4 & 0.224(0.153) \\
       & R42 &            &        & 34.316 & 5.886 & 0 &  5 & 0.163(0.120) \\
       & R43 &            &        & 34.530 & 5.742 & 0 &  3 & 0.220(0.134) \\
       & R44 &            &        & 34.917 &-0.059 & 3 &  4 & 0.325(0.139) \\
       & R45 &            &        & 36.549 & 6.027 & 3 &  3 & 0.183(0.118) \\
       & R46 &            &        & 36.627 & 0.197 & 3 &  4 & 0.202(0.139) \\
       & R47 &            &        & 36.812 & 5.648 & 0 &  3 & 0.155(0.074) \\
       & R48 &            &        & 37.079 & 6.103 & 0 &  4 & 0.171(0.134) \\
       & R49 &            &        & 37.233 & 0.270 & 0 &  3 & 0.241(0.165) \\
       & R50 &            &        & 37.698 & 3.499 & 1 &  6 & 0.477(0.146) \\
       & R51 &            &        & 39.368 & 6.070 & 0 &  5 & 0.192(0.098) \\
       & R52 &            &        & 39.619 & 0.380 & 0 &  5 & 0.242(0.067) \\
       & R53 &            &        & 41.400 & 5.806 & 0 &  4 & 0.176(0.081) \\
       & R54 &            &        & 49.645 & 6.054 & 3 &  4 & 0.160(0.065) \\
       & R55 &            &        & 53.073 & 6.031 & 0 &  4 & 0.189(0.105) \\
       &     &            &        &        &       &   &    &      \\
\end{tabular}
\end{center}
\caption{All the invariant sets found directly from turbulent DNS data
  (from series A above the separating blank line and from series B
  below). `Frequency' is the number of times the solution was
  extracted (Series A runs). There is one steady {\bf E}quilibrium,
  $c$ is the phase speed of the {\bf T}ravelling waves found, $T$ is
  the {\bf P}eriod of periodic and {\bf R}elative periodic orbits
  which also either have a shift $s$ and/or shift $m$.  $N$ is the
  number of unstable directions and $\sum_{j=1}^N \Re e (\lambda_j)$
  the sum of the real parts of all the unstable eigenvalues.  }
\end{table}

%
%
%
%
\begin{table}
\begin{center}
\begin{tabular}{@{}crccccrrc@{}}
& UPO & \hspace*{0.25cm}frequency\hspace*{0.25cm} 
&\hspace*{0.125cm} $c$\hspace*{0.125cm}& \hspace*{0.25cm} $T$\hspace*{0.25cm} 
&\qquad $-s$& \qquad $-m$ 
&\hspace*{0.25cm} $N$ & \qquad $\sum_{j=1}^{N} \Re
  e(\lambda_j)$\,(\,max$\, \Re e(\lambda_j$)\,) \\
& & & & & & & \\ 
$Re=60$&E1  &    4   &        &       &       & 0 & 14 & 5.053(0.858)\\
       &T1  & high   & 0.0019 &       &       & 0 &  4 & 0.139(0.064)\\
       &T3  & high   & 0.0124 &       &       & 0 & 17 & 3.377(0.684)\\
       &T4  &    1   & 0.0082 &       &       & 0 &  3 & 0.257(0.178)\\
       &R7  & high   &        & 2.472 & 0.036 & 0 &  9 & 0.911(0.214)\\
       &R8  &    1   &        & 1.638 & 0.022 & 0 & 14 & 2.903(0.681)\\
       &    &        &        &       &       &   &    &             \\
       &R56 &        &        & 16.326& 0.588 & 2 &  6 & 0.609(0.139)\\
       &R57 &        &        & 17.909& 5.802 & 0 &  7 & 0.805(0.169)\\
       &R58 &        &        & 20.546& 0.659 & 2 &  8 & 0.529(0.168)\\
       &    &        &        &       &       &   &    &     \\
$Re=80$&T1  & high   & 0.0115 &       &       & 0 &  6 & 0.360(0.105)\\
       &T3  &    8   & 0.0154 &       &       & 0 & 21 & 5.588(0.958)\\
       &T5  &    3   & 0.0831 &       &       & 0 & 20 & 4.183(0.658)\\
       &R7  &   10   &        & 2.299 & 0.054 & 0 & 13 & 1.326(0.181)\\
       &R8  &    2   &        & 1.705 & 0.028 & 0 & 18 & 4.318(1.105)\\
       &R9  &    1   &        & 2.150 & 0.032 & 0 & 19 & 3.987(1.026)\\
       &R10 &    1   &        & 1.280 & 0.020 & 0 & 20 & 5.310(0.878)\\
       &R11 &    1   &        & 2.443 & 0.031 & 0 & 10 & 0.704(0.277)\\
       &R12 &    1   &        & 2.095 & 0.034 & 0 & 11 & 3.083(1.004)\\
       &R13 &    1   &        &15.285 & 0.181 & 0 &  8 & 0.409(0.131)\\
       &    &        &        &       &       &   &    &             \\
       &R59 &        &        &15.667 & 0.397 & 1 & 11 & 0.949(0.176)\\
       &R60 &        &        &16.071 & 0.462 & 1 & 11 & 1.136(0.248)\\
       &    &        &        &       &       &   &    &     \\
$Re=100$&T1 & high   & 0.0155 &       &       & 0 & 10 & 0.646(0.122)\\
        &T3 &    7   & 0.0179 &       &       & 0 & 25 & 6.491(1.042)\\
        &T4 &    5   & 0.0118 &       &       & 0 &  3 & 0.684(0.370)\\
        &T5 &    6   & 0.0691 &       &       & 0 & 28 & 6.238(0.689)\\
        &P4 &    1   &        & 1.185 &       & 0 & 16 & 7.376(1.201)\\
        &R7 &    4   &        & 1.971 & 0.030 & 0 & 15 & 1.933(0.326)\\
        &R11&    3   &        & 2.262 & 0.001 & 0 &  9 & 0.905(0.385)\\
        &R12&    1   &        & 1.902 & 0.029 & 0 & 14 & 3.953(1.244)\\
        &R14&    5   &        & 4.526 & 0.071 & 0 &  8 & 0.428(0.105)\\
        &R15&    2   &        & 1.984 & 0.122 & 0 & 16 & 3.110(0.556)\\
        &R16&    2   &        & 1.938 & 0.121 & 0 &  6 & 0.945(0.270)\\
        &R17&    1   &        & 3.827 & 0.008 & 0 & 16 & 2.894(0.818)\\
        &R61&    1   &        & 1.344 & 0.090 & 0 & 21 & 4.997(0.476)\\
        &   &        &        &       &       &   &    &     \\
\end{tabular}
\end{center}
\caption{All the invariant sets found directly from turbulent DNS data
  (from series A above the separating blank line and from series B
  below: none were found at $Re=100$ in the series B runs despite the
  numerical gap between $R17$ and $R61$). `Frequency' is the number of
  times the solution was extracted. Solutions listed under each $Re$
  indicate those actually extracted at that $Re$ - hence multiple
  entries. `Frequency' is the number of times the solution was
  extracted (Series A runs). There is one steady {\bf E}quilibrium,
  $c$ is the phase speed of the {\bf T}ravelling waves found, $T$ is
  the {\bf P}eriod of periodic and {\bf R}elative periodic orbits
  which also either have a shift $s$ and/or shift $m$.  $N$ is the
  number of unstable directions and $\sum_{j=1}^N \Re e (\lambda_j)$
  the sum of the real parts of all the unstable eigenvalues.  }
\end{table}


\subsection{Recurrent structures found}

\subsubsection{Re=40}

\begin{figure}
\begin{center}\setlength{\unitlength}{1cm}
\begin{picture}(14,5)
\put(0,0){\epsfig{figure=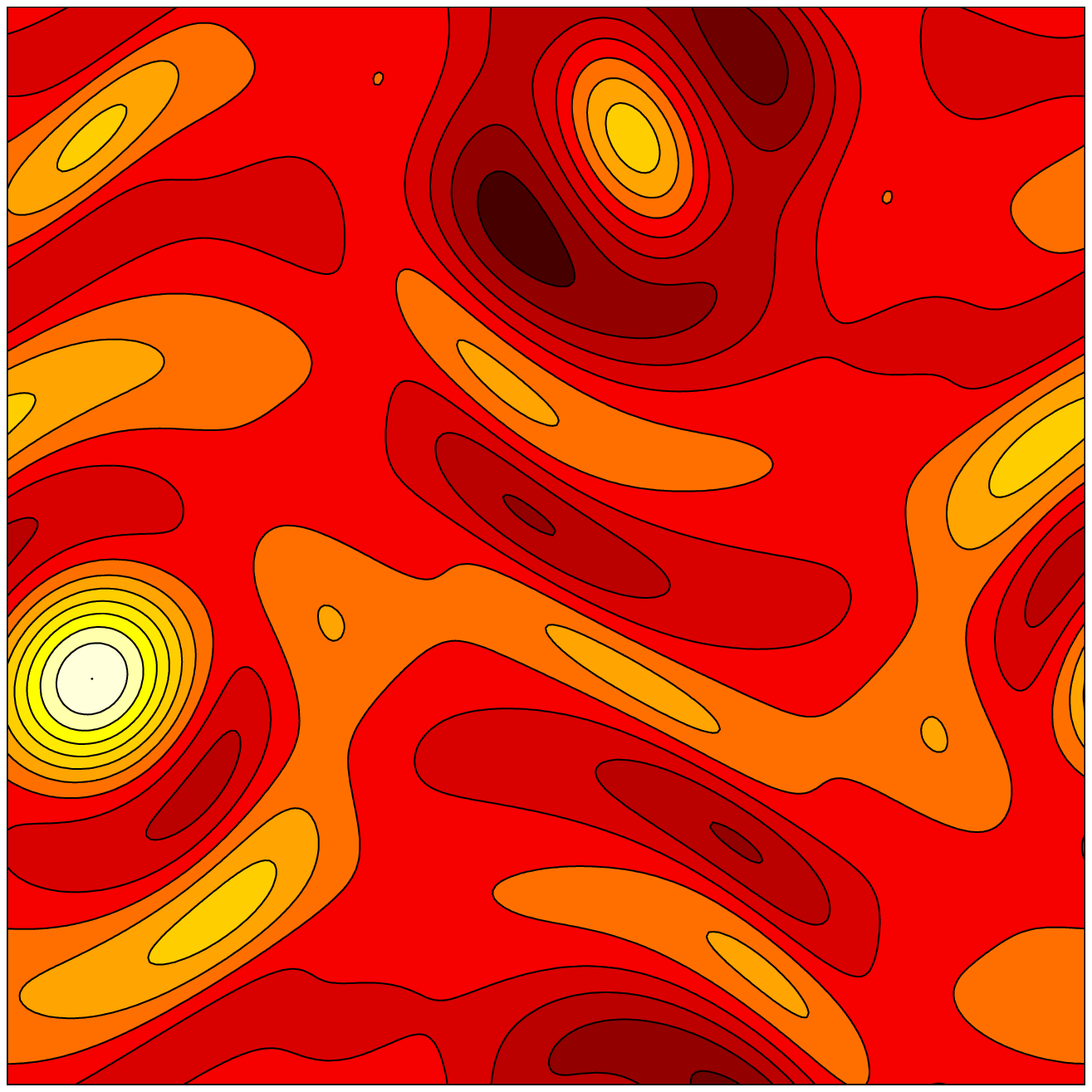,width=4.25cm,height=4.25cm,clip=true}}
\put(4.5,0){\epsfig{figure=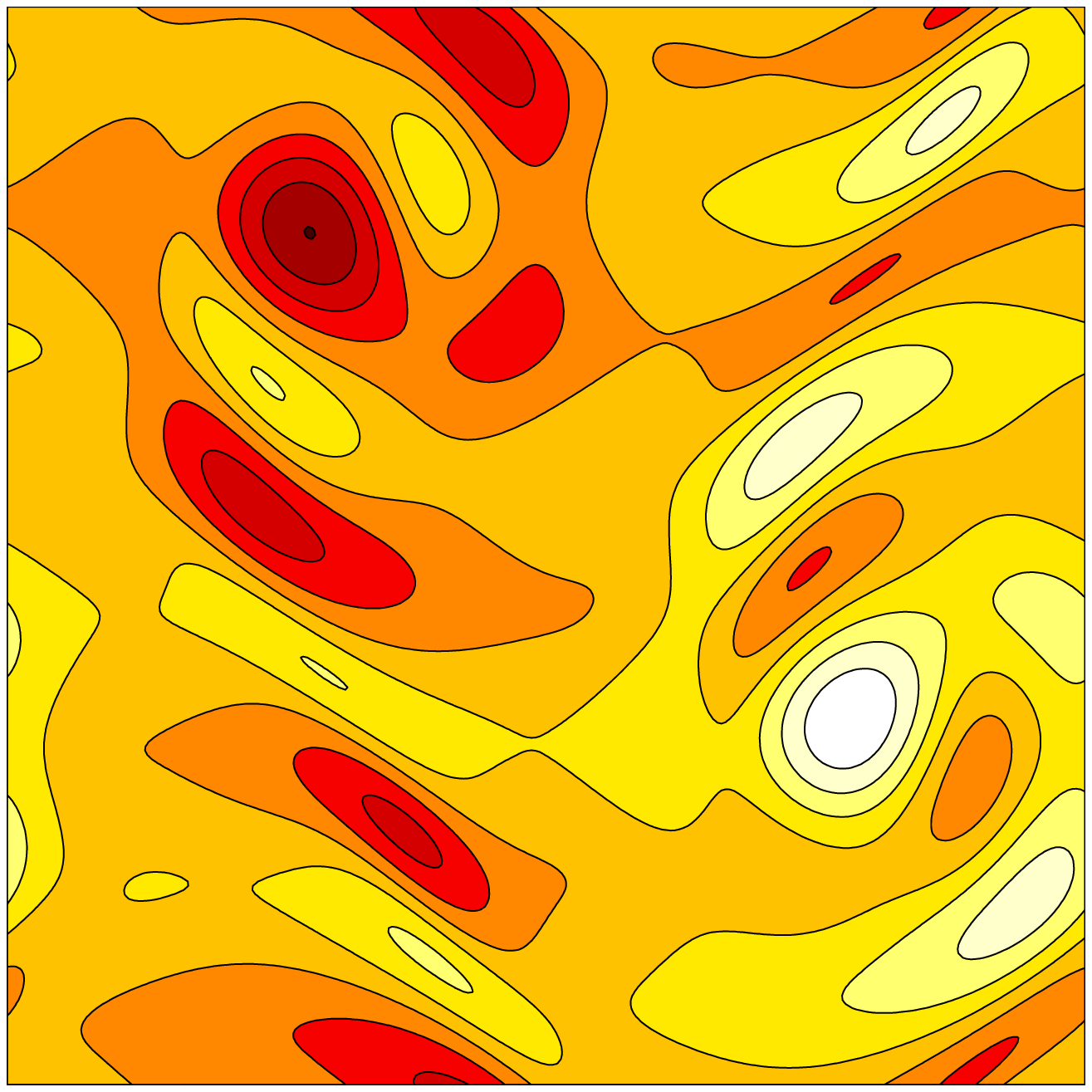,width=4.25cm,height=4.25cm,clip=true}}
\put(9,0){\epsfig{figure=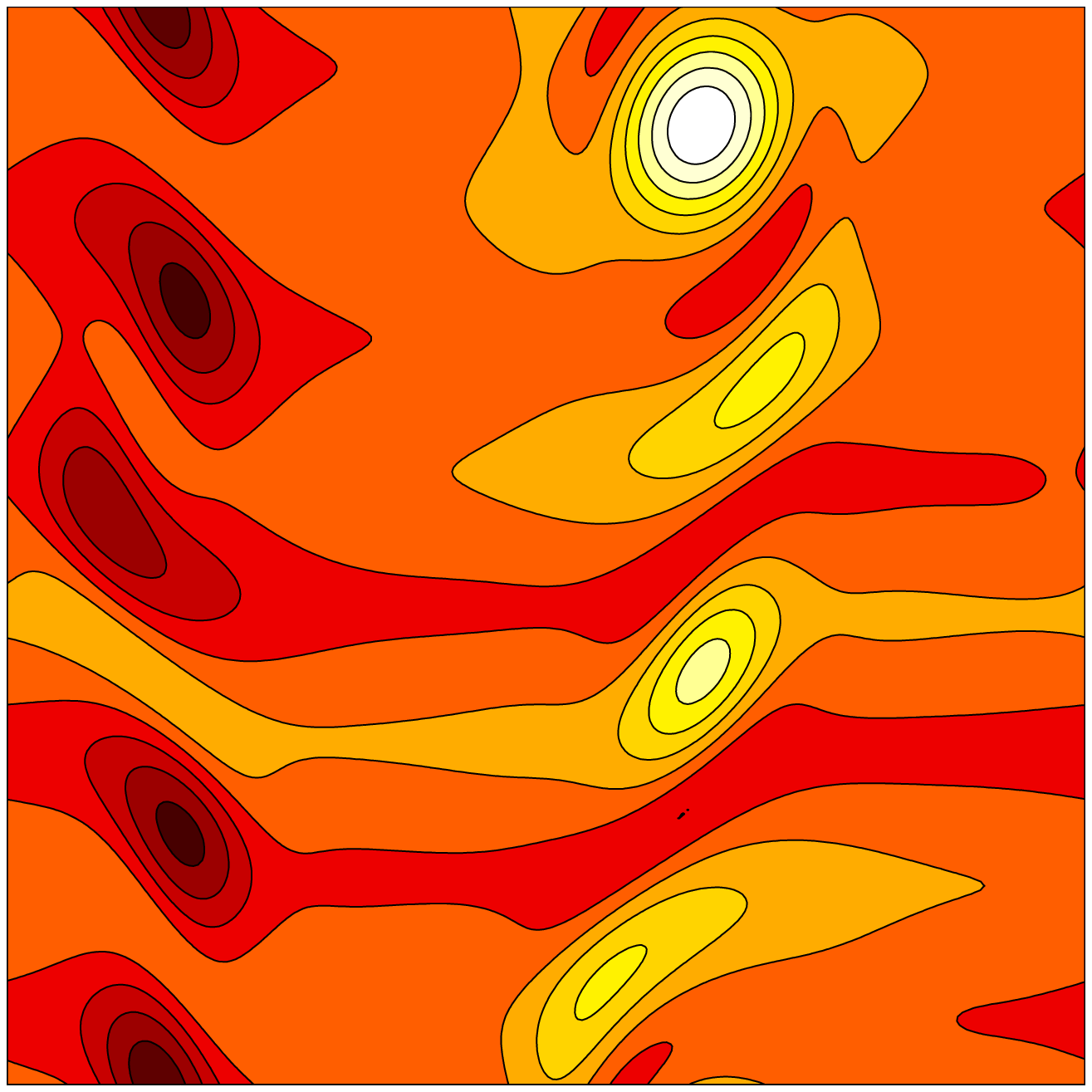,width=4.25cm,height=4.25cm,clip=true}}
\end{picture}
\end{center}
\caption{The steady solution $E1$ (left) and the travelling waves $T1$
  (middle) and $T2$ (right) at $Re=40$. Vorticity is contoured using
  15 contours between -7.2 (dark,red) and 12.1 (light,white) (range is
  $-6.5\leq \omega \leq 12.1$ for $E1$, $-7.2 \leq \omega \leq 7.2$
  for $T1$ and $-6.8 \leq \omega \leq 10.4$ for $T2$). }
\label{E1T1T2}
\end{figure}

Table 2 lists the recurrent structures found at $Re=40$. The
equilibrium flow $E1$ (see figure \ref{E1T1T2}), which was found many
times in the series A runs, corresponds to the $\R$-symmetric steady
state which bifurcates off the basic solution at $Re=9.9669$ as shown
in figure \ref{D_Re}. $E1$ loses stability at about $Re=15$ to the
travelling wave $T1$ or later via $P1$ in the $\R$-symmetric subspace
for a $Re \in (30,31)$ ($P2$ which is $\R$-symmetric and $P3$ which is
not bifurcate at yet higher $Re$ from $E1$). All three flows, $E1$,
$P1$ and $T1$ are found to be repeatedly visited by the (series A) DNS
indicating the strong influence of the $\R$-symmetric subspace on the
`turbulent' dynamics despite them all being unstable (e.g. $E1$ has 9
unstable directions at $Re=40$; see Table 2). However, a further 47
recurrent flows were also identified from the DNS: another travelling
wave $T2$, two further periodic orbits $P2$ and $P3$, and 44 relative
periodic orbits, $R1-R6$ and $R18-R55$ (note all have a non-zero shift
$s$ and some also a non-zero integer $m$). A priori, we expected to
find mainly small period recurrent structures due to the method of
extraction. Longer periods mean more time for the turbulent trajectory
to diverge away from the unstable recurrent flow and hence a higher
probability for a) the episode to escape detection as a nearly
recurrent flow and b) even if detected, for GMRES to fail to converge
due to the quality of the initial approximation. This seems borne out
by the periodic orbits found but not for the relative period orbits
where the majority have a period over 20 and some over 50 time
units. That such long period structures exist and were `extractable'
from the DNS frankly was a surprise and begs the question whether our
`long' runs of $10^5$ time units (now known to be only a factor of
$O(1000)$ longer than some recurrent flows) were actually really long
enough to capture all the structures possible. This issue will be
raised again below.

\begin{figure}
\centerline{\includegraphics[scale=0.8]{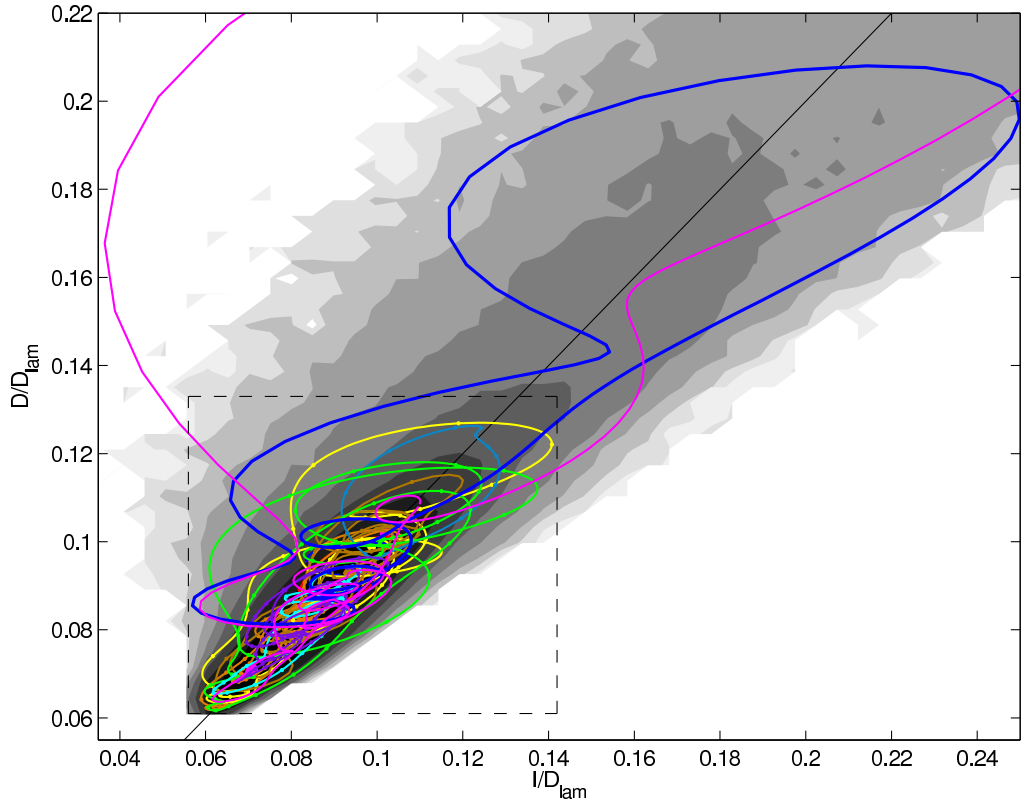}}
\caption{ The normalised dissipation $D(t)/D_{lam}$ verses $I/D_{lam}$
  for a small collection of the recurrent flows found with the pdf of
  the DNS turbulence plotted in the background (11 shades at levels
  $10^{\alpha}$ where $\alpha={-5,-4.5,\ldots,-0.5,0}$). Plotted are
  $E1$ ($D/D_{lam}=I/D_{lam}=0.102$), $T1$ (0.071), $T2$ (0.070),
  $P1$, $P2$, $R1-6$, $R18$, $R26$ (large blue orbit) and $R50$ (the
  even larger magenta orbit). The dashed box is used to show more
  recurrent structures in figure \ref{DvsI_40_multi}.}
\label{DvsI_40_sample}
\end{figure}

With so many recurrent flows found, it becomes impractical to display
and characterise each flow separately. Table 2 lists some key
characteristics along with their stability information (all are
unstable but none with more than 9 unstable directions out of 22,428
possible directions). One useful projection used by Kawahara \& Kida
(2001), however, is the `energy out ($D(t)$) verses energy in
($I(t)$)' plot which is shown in figure \ref{DvsI_40_sample} (both
quantities normalised by $D_{lam}$). The line $D=I$ corresponds to
dissipation exactly balancing energy input which has to be the case
over all times for equilibria and travelling waves (which are just
equilibria in an appropriate Galilean frame): these are therefore just
points on this line in this plot. Figure \ref{DvsI_40_sample} shows
how a representative subset of these recurrent flows look when
compared with the joint dissipation-input probability density function
(pdf) of the DNS. The darkest shading makes it clear that the DNS
stays predominantly in the region ${0.055\,<\, I/D_{lam}\,<\,0.115,\,
  0.06\,<\,D/D_{lam}\,<\, 0.11}$. The recurrent flows shown are also
dominantly concentrated in this region although there are two relative
periodic orbits shown - $R26$ and $R50$ - which have large dissipation
episodes (it's worth emphasizing that the basic state would be
represented by the point (1,1) in this plot so the turbulent flow
adopts a much reduced dissipative state). Since this $D$ verses $I$
plot is such a drastic projection of the dynamics, the fact that two
flows look close there doesn't necessarily mean they are close in the
full phase space. However, because all the recurrent flows discussed
here have been extracted from turbulent DNS, this conclusion
nevertheless seems reasonable.

\begin{figure}
\centerline{\includegraphics[scale=0.8]{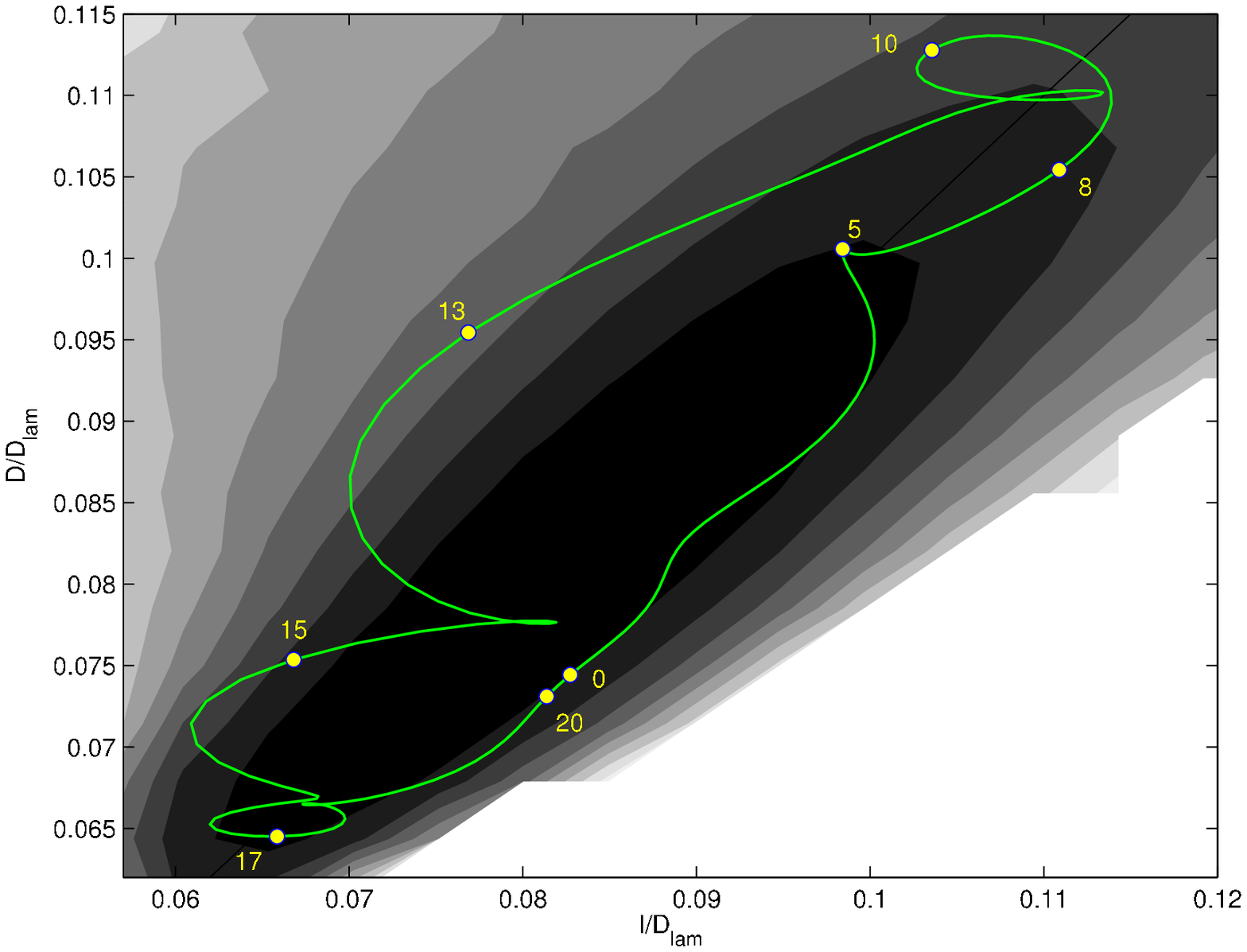}}
\caption{ The normalised dissipation $D(t)/D_{lam}$ verses $I/D_{lam}$
  for the relative periodic orbit $R25$ at Re=$40$ with the pdf of the
  DNS turbulence plotted in the background (11 shades at levels $10^{\alpha}$ where
  $\alpha={-5,-4.5,\ldots,-0.5,0}$). The labels refer to times along the orbit at
  which snapshots are shown in figure \ref{plot_R25}.}
\label{DvsI_R25}
\end{figure}

\begin{figure}
\begin{center}\setlength{\unitlength}{1cm}
\begin{picture}(14,7)
\put(0,3.4)   {\epsfig{figure=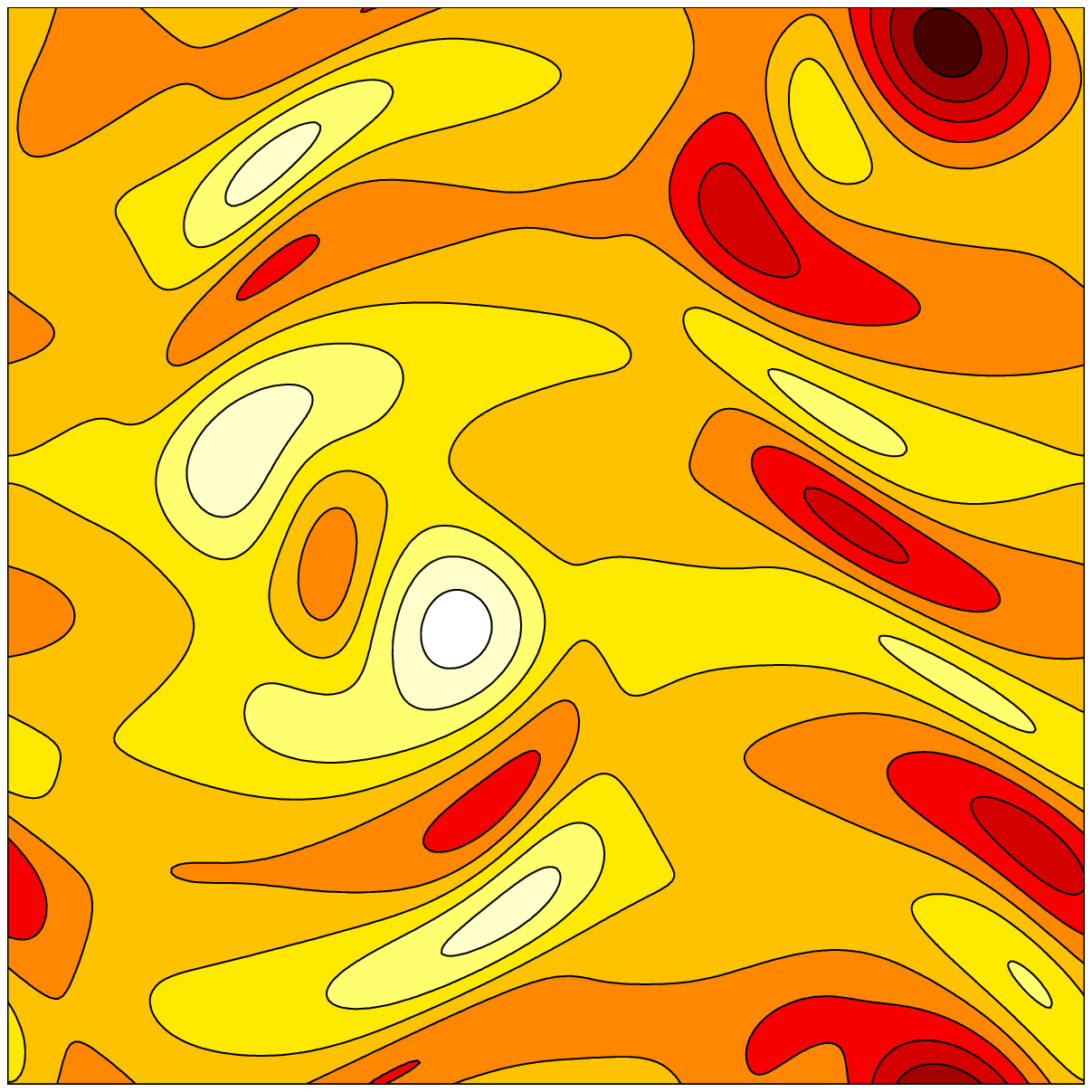,width=3.2cm,height=3.2cm,clip=true}}
\put(3.4,3.4) {\epsfig{figure=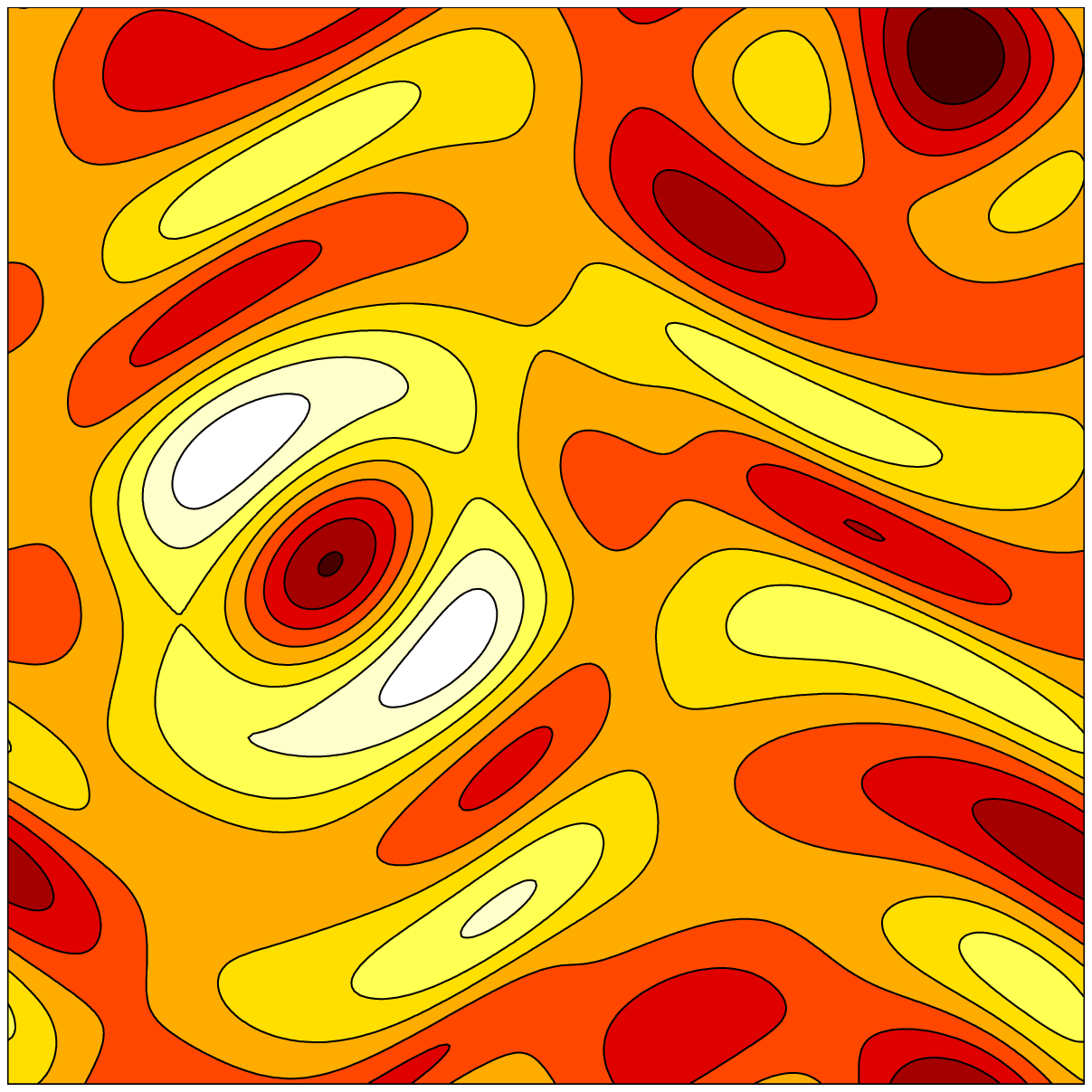,width=3.2cm,height=3.2cm,clip=true}}
\put(6.8,3.4) {\epsfig{figure=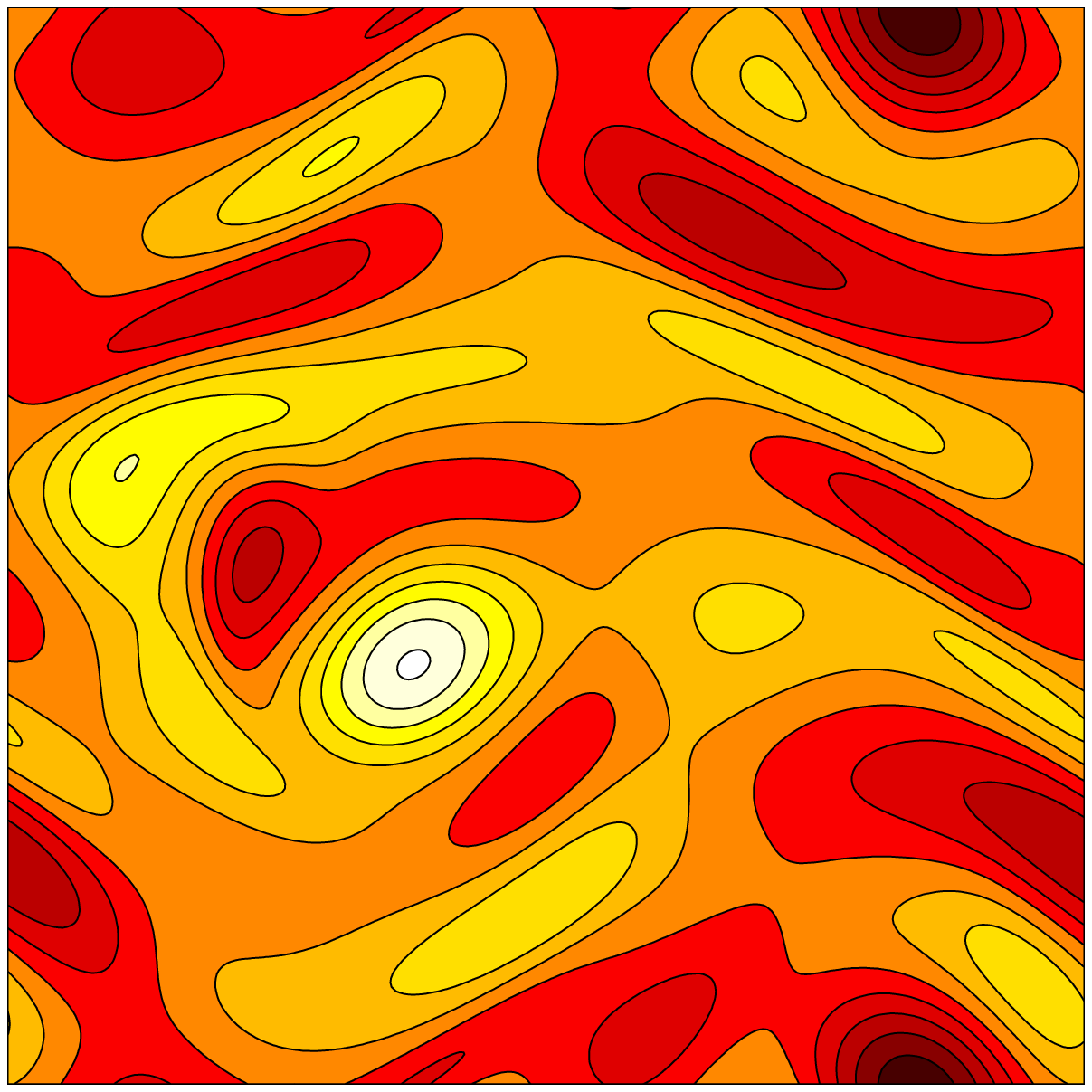,width=3.2cm,height=3.2cm,clip=true}}
\put(10.2,3.4){\epsfig{figure=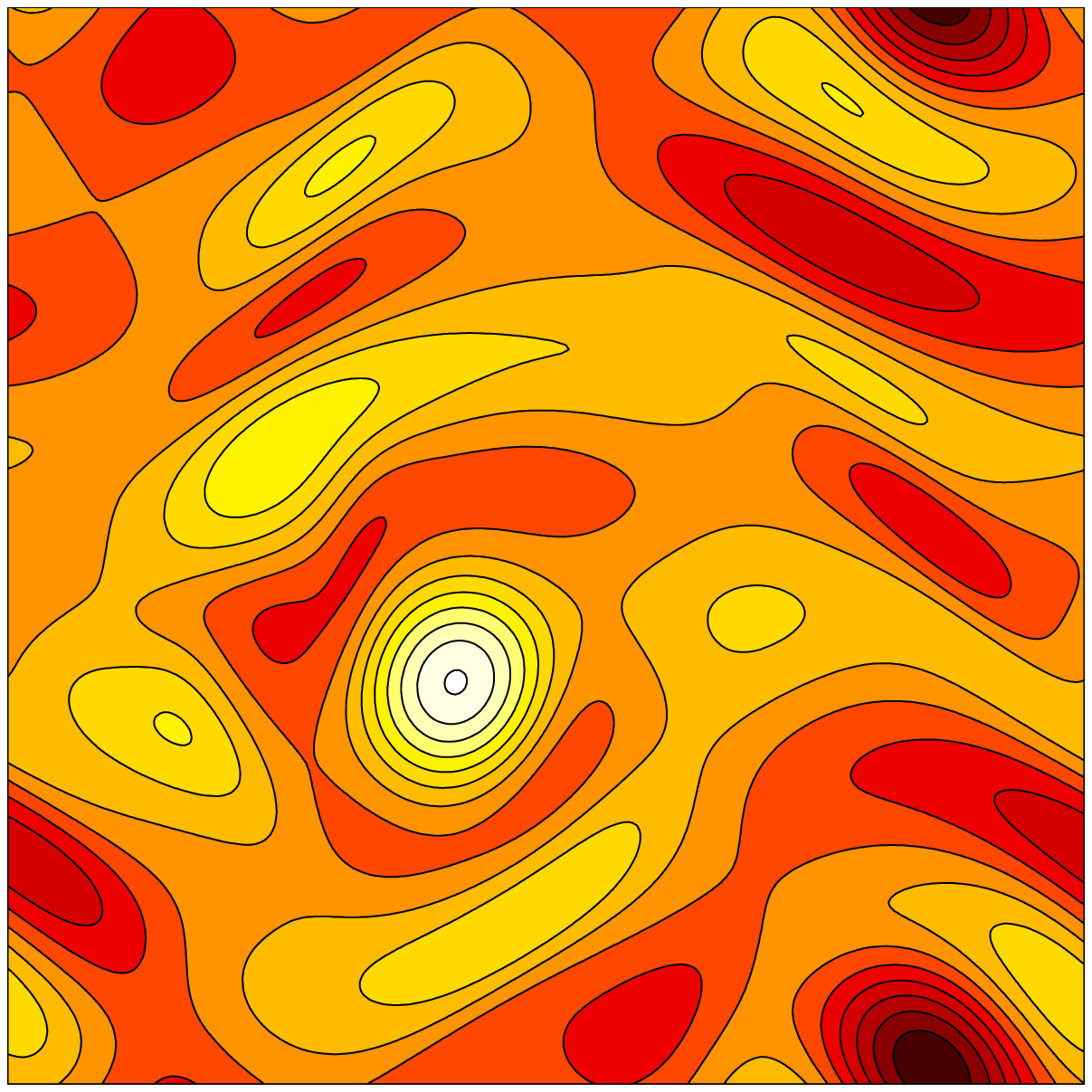,width=3.2cm,height=3.2cm,clip=true}}
\put(0,0)     {\epsfig{figure=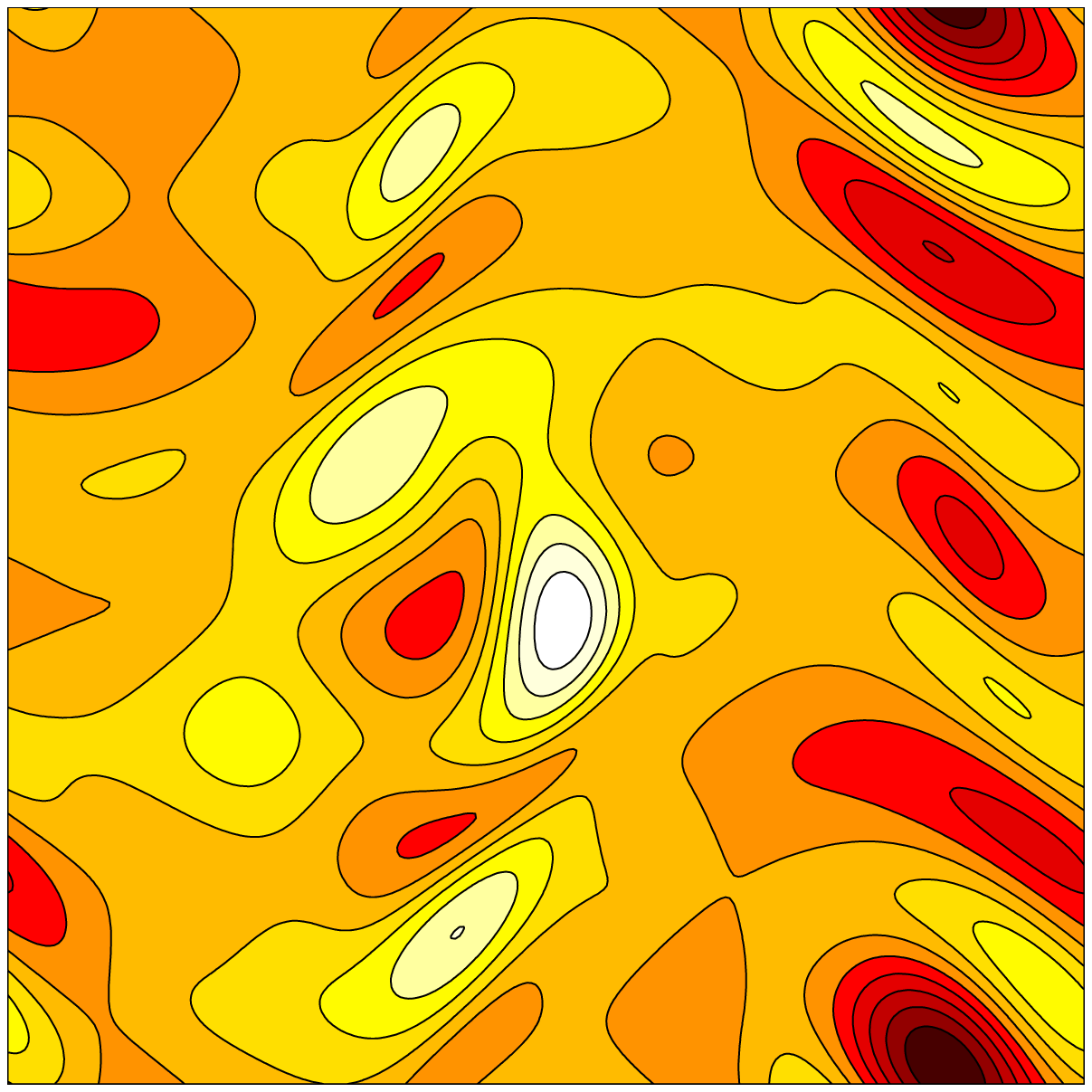,width=3.2cm,height=3.2cm,clip=true}}
\put(3.4,0)   {\epsfig{figure=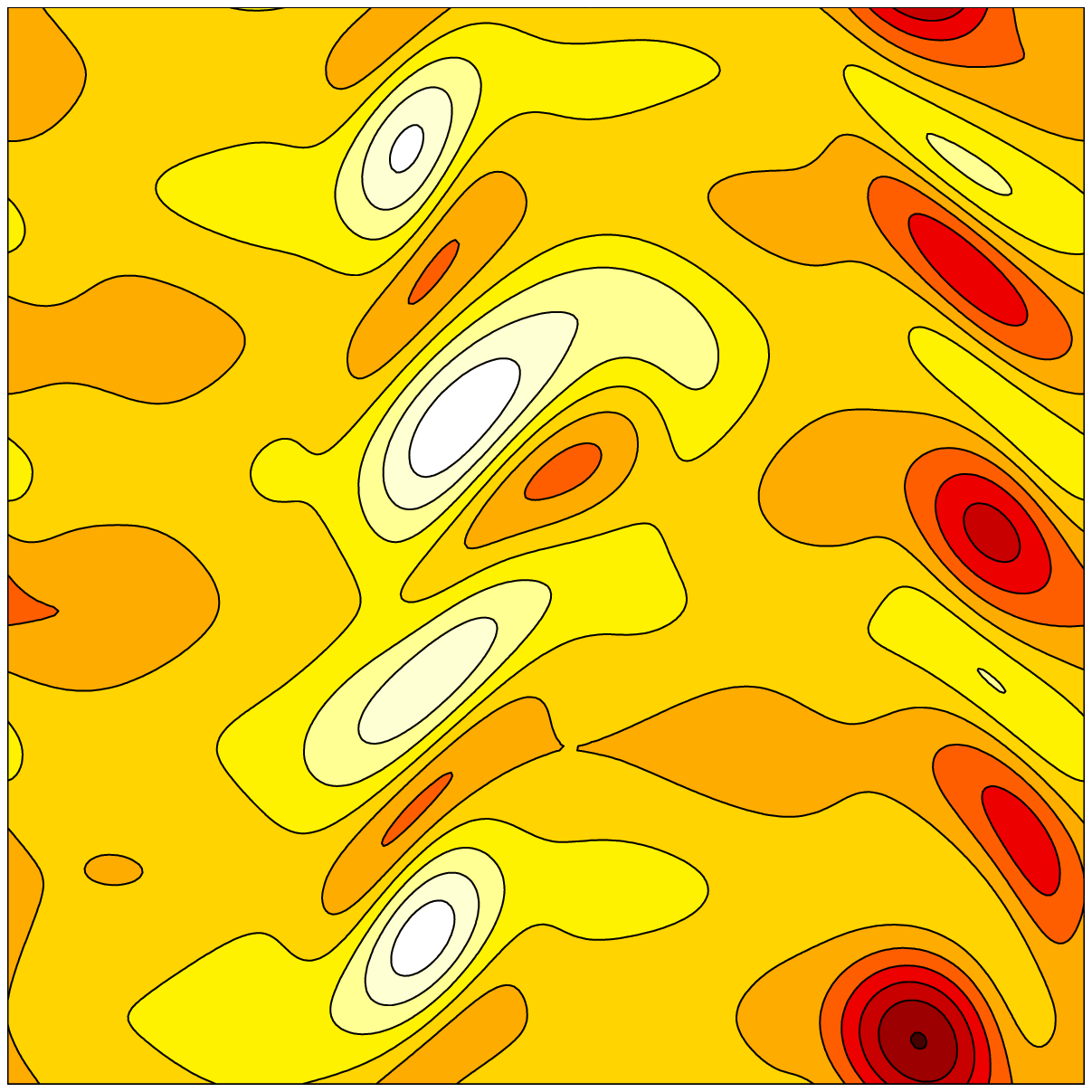,width=3.2cm,height=3.2cm,clip=true}}
\put(6.8,0)   {\epsfig{figure=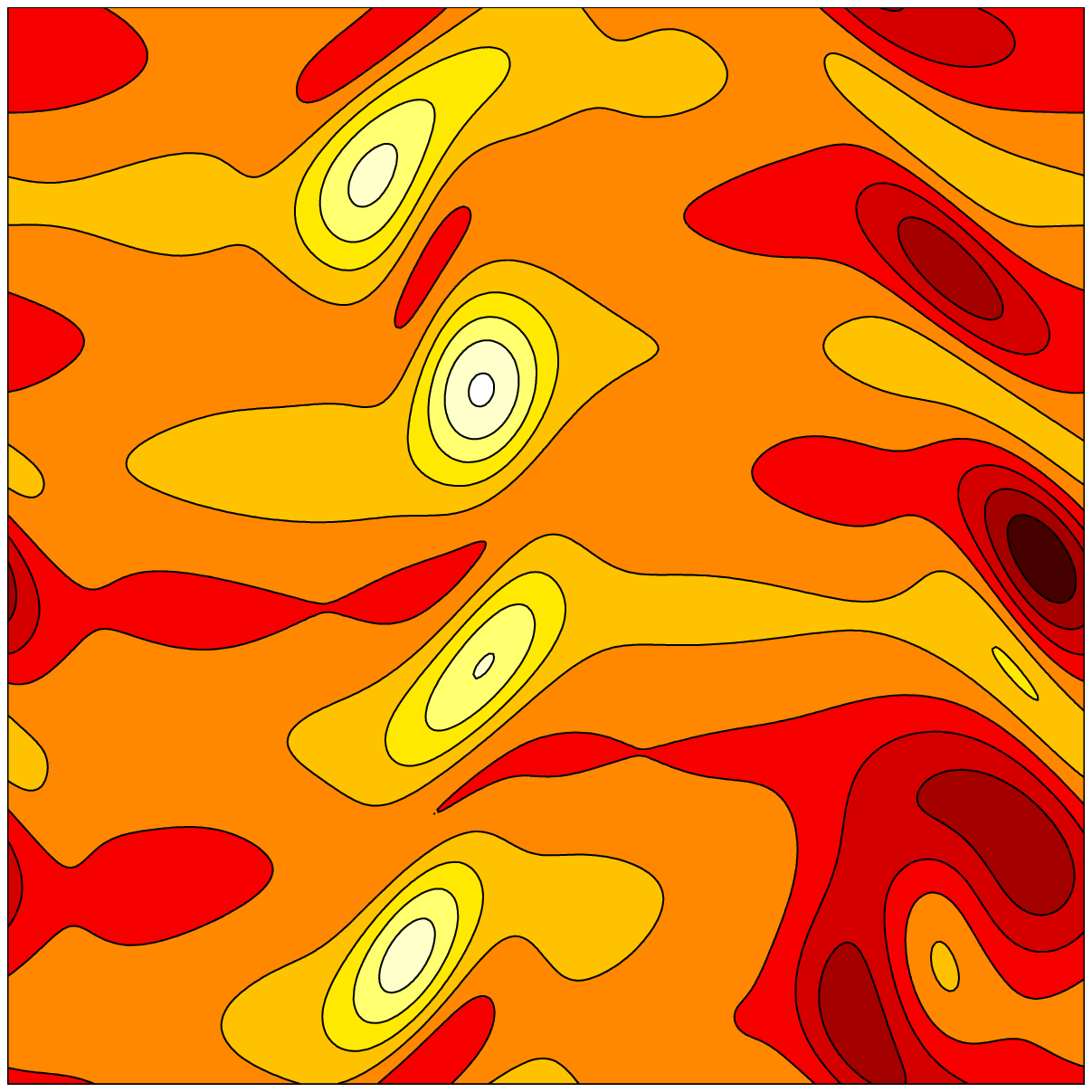,width=3.2cm,height=3.2cm,clip=true}}
\put(10.2,0)  {\epsfig{figure=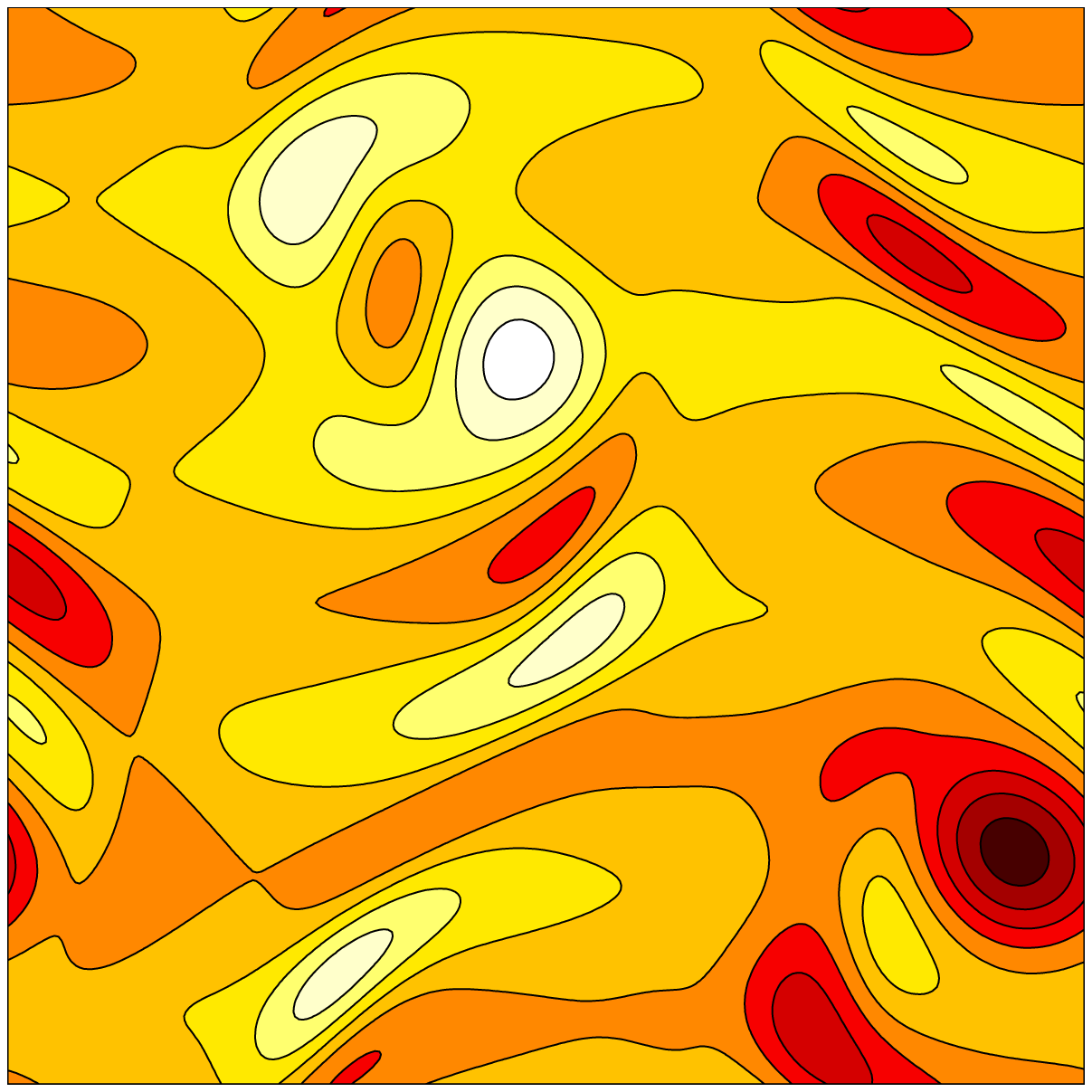,width=3.2cm,height=3.2cm,clip=true}}
\end{picture}
\end{center}
\caption{A time sequence of vorticity plots for $R25$ at $Re=40$ at
  times (running left to right across the top and then bottom)
  $t=0,5,8,10,13,15,17,20$ (marked as dots on figure
  \ref{DvsI_R25}). The period is 20.2 so the flow in the bottom right
  is nearly the same as the top left {\em except} for shifts in $x$
  and $y$. For all, 15 contours are plotted from -12 to 12.}
\label{plot_R25}
\end{figure}

\begin{figure}
\centerline{\includegraphics[scale=0.7]{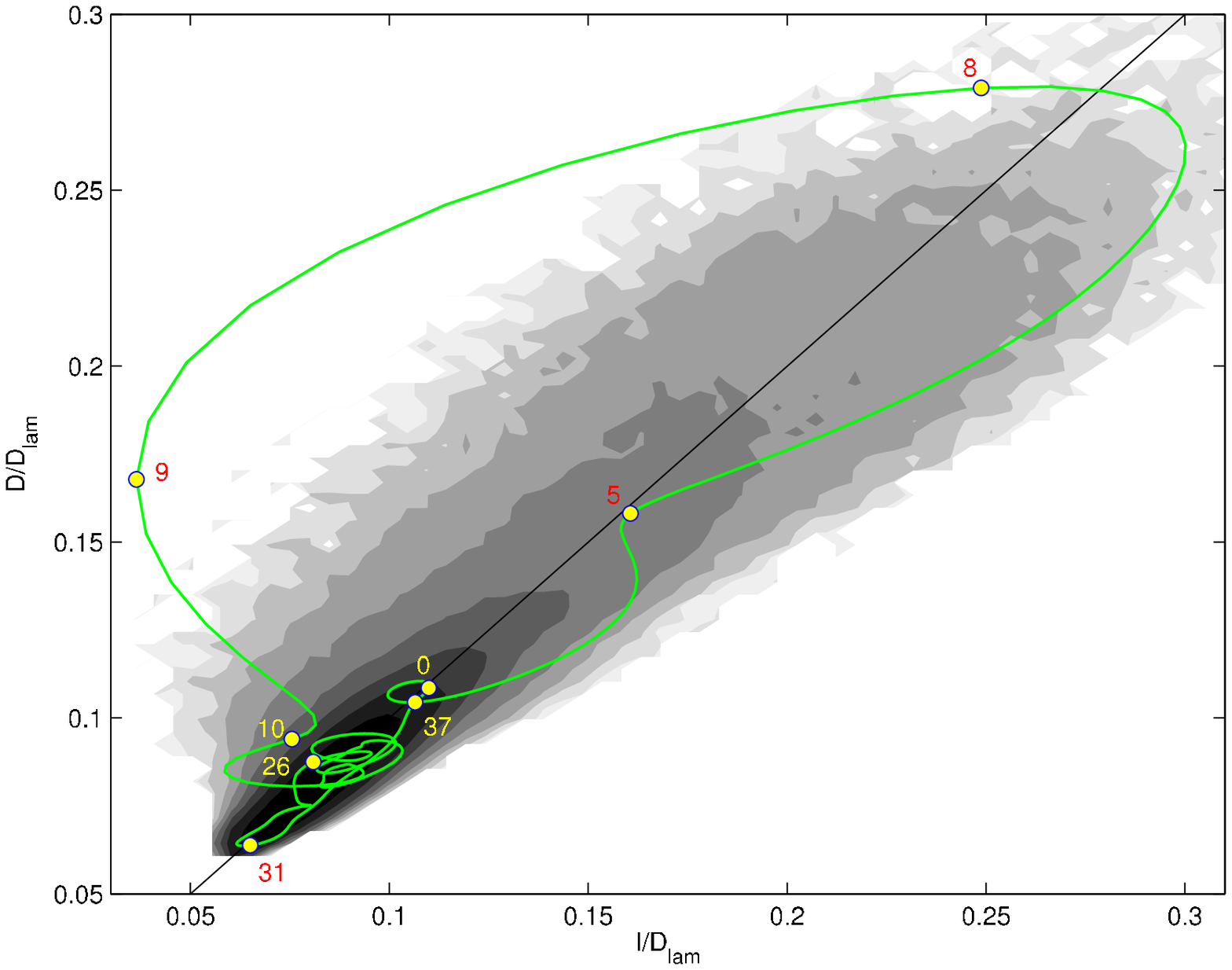}}
\caption{ The normalised dissipation $D(t)/D_{lam}$ verses $I/D_{lam}$
  for the relative periodic orbit $R50$ at Re=$40$ with the pdf of the
  DNS turbulence plotted in the background (11 shades at levels $10^{\alpha}$ where
  $\alpha={-5,-4.5,\ldots,-0.5,0}$). The labels refer to times along the orbit at
  which snapshots are shown in figure \ref{plot_R50}.}
\label{DvsI_R50}
\end{figure}

\begin{figure}
\begin{center}\setlength{\unitlength}{1cm}
\begin{picture}(14,7)
\put(0,3.4)   {\epsfig{figure=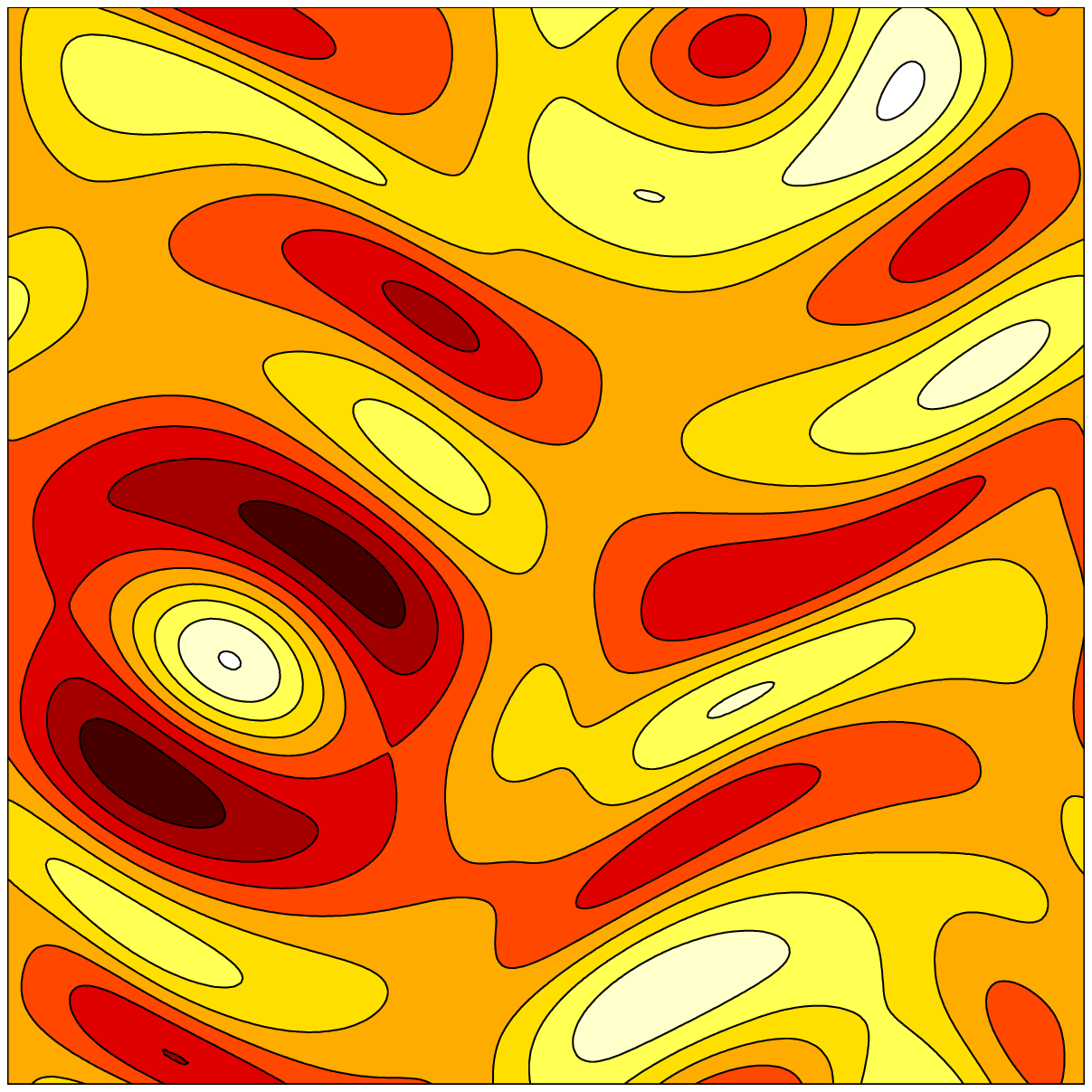,width=3.2cm,height=3.2cm,clip=true}}
\put(3.4,3.4) {\epsfig{figure=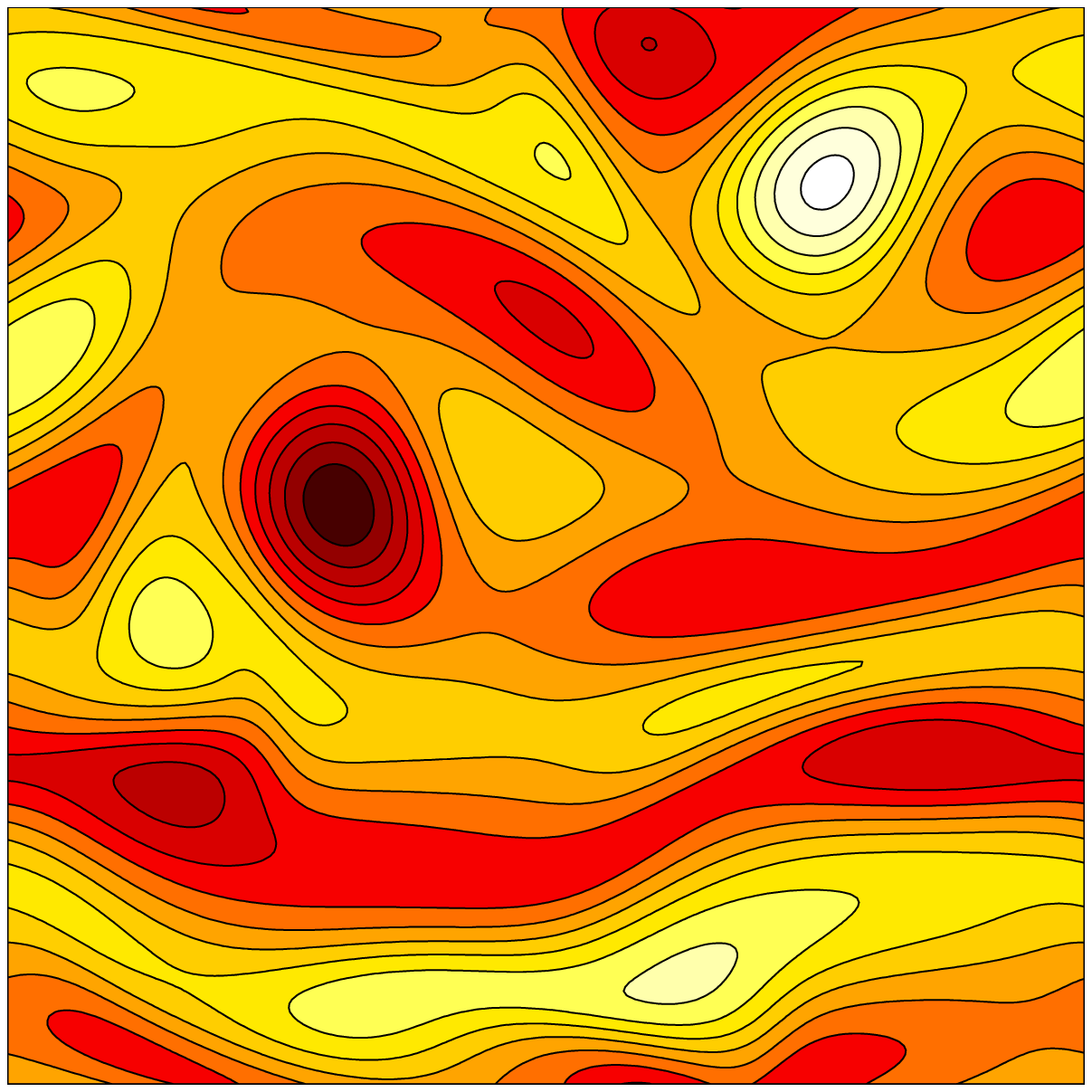,width=3.2cm,height=3.2cm,clip=true}}
\put(6.8,3.4) {\epsfig{figure=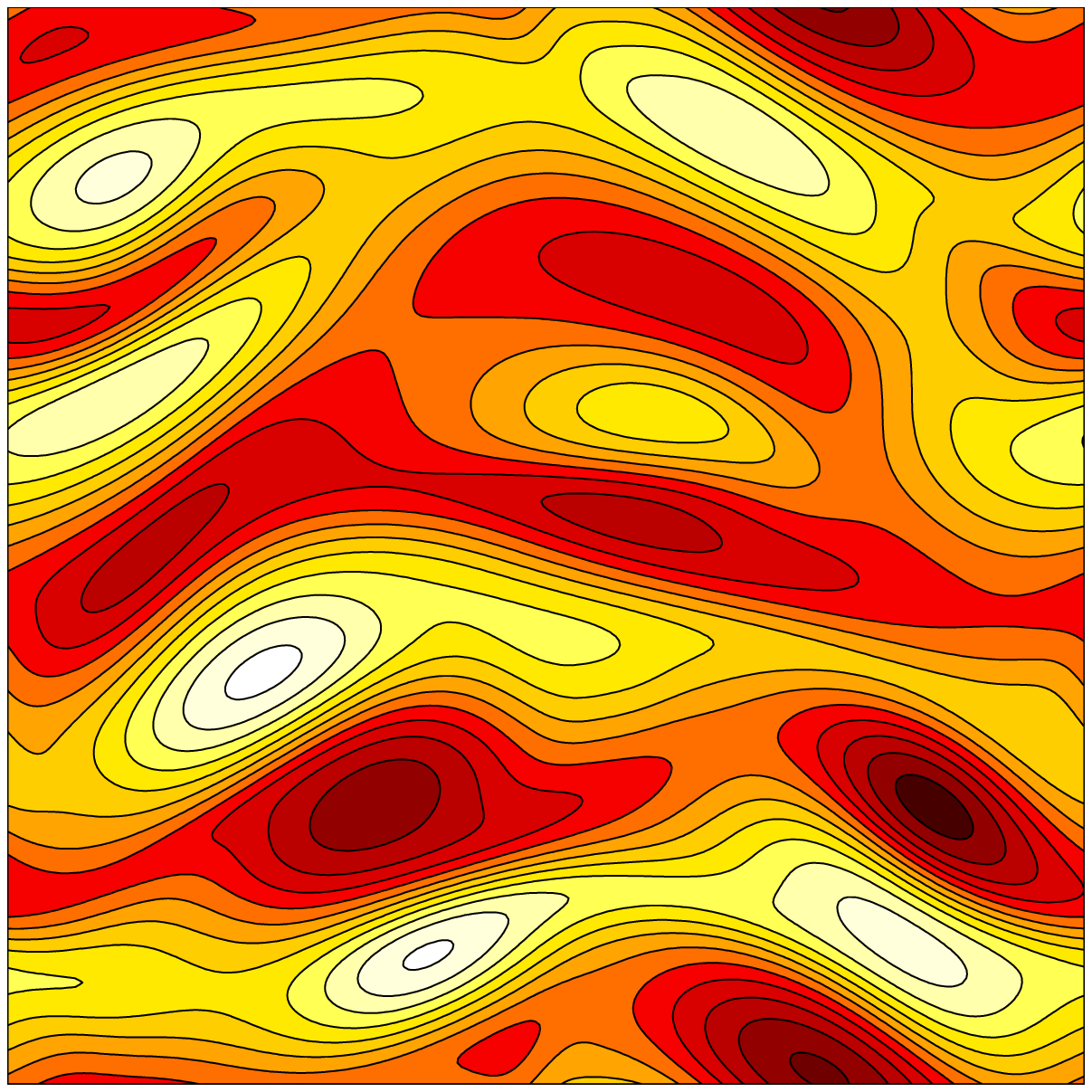,width=3.2cm,height=3.2cm,clip=true}}
\put(10.2,3.4){\epsfig{figure=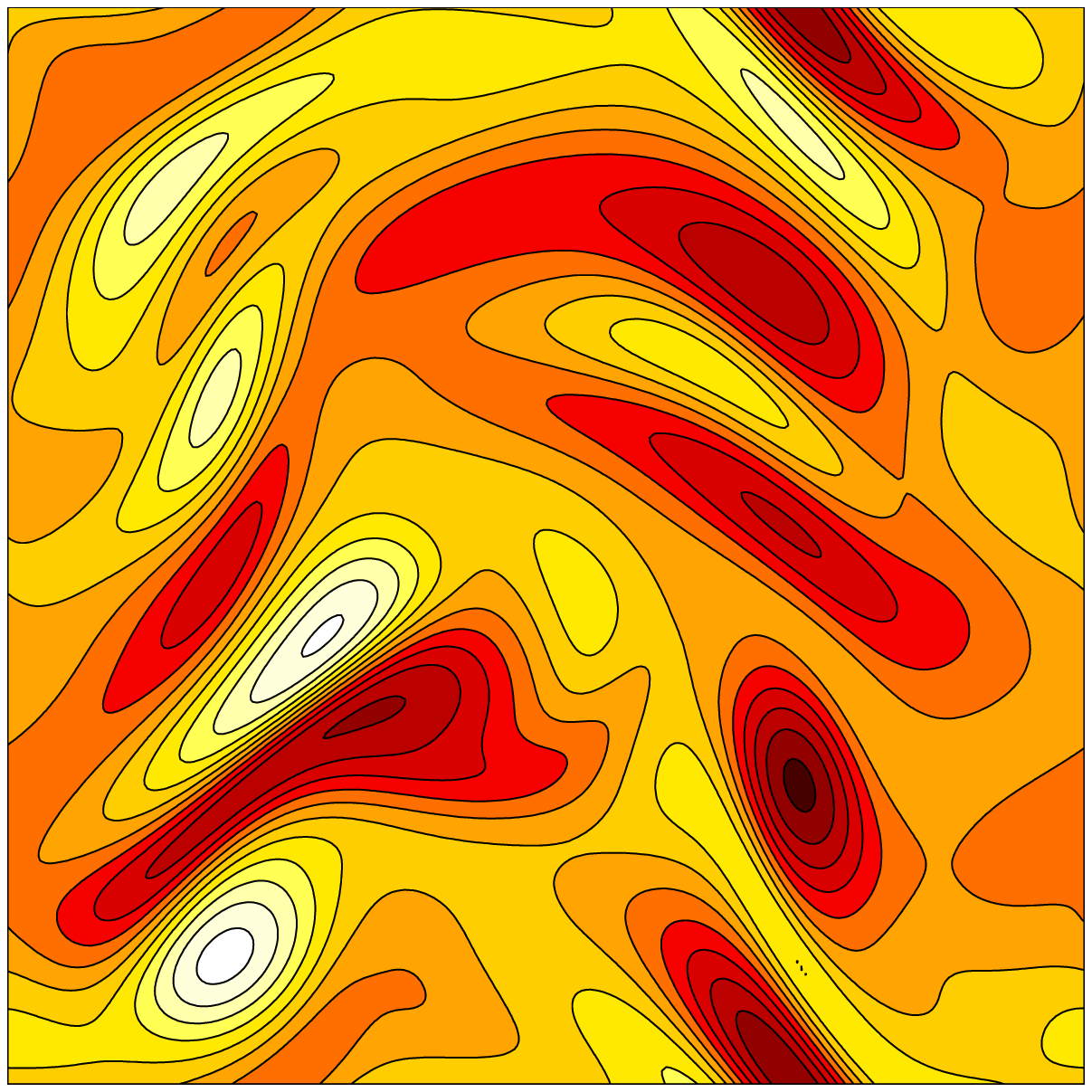,width=3.2cm,height=3.2cm,clip=true}}
\put(0,0)     {\epsfig{figure=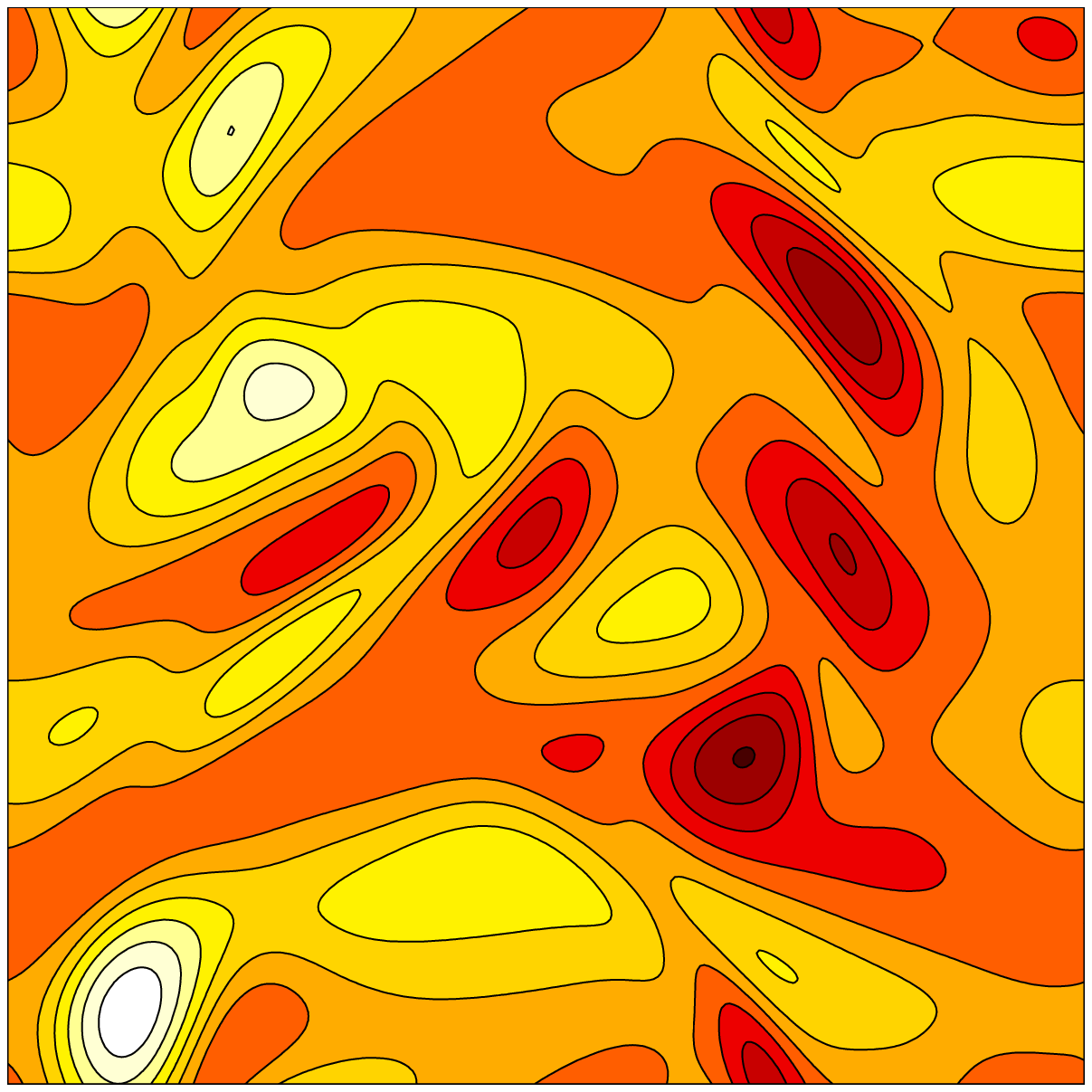,width=3.2cm,height=3.2cm,clip=true}}
\put(3.4,0)   {\epsfig{figure=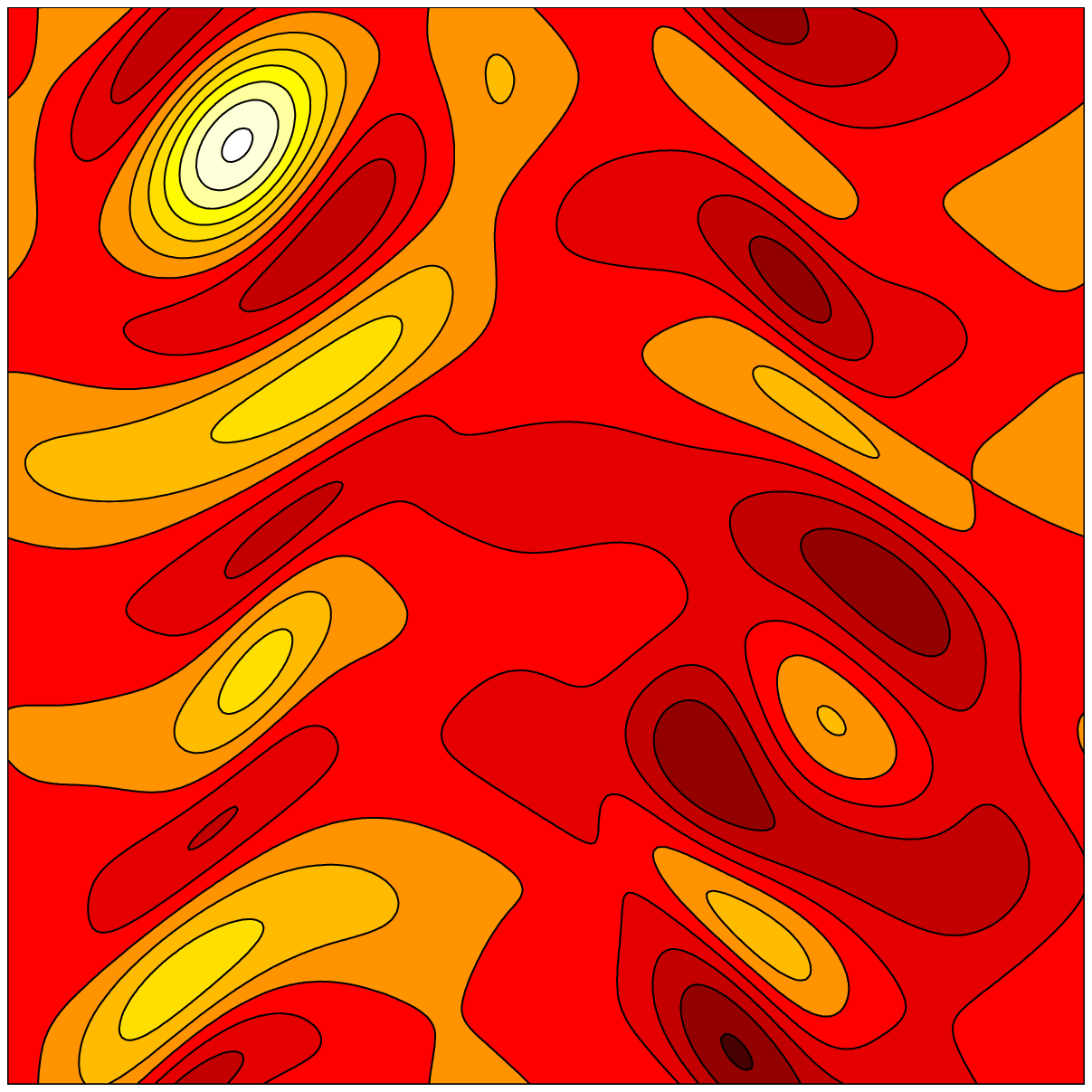,width=3.2cm,height=3.2cm,clip=true}}
\put(6.8,0)   {\epsfig{figure=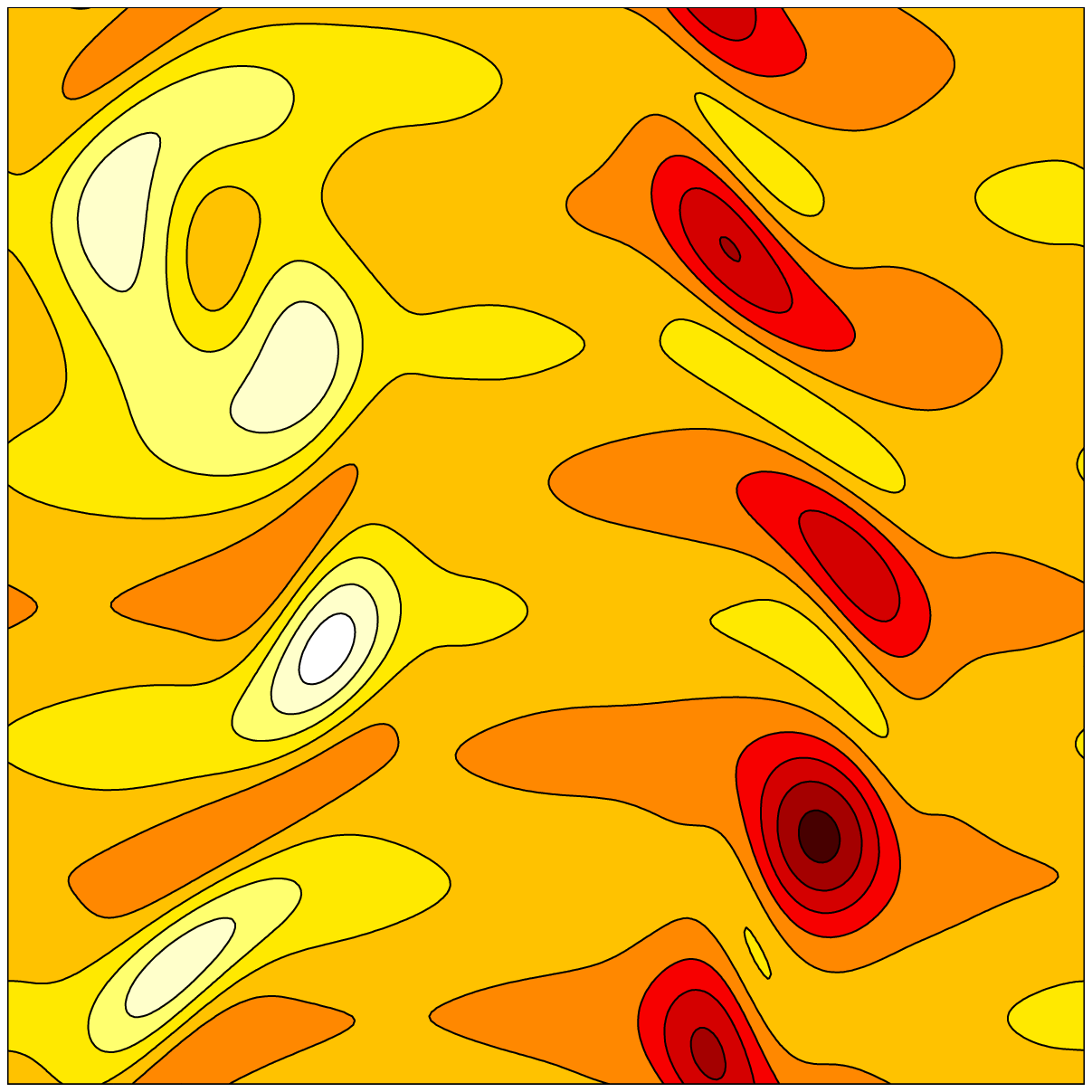,width=3.2cm,height=3.2cm,clip=true}}
\put(10.2,0)  {\epsfig{figure=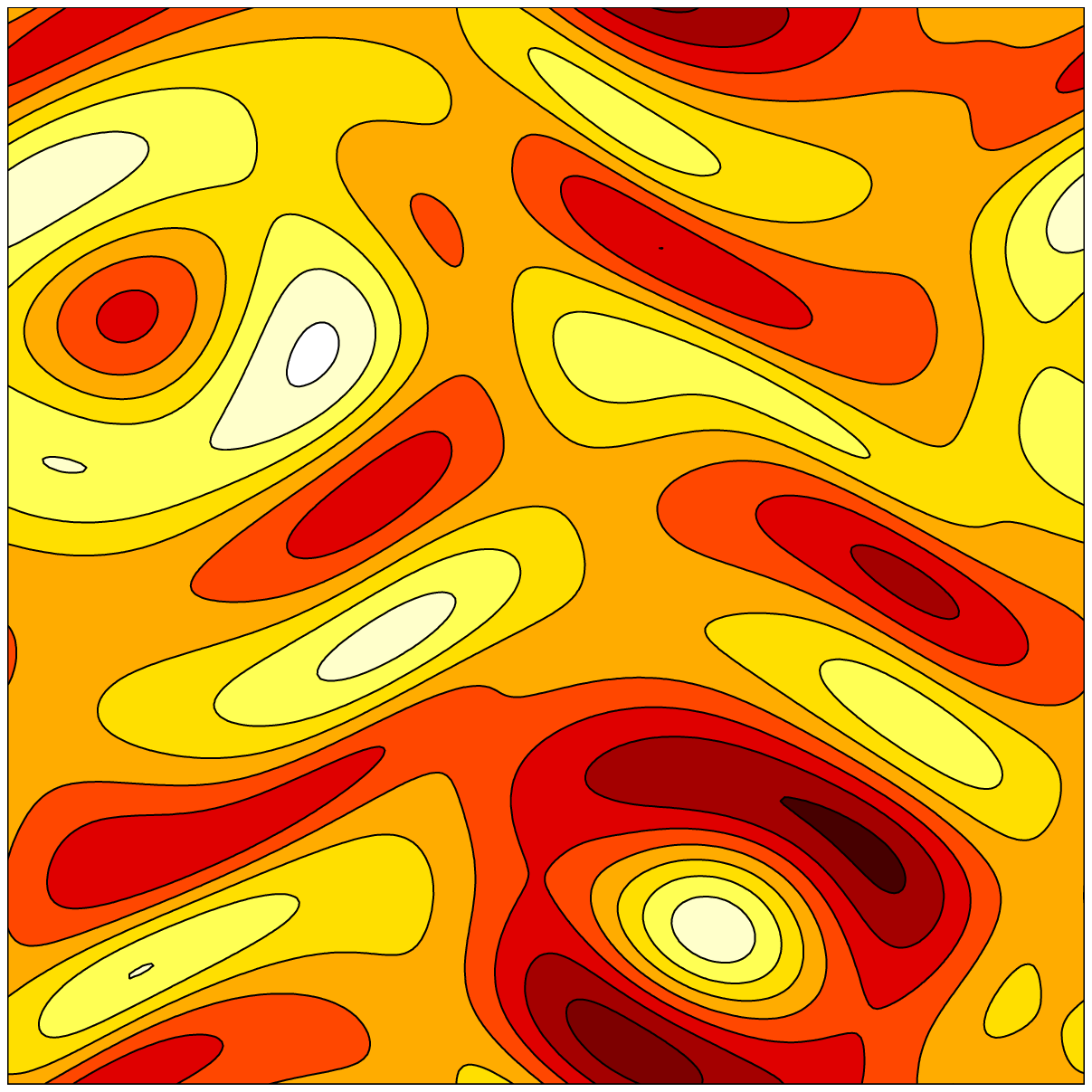,width=3.2cm,height=3.2cm,clip=true}}
\end{picture}
\end{center}
\caption{A time sequence of vorticity plots for $R50$ at $Re=40$ at
  times (running left to right across the top and then bottom)
  $t=0,5,8,9,10,26,31,37$ (marked as dots on figure
  \ref{DvsI_R50}). The period is 37.7 so the flow in the bottom right
  is nearly the same as the top left {\em except} for shifts in $x$
  and $y$. For all, 15 contours are plotted from -12 to 12.}
\label{plot_R50}
\end{figure}

In figure \ref{DvsI_R25} we focus on one typical `embedded' relative
periodic orbit $R25$ which stays within the central region of the DNS
joint pdf. There is a clear temporal cycle where the energy input
increases (exceeding the dissipation) and then decreases (now exceeded
by the dissipation).  Plotting the associated vorticity fields over
this cycle - figure \ref{plot_R25} - shows the character of the
flow. At the dissipation low point (time 17 in figures \ref{DvsI_R25}
and \ref{plot_R25}), the vorticity is concentrated into weak
$y$-aligned patches which are separated from each other whereas at the
high dissipation point (time 8), the vorticity seems to be undergoing
a shearing episode with only one stronger vortex recognisable. These
two extremes bear more than a passing resemble to either $T1$
($D/D_{lam}=0.071$) or $T2$ (0.071) and $E1$ (0.102) respectively
suggesting that $R25$ is probably a closed trajectory linking their
neighbourhoods. $R50$, in contrast, undergoes a large high dissipation
excursion as shown more completely in figure \ref{DvsI_R50}. The
associated vorticity fields - see figure \ref{plot_R50} - show similar
structures to $R25$ when in the same part of $(I,D)$ space (compare
t=5 for $R25$ with t=0 for $R50$, and t=15 for $R25$ and t=31 for
$R50$) but $R50$ exhibits intense shearing too and vortex break-up at
times 5, 8 and 9.  $R50$ clearly reflects an important but infrequent
aspect of the turbulent dynamics as indicated by the fact that the
joint $(D,I)$ pdf of the DNS stretches to such high values of the
dissipation. Whether we have extracted enough of such recurrent
structures to capture this episodic behaviour is of course a key issue
for this study and will be discussed in \S \ref{4.4}

\begin{figure}
\begin{center}\setlength{\unitlength}{1cm}
\begin{picture}(14,13)
\put(0,6.5)     {\epsfig{figure=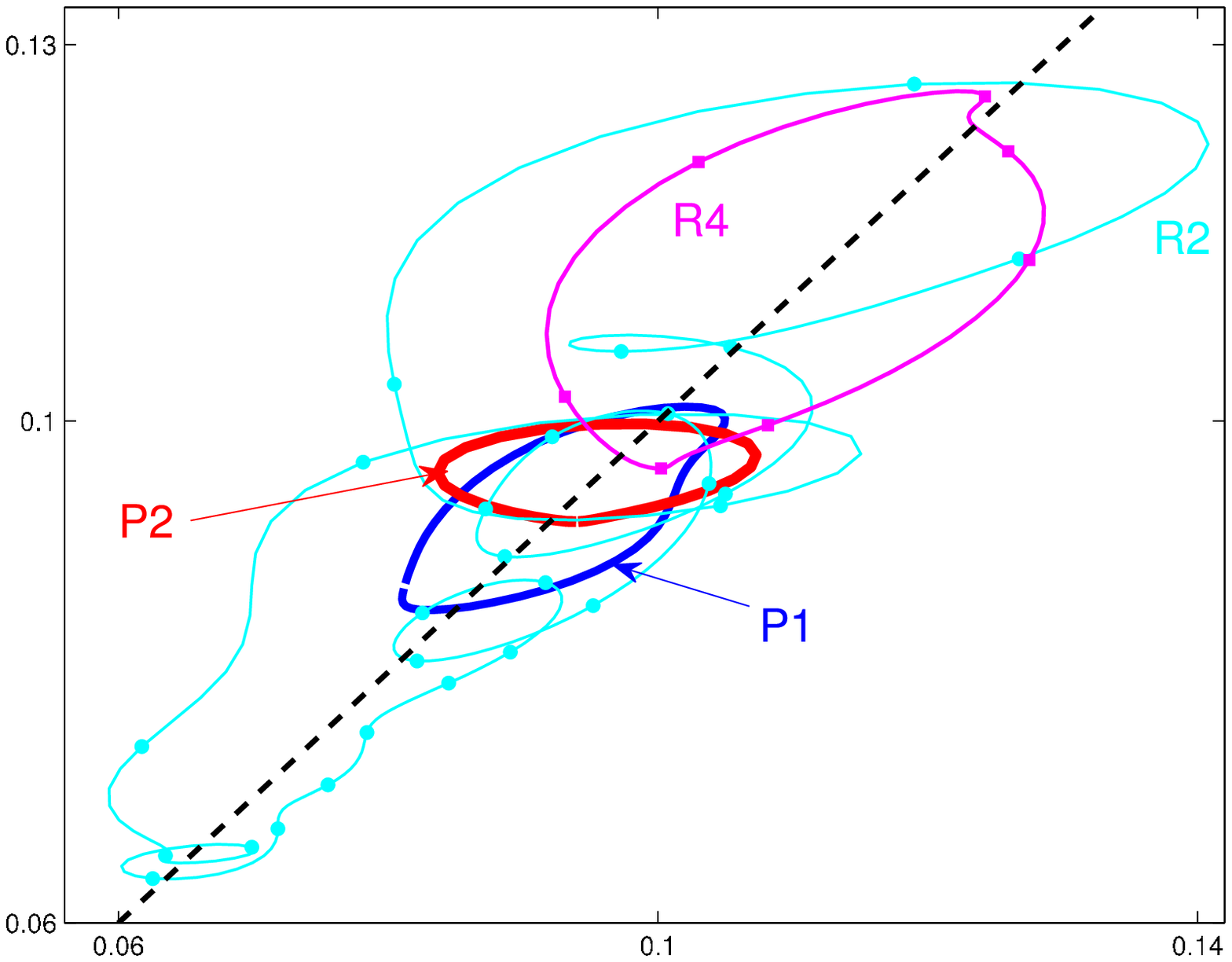,width=6.5cm,height=6.25cm,clip=true}}
\put(6.75,6.5)   {\epsfig{figure=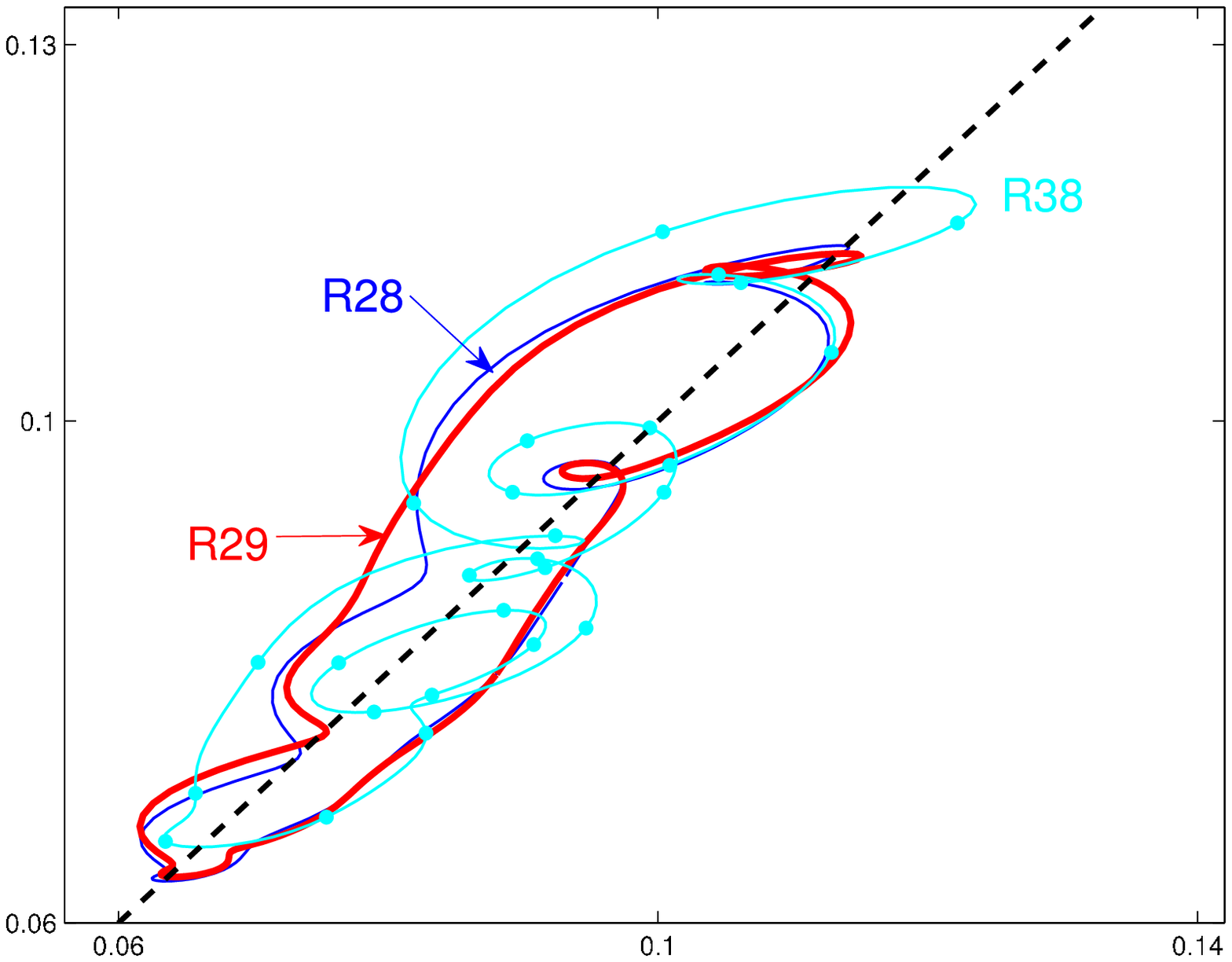,width=6.5cm,height=6.25cm,clip=true}}
\put(0,0)   {\epsfig{figure=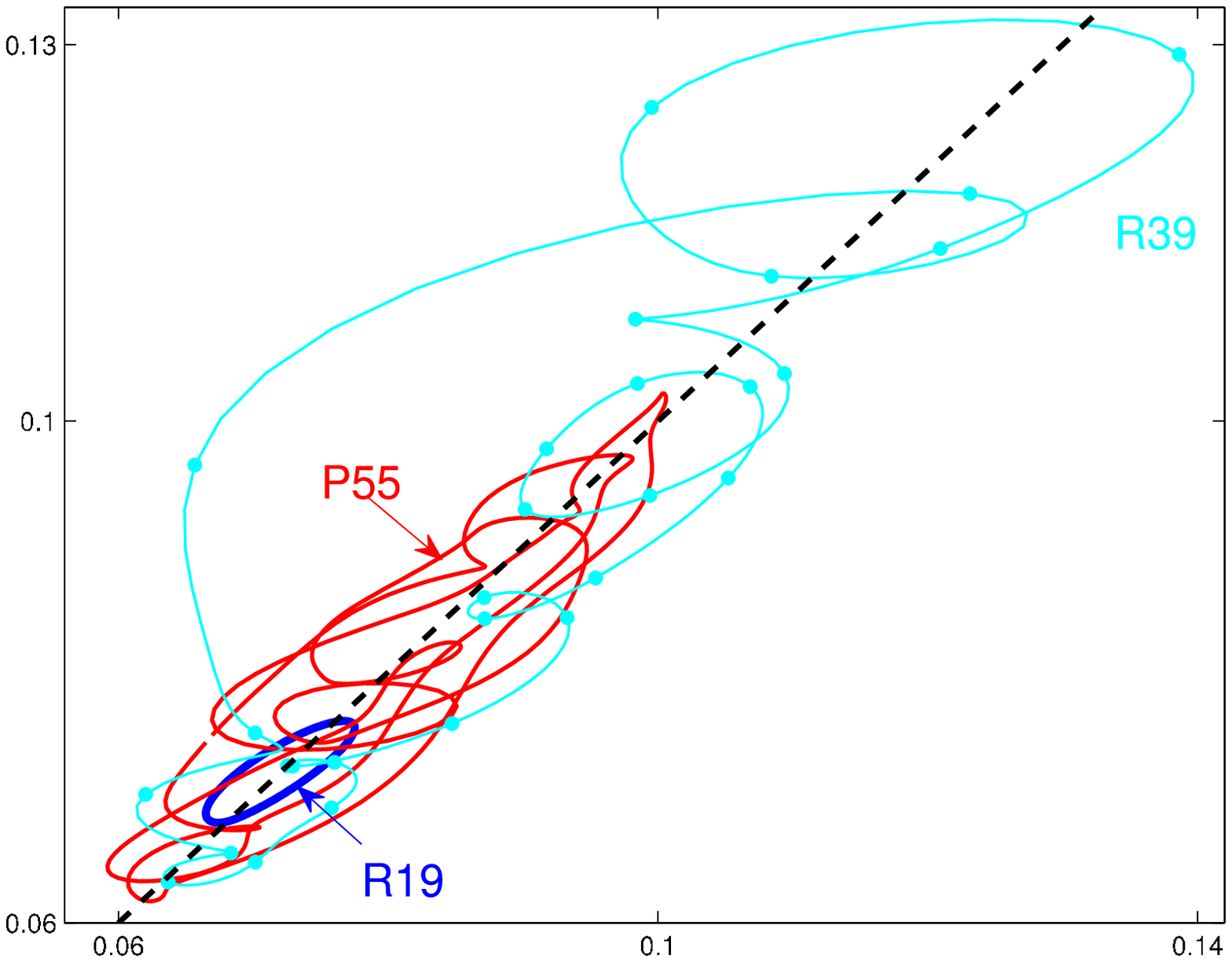,width=6.5cm,height=6.25cm,clip=true}}
\put(6.75,0) {\epsfig{figure=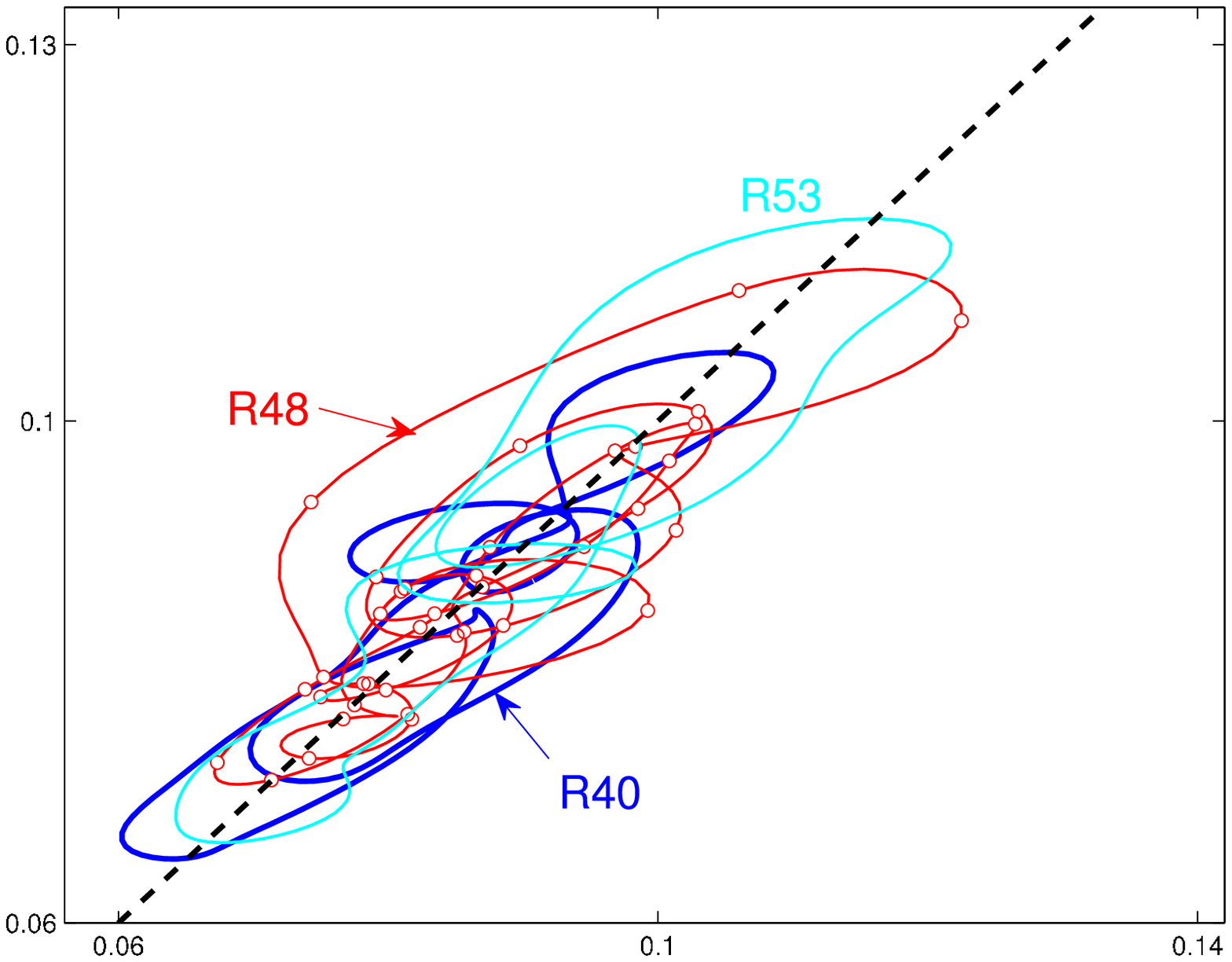,width=6.5cm,height=6.25cm,clip=true}}
\end{picture}
\end{center}
\caption{A sampling of the recurrent structures found at $Re=40$
  plotted over the zoom-in box shown in figure
  \ref{DvsI_40_sample}. Upper left shows $P1$,$P2$,$R2$ and $R4$; Upper
  right shows $R28$,$R29$ and $R38$; Lower left shows $R19$, $R39$ and
$R55$; and lower right shows $R40$, $R48$ and $R53$. In all cases,
  symbols added to lines are to assist in their distinction and are
  placed 1 time unit apart to indicate speed of flow.}
\label{DvsI_40_multi}
\end{figure}

Figure \ref{DvsI_40_multi} is an attempt to show more of the recurrent
structures found by zooming in on the central dashed box drawn in
figure \ref{DvsI_40_sample}.  This illustrates the intricacy of most
of the flows found - many of the relative periodic orbits trace
complicated $D-I$ curves whereas, in contrast, the periodic orbits are
simple loops. Another key observation is that some relative periodic
orbits look very similar - e.g. $R28$ and $R29$ (and other pairings
not shown). This, of course, resonates with the mental picture one has
of periodic orbits being dense in a chaotic attractor. In fact, the
consecutive numbering of $R28$ and $R29$ indicates that these relative
periodic orbits were found concurrently from the DNS confirming their
proximity in phase space. Also it is clear than some relative periodic
orbits look like merged versions of two shorter orbits (not shown) again
consistent with low-dimensional dynamical systems thinking.

%
%
%

%
%

%
%

%
%

%
%

%
%

\subsubsection{Re=60, 80 \& 100}

\begin{figure}
\centerline{\includegraphics[scale=0.7]{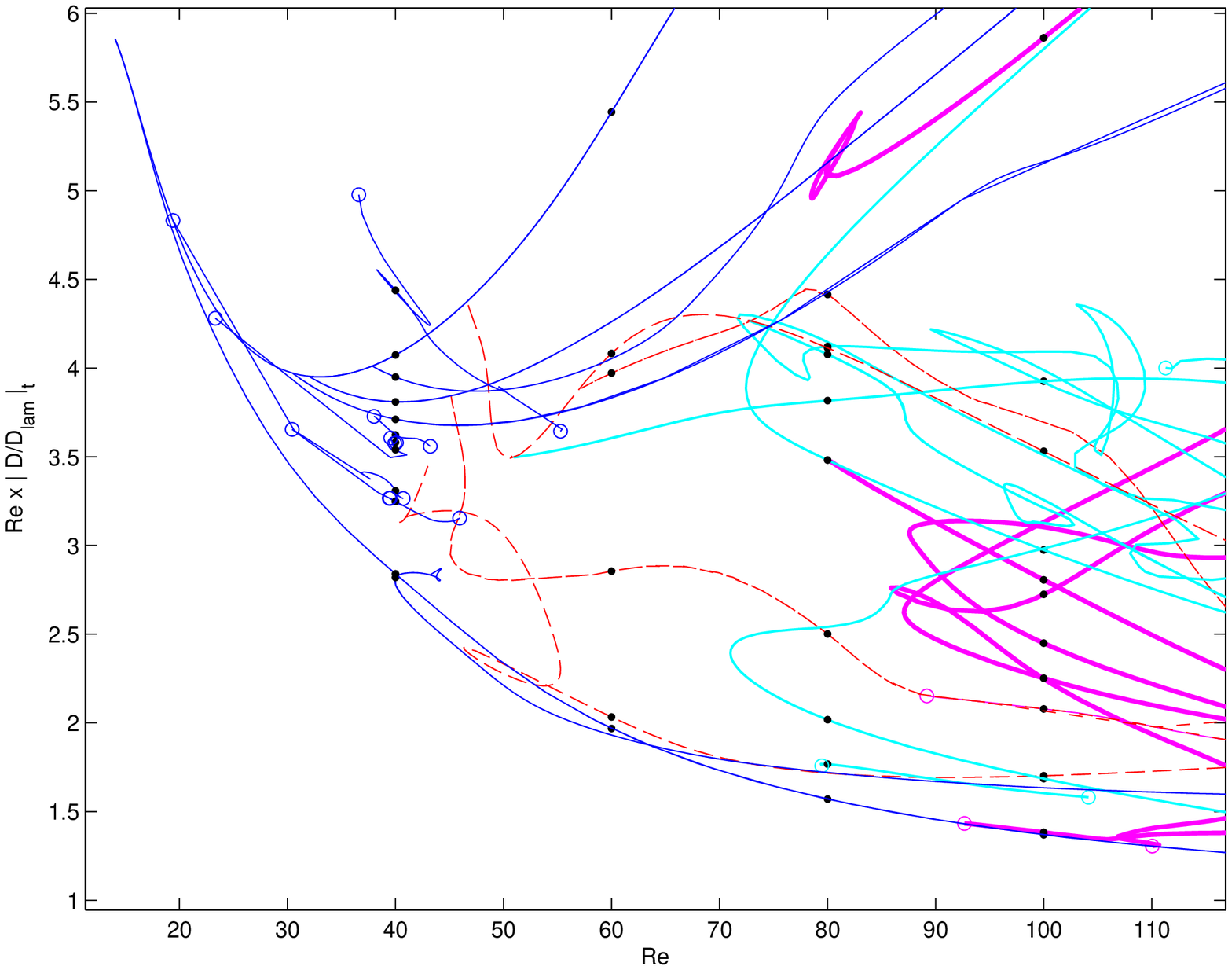}}
\caption{The scaled, time-averaged dissipation $Re\langle D/D_{lam}
  \rangle_t$ verses $Re$ for recurrent flows discovered by the series
  A runs. Blue thin lines trace recurrent flows found at $Re=40$, red
  thin dashed lines new recurrent flows found at $Re=60$, thick cyan
  lines those found at $Re=80$ and very thick magenta lines those at
  $Re=100$. The black dots identify the subset of solutions which were
  identified by processing the dns runs at the respective $Re$ as
  opposed to just being continued up or down from other $Re$
  (e.g. there are 6 dots at $Re=60$ corresponding to $E1,T1,T3,T4,R7$
  and $R8$: $R56$-$R58$ were discovered in the series B runs, and none of the
  red dashed lines join dots at $Re=40$). Open circles indicate limits
  beyond which a solution branch could not be continued. The situation
  is clearly complicated with solutions seemingly dynamically
  important at some $Re$ but not at others. }
\label{Contin_D}
\end{figure}

At higher $Re$, we managed to extract far fewer recurrent flows from
the DNS. There are certainly reasons to expect this, most notably that
the recurrent flows present should become more unstable and it is
therefore harder to find good guesses from the DNS.  And there is also
the fact that the `turbulence' should explore more of phase space and
therefore close visits to simple invariant sets should become rarer.
However, the sharp drop in the number of recurrent flows found - see
Table 3 - was still a surprise. In keeping with the philosophy of this
work, only recurrent flows extracted from the DNS {\em at that} $Re$
are listed in Table 3 under the relevant $Re$ heading. This then says
nothing about whether a certain recurrent flow found at one $Re$ might
not exist at another.  To explore this a little, we carried out some
branch continuation (see appendix A) on the recurrent flows extracted
from the series A DNS while the runs and analysis for the series B DNS
were progressing. The results are shown in figure \ref{Contin_D}
colour coded to group recurrent flows found at the same $Re$ and with
black dots indicating branches detected at a given $Re$ (note the
rescaled dissipation measure on the ordinate to make the plot
clearer). For example, the $T3$ branch is shown as a dashed red line
(second red dashed line up from the $Re$ axis) as it was first found
at $Re=60$ and it  bears three black dots marked at $Re=60, 80$
and $100$ as $T3$ was extracted from the DNS at all three $Re$. This
type of analysis can indicate the bifurcation structure - e.g.  $T3$
clearly bifurcates off a recurrent flow found at $Re=40$ - but is very
time-consuming to pursue through to completion as branches can become
difficult to continue and interpret (note the number of open circles
in figure \ref{Contin_D} which indicate where the branch continuation
procedure stagnated for some reason). This aside, the overriding
impression is one of simple invariant sets proliferating with
increasing $Re$. Notably, only two recurrent flows found at $Re=40$
are also extracted from the $Re=60$ DNS - $E1$ (the highest blue line
with a dot at $Re=60$) and $T1$. $E1$ seems to lose dyanmical
importance for yet higher $Re$ but $T1$ is found for all four $Re$
studied here.

\begin{figure}
\begin{center}\setlength{\unitlength}{1cm}
\begin{picture}(14,5)
\put(0,0){\epsfig{figure=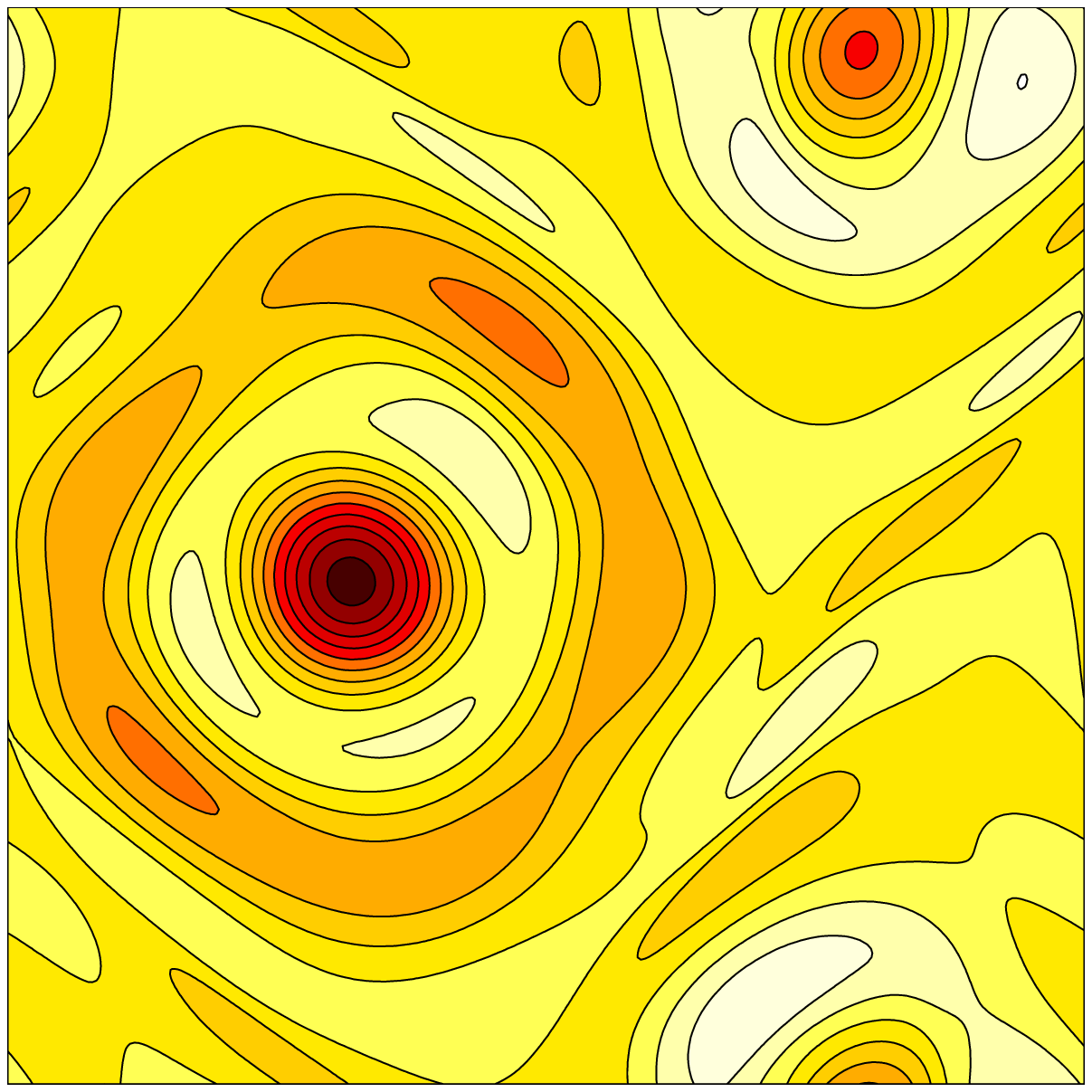,width=4.25cm,height=4.25cm,clip=true}}
\put(4.5,0){\epsfig{figure=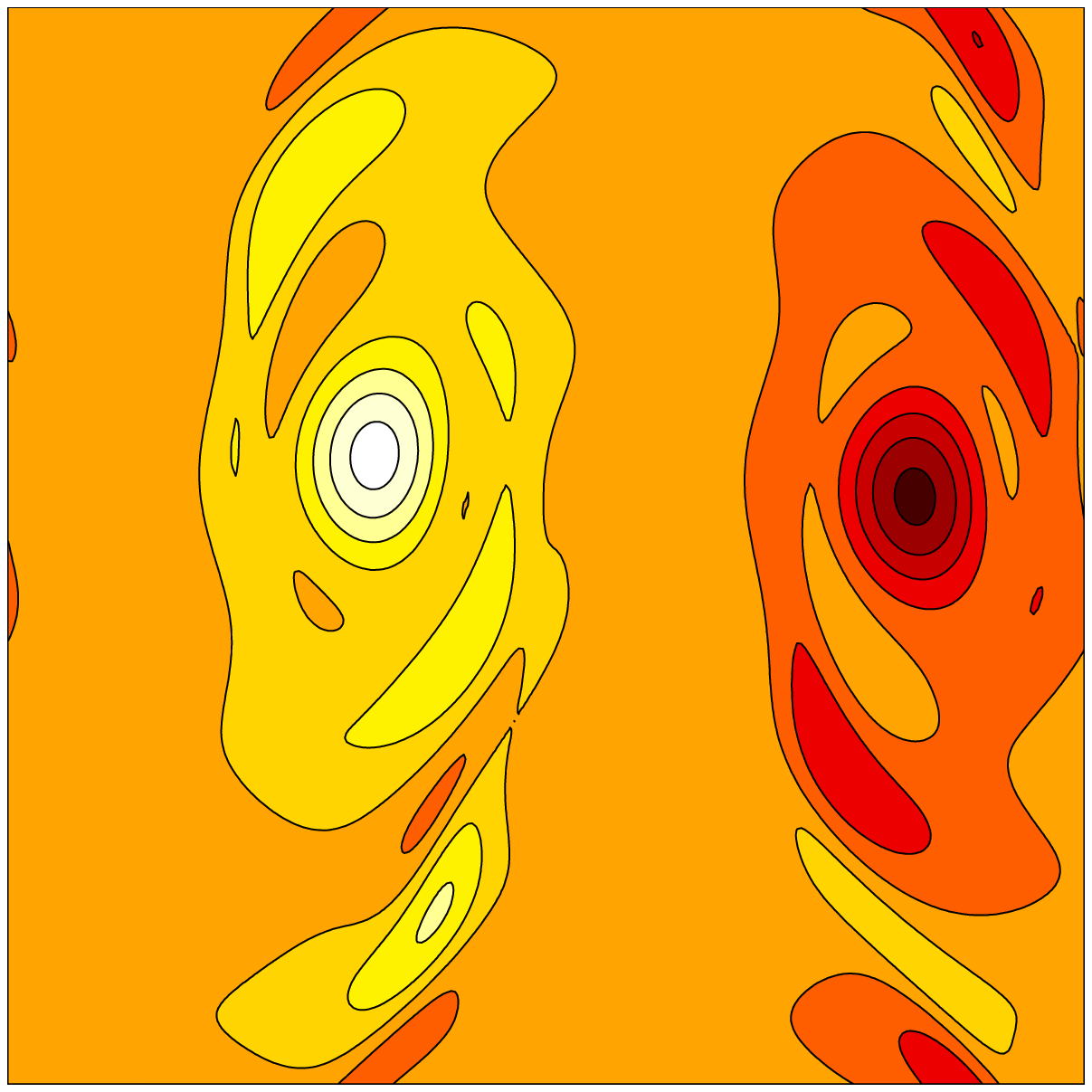,width=4.25cm,height=4.25cm,clip=true}}
\put(9,0){\epsfig{figure=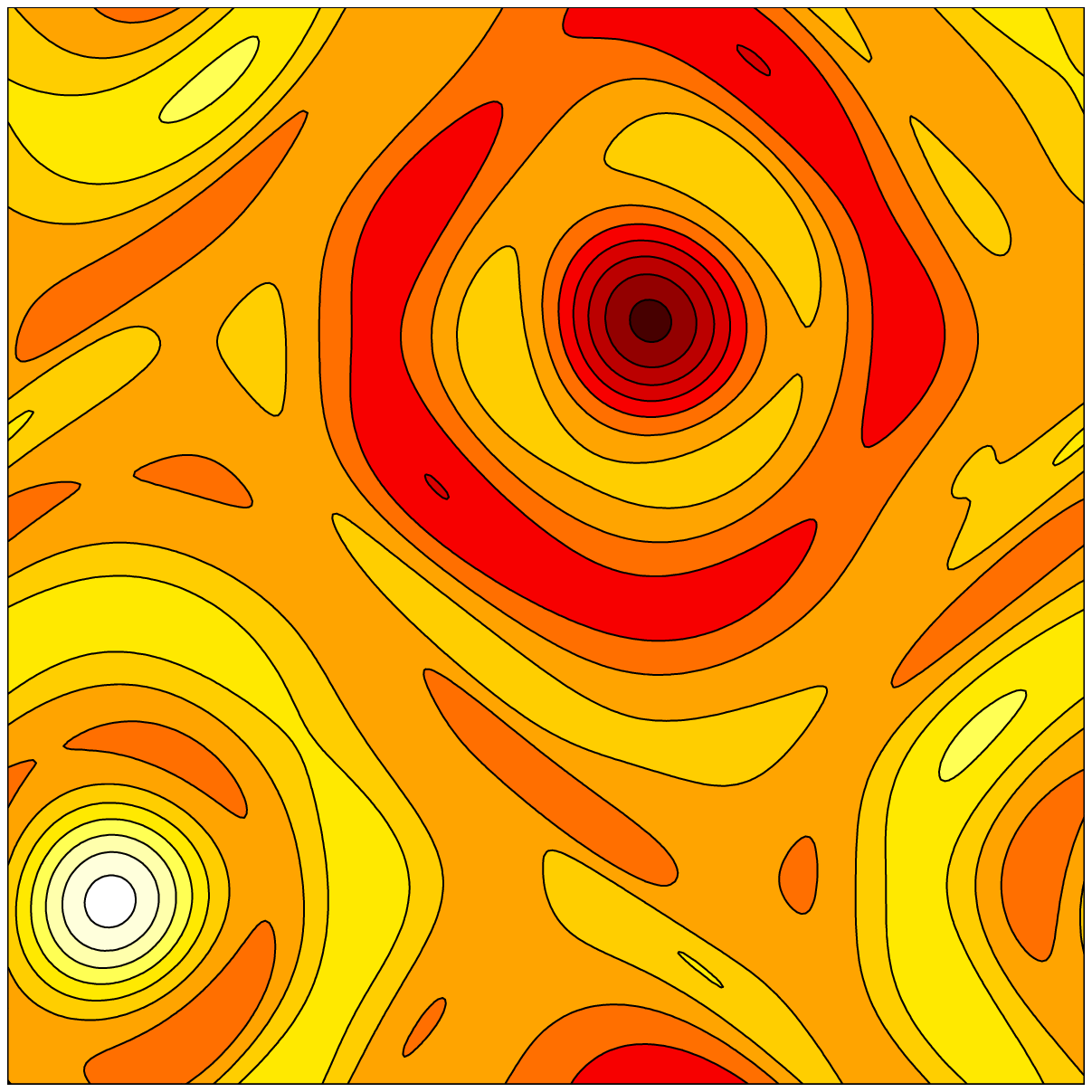,width=4.25cm,height=4.25cm,clip=true}}
\end{picture}
\end{center}
\caption{The travelling waves $T3$ (left), $T4$ (middle) and $T5$
  (right) at $Re=100$. Vorticity is contoured using 15 contours
  between -20 (dark,red) and 15 (light,white) (range is $-18.9\leq
  \omega \leq 8.03$ for $T3$, $-11.5 \leq \omega \leq 11.5$ for $T4$
  and $-13.7\leq \omega \leq 13.7$ for $T5$). }
\label{T3T4T5}
\end{figure}

\begin{figure}
\centerline{\includegraphics[scale=0.7]{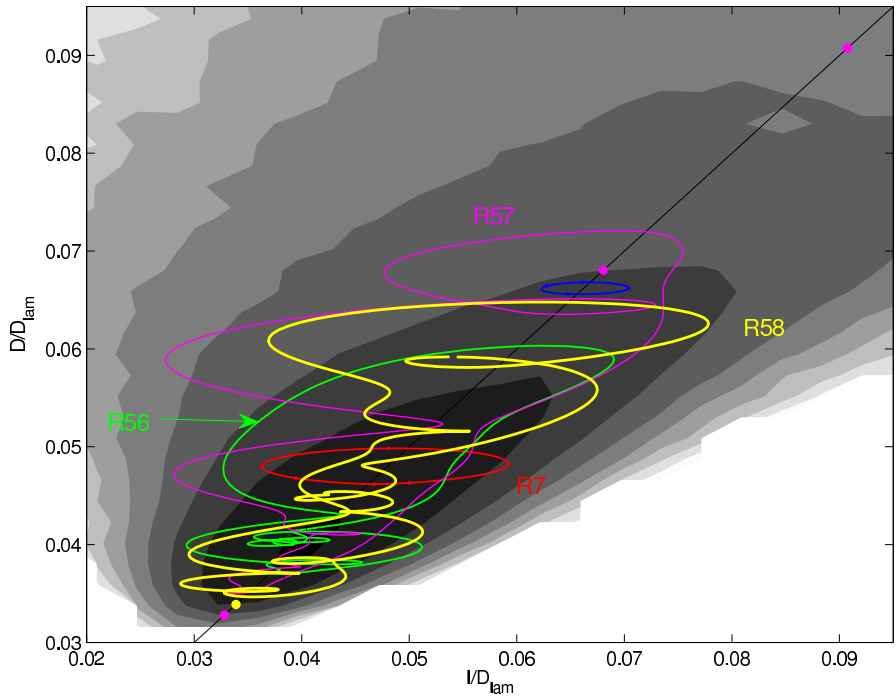}}
\caption{ The normalised dissipation $D(t)/D_{lam}$ verses $I/D_{lam}$
  for the recurrent flows found at $Re=60$ with the pdf of
  the DNS turbulence plotted in the background (11 shades at levels
  $10^{\alpha}$ where $\alpha={-5,-4.5,\ldots,-0.5,0}$). Plotted are
  $E1$ ($D/D_{lam}=I/D_{lam}=0.091$), $T1$ (0.033), $T3$ (0.068),
  $T4$ (0.034), $R7$, $R8$, $R56-8$.}
\label{DvsI_60}
\end{figure}

\begin{figure}
\centerline{\includegraphics[scale=0.7]{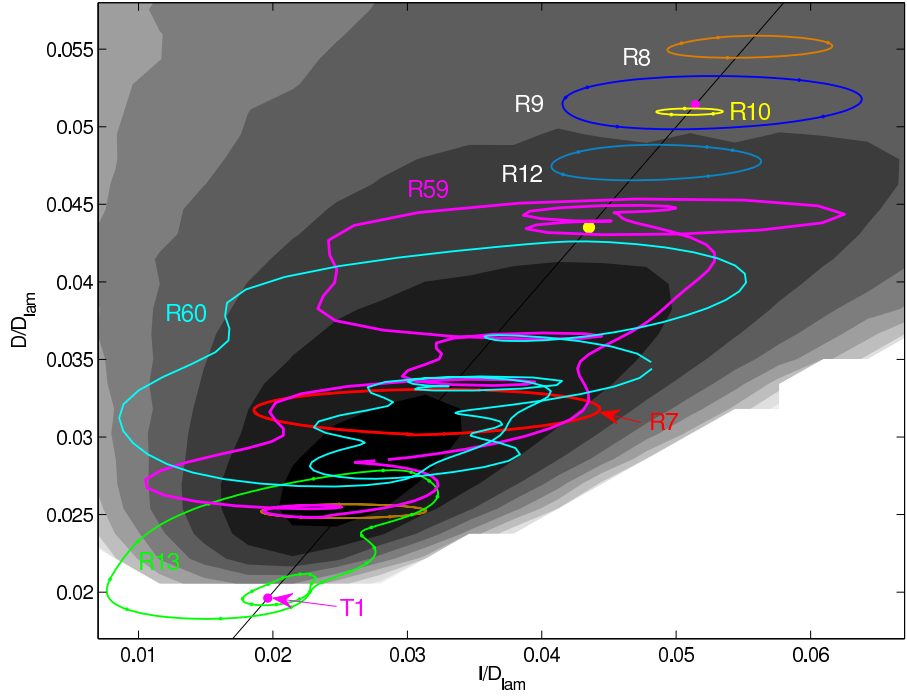}}
\caption{ The normalised dissipation $D(t)/D_{lam}$ verses $I/D_{lam}$
  for the recurrent flows found at $Re=80$ with the pdf of the DNS
  turbulence plotted in the background (11 shades at levels
  $10^{\alpha}$ where $\alpha={-5,-4.5,\ldots,-0.5,0}$). Plotted are
  $T1$ ($D/D_{lam}=I/D_{lam}=0.0196$), $T3$ (0.0514), $T5$ (0.0435,
  yellow dot),
  $R7-R13$ and $R59-R60$.}
\label{DvsI_80}
\end{figure}

\begin{figure}
\centerline{\includegraphics[scale=0.7]{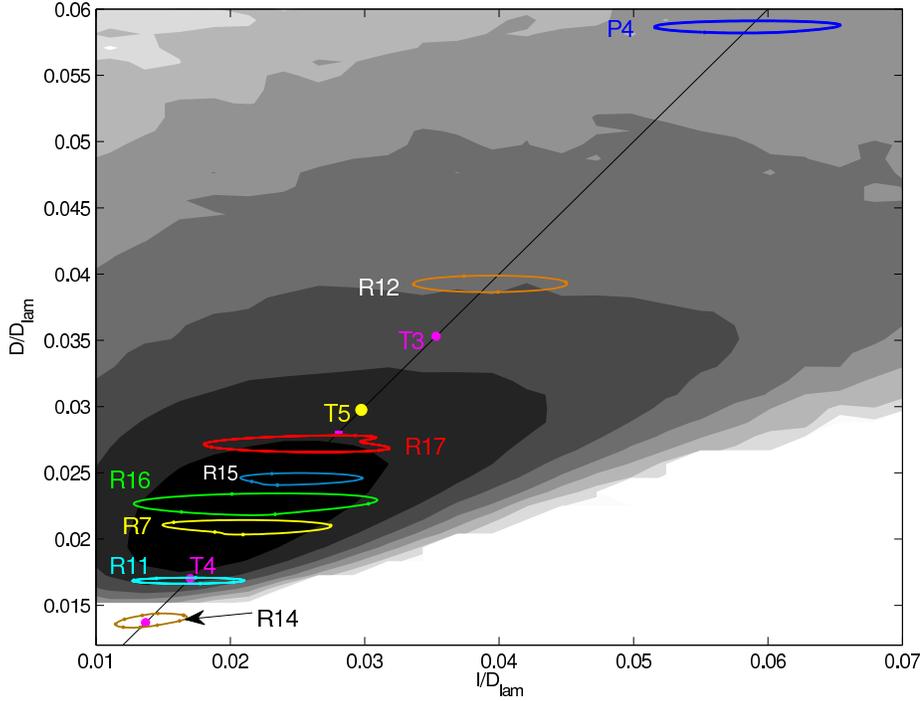}}
\caption{ The normalised dissipation $D(t)/D_{lam}$ verses $I/D_{lam}$
  for the recurrent flows found at $Re=100$ with the pdf of the DNS
  turbulence plotted in the background (11 shades at levels
  $10^{\alpha}$ where $\alpha={-5,-4.5,\ldots,-0.5,0}$). Plotted are
  $T1$ ($D/D_{lam}=I/D_{lam}=0.0137$), $T3$ (0.0353), $T4$ (0.0170), $T5$ (0.0297,
  yellow dot), $P4$,
  $R7$, $R11$, $R12$, $R14-R17$ and $R61$.}
\label{DvsI_100}
\end{figure}

\begin{figure}
\centerline{\includegraphics[scale=0.7]{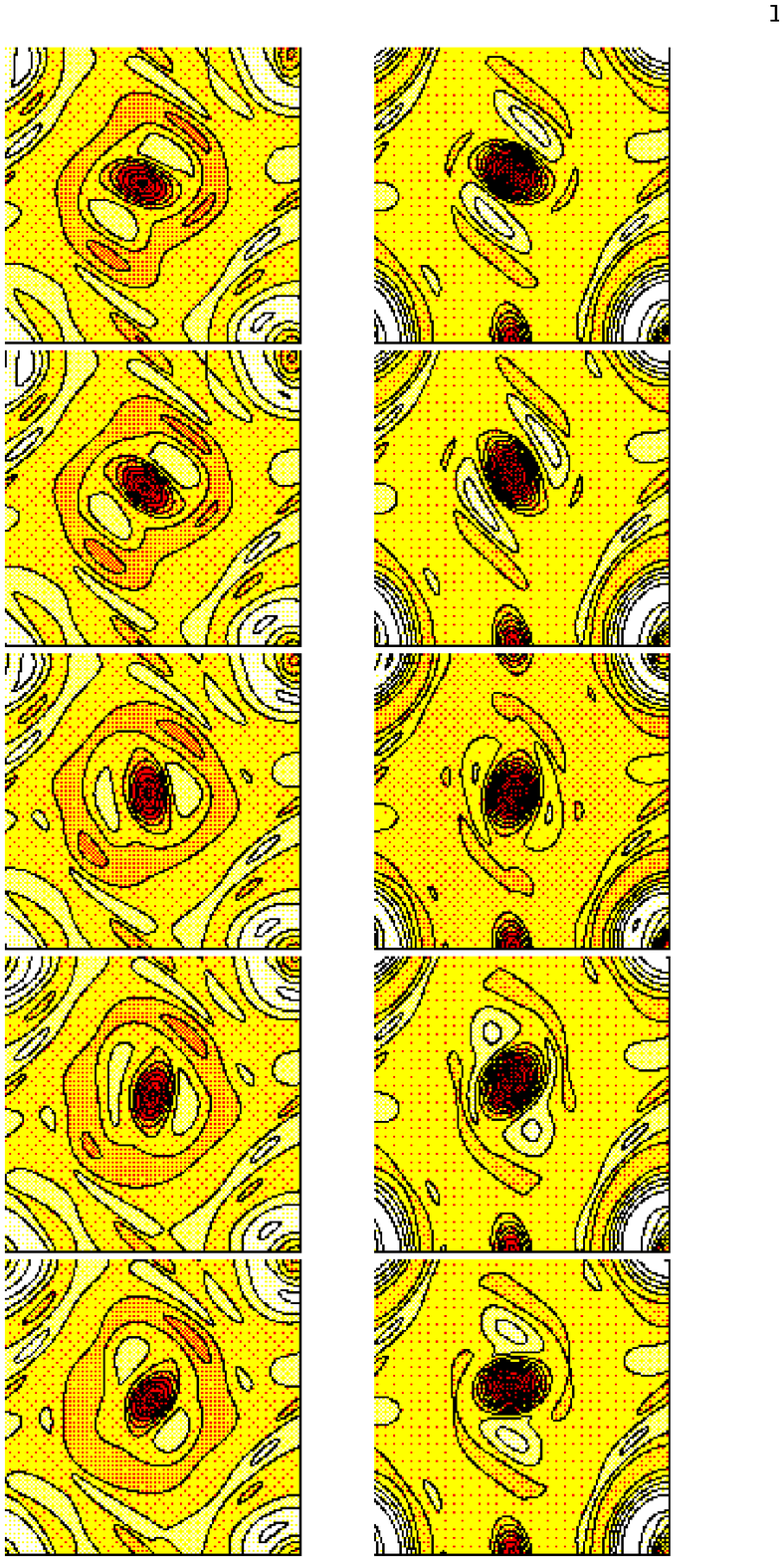}}
\caption{(Low resolution version  for the arXiv) The dns trajectory (left) synchronised with the subsequently
  converged periodic orbit $P4$ (right) at $Re=100$. Time proceeds
  downwards with snapshots at $t_0,t_0+0.2,t_0+0.6,t_0+0.8$ and
  $t_0+1.0$ (the period of $P4$ is 1.185). The vorticity
  scale ranges from -26 (dark red) to  12 (light white). }
\label{P4}
\end{figure}

Figure \ref{T3T4T5} shows the new travelling waves found and figures
\ref{DvsI_60} to \ref{DvsI_100} the $D-I$ plots where now {\em all}
the recurrent flows found at the respective $Re$ are marked. Again
most sit in the $D-I$ region where the DNS spends the majority of its
time although as at $Re=40$ there are some outliers (e.g. $E1$ at
$Re=60$, $R8$ at $Re=80$, and $P4$ and $R14$ at $Re=100$). That $R14$
actually appears outside the footprint of the DNS pdf at first looks
erroneous but is in fact merely an indication that when the
`turbulence' approached $R14$ in phase space, it maintained higher
(global) dissipation and energy input than $R14$. This can occur when
part of the domain resembles $R14$ while the rest does not and
exhibits enhanced dissipation. A good example of this is shown in
figure \ref{P4} which details the turbulent episode which signalled
the presence of $P4$ (left column) alongside the successfully
converged periodic orbit $P4$ (right column). Visually, the eye is
drawn to the centre of the domain where in both columns an isolated
vortex is clearly seen rotating in a clockwise fashion. However, the
corners are just as significant in that they also indicate an isolated
vortex, yet this is stronger with higher gradients (and hence larger
dissipation) for $P4$ than the DNS signal. Plotting the two time
sequences on a $D-I$ plot shows $P4$ as a closed loop much higher up
the $D=I$ line than the DNS (not shown).

%
%

%
%

\section{Recurrent flows as a turbulent alphabet}\label{4.4}

Given the sets of recurrent flows extracted at each $Re$, the question
is then how to use them to predict properties of the turbulence
encountered. Periodic Orbit Theory advocates a weighted expansion of
the recurrent flows such that
\beq
\Gamma_{prediction}^N:= \frac{\sum_{i=1}^N w_i \Gamma_i}{\sum_{i=1}^N w_i}
\label{expansion}
\eeq
where $\Gamma$ is any property such the mean dissipation rate, the
mean profile or a pdf, $N$ the total
number of recurrent flows and the weights are
\beq w_i=\left[ \prod_{\lambda^{(i)}_k \neq 0} |\,
  (1-e^{\lambda^{(i)}_kT^{(i)}})| \, \right]^{-1} 
\label{proper}
\eeq
(e.g. Cvitanovic 1995, Lan 2010).  Here $\lambda^{(i)}_k$ is the $k$th
eigenvalue of the linearised operator around the $i$th recurrent flow
of period $T^{(i)}$ (the associated Floquet multiplier would be
$\mu^{(i)}_k:=e^{\lambda^{(i)}_k T^{(i)}}$).  Calculating the weights
in (\ref{proper}) represents a considerable challenge in such a
high-dimensional system.  Moreover, as already indicated, this
expression is derived under special conditions not satisfied by the
Navier-Stokes equations (e.g. hyperbolicity) as well as being derived
only for (unstable) periodic orbits rather than (unstable) relative
periodic orbits (the latter do not exist in very low dimensional
systems). A modified theory is being developed (Cvitanovic 2012) to
include them but here, in the spirit of what follows, we brush over
this subtlety to treat relative periodic orbits just like periodic
orbits (see also Lopez et al. 2005). Given these issues, we proceed in
a more pragmatic fashion in keeping with the previous suggestive work
of Zoldi \& Greenside (1998) and Kazantsev (1998,2001) (see also
Dettmann \& Morriss 1997). These authors proposed and tested a
strategy of developing weights based upon {\em only} the unstable
eigenvalues associated with the recurrent flow (i.e. $\Re
e(\lambda_k^{(i)})>0$). In particular, Zoldi \& Greenside (1998)
advocated the `escape-time' weighting
\beq 
w_i \propto \frac{1}{\sum_{k \in \K^{(i)}} Re(\lambda_k^{(i)})}
\qquad {\rm protocol \, 1} \eeq
where $\K^{(i)}$ is the set of $k$ such that $\Re e (\lambda_k^{(i)})
>0$ since this efficiently captures how unstable the recurrent flow is
and inversely correlates this with how long the turbulent trajectory
should spend in its vicinity. Kazantsev (1998, 2001) argued that this
formula should be modified to reflect the fact that longer period
orbits have a greater `presence' in phase space than shorter period
orbits and added the period $T^{(i)}$ to the numerator
\beq
w_i \propto \frac{T^{(i)}}{\sum_{k \in \K^{(i)}} Re(\lambda_k^{(i)})} 
\qquad {\rm protocol \, 2}
\eeq
Significantly, this protocol suppresses any contribution from
equilibria or travelling waves. We now test both of these protocols
together with a `control' choice of `no weighting' so just
\beq
w_i \propto 1 \qquad {\rm protocol \, 3}.
\eeq
The key measures we use to characterise the 2D turbulence
simulated here are pdfs of $E(t)$ and $D(t)$ together with the
profiles $\bar{u}(y)$, $u_{rms}(y)$ and $v_{rms}(y)$ (the pdf of
$I(t)$ was also considered but adds little information to that
provided by the pdf for  $D(t)$). 

\subsection{$Re=40$}

\begin{figure}
\begin{center}\setlength{\unitlength}{1cm}
\begin{picture}(13.5,7.25)
\put(0,0){\epsfig{figure=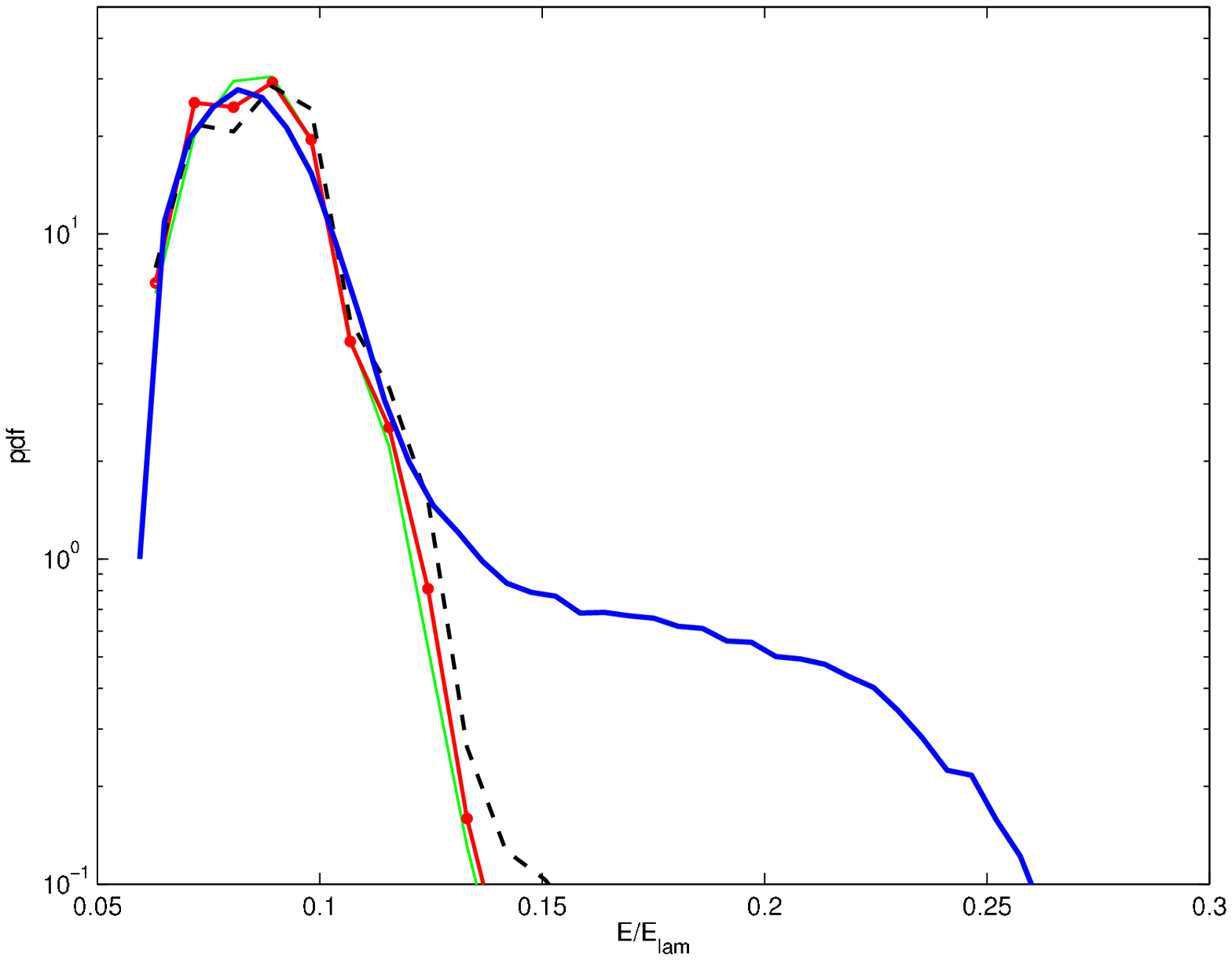,width=6.5cm,height=6cm,clip=true}}
\put(6.75,0){\epsfig{figure=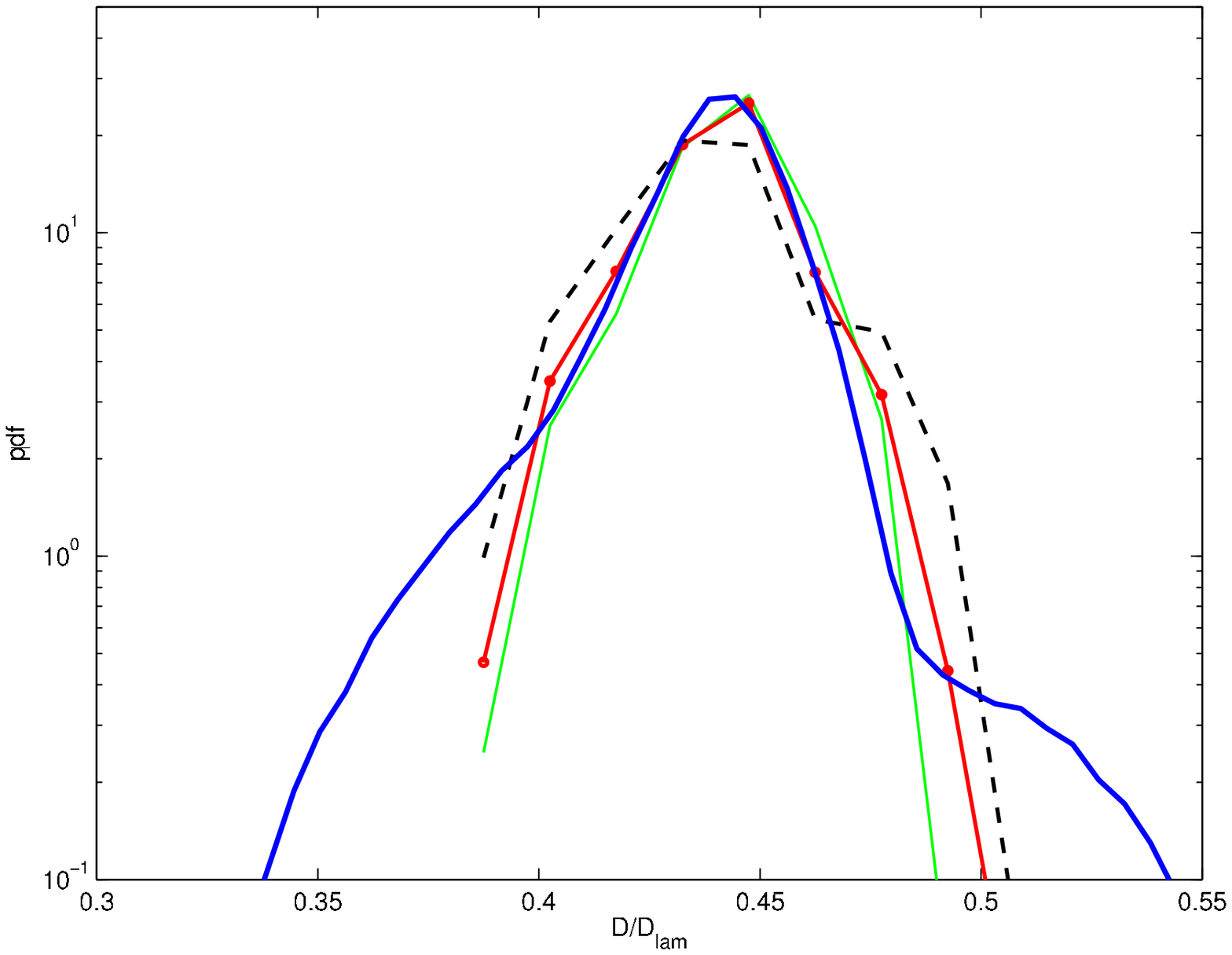,width=6.5cm,height=6cm,clip=true}}
\end{picture}
\end{center}
\caption{The probability density functions for $E(t)/E_{lam}$ and
  $D(t)/D_{lam}$ from DNS (blue thick line) and predictions using
  weighting protocol 1 (red, thick line with dots), 2 (green, thin
  line) and 3 (black thick dashed) at $Re=40$. 40 bins were used to
  calculate the pdfs for the recurrent flows and 100 bins for the DNS
  due to its greater range. These choices gave the best balance of
  resolution with the data available. }
\label{pdf_40}
\end{figure}

\begin{figure}
\begin{center}\setlength{\unitlength}{1cm}
\begin{picture}(13.5,7.25)
\put(0,0){\epsfig{figure=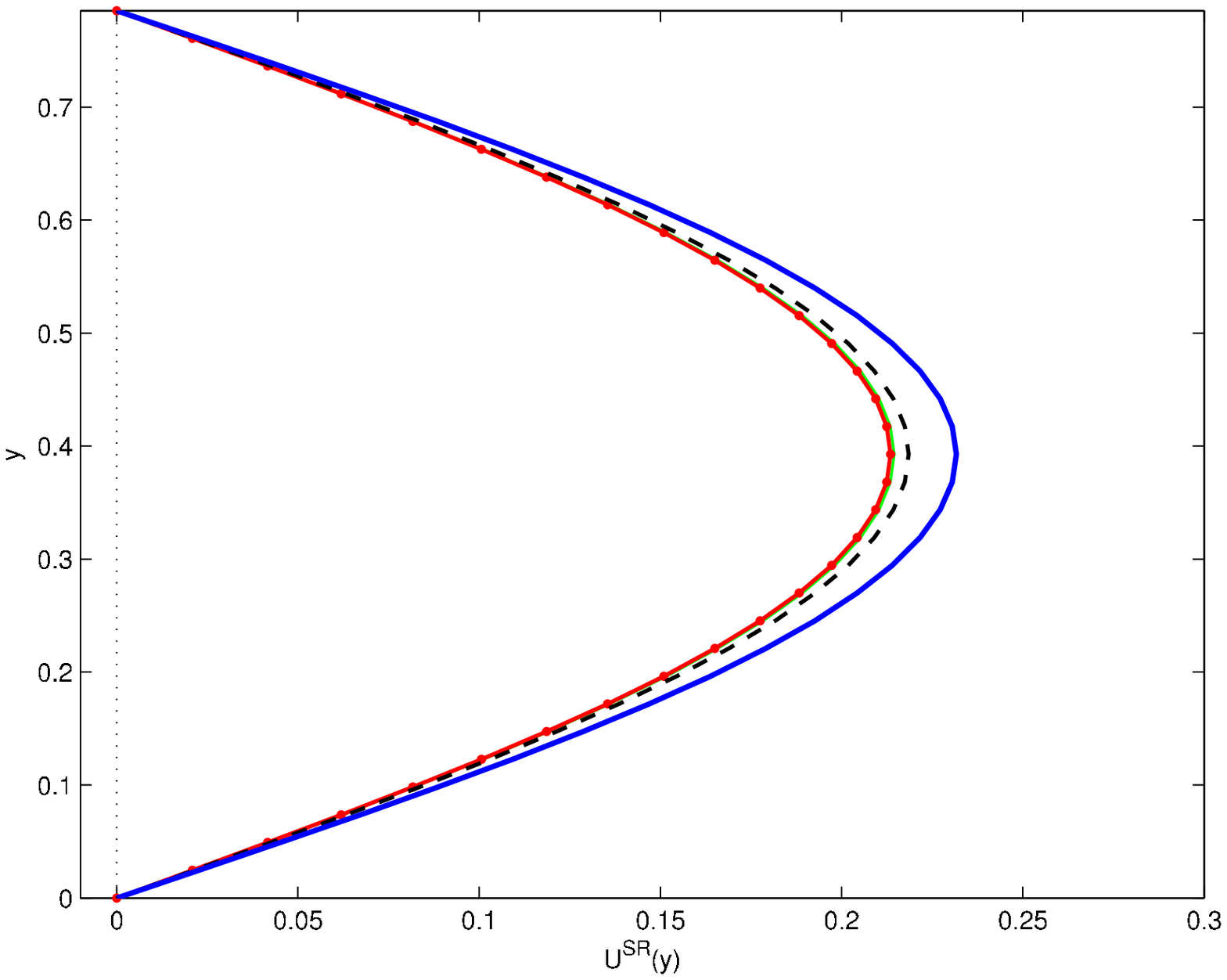,width=6.5cm,height=7cm,clip=true}}
\put(6.75,0){\epsfig{figure=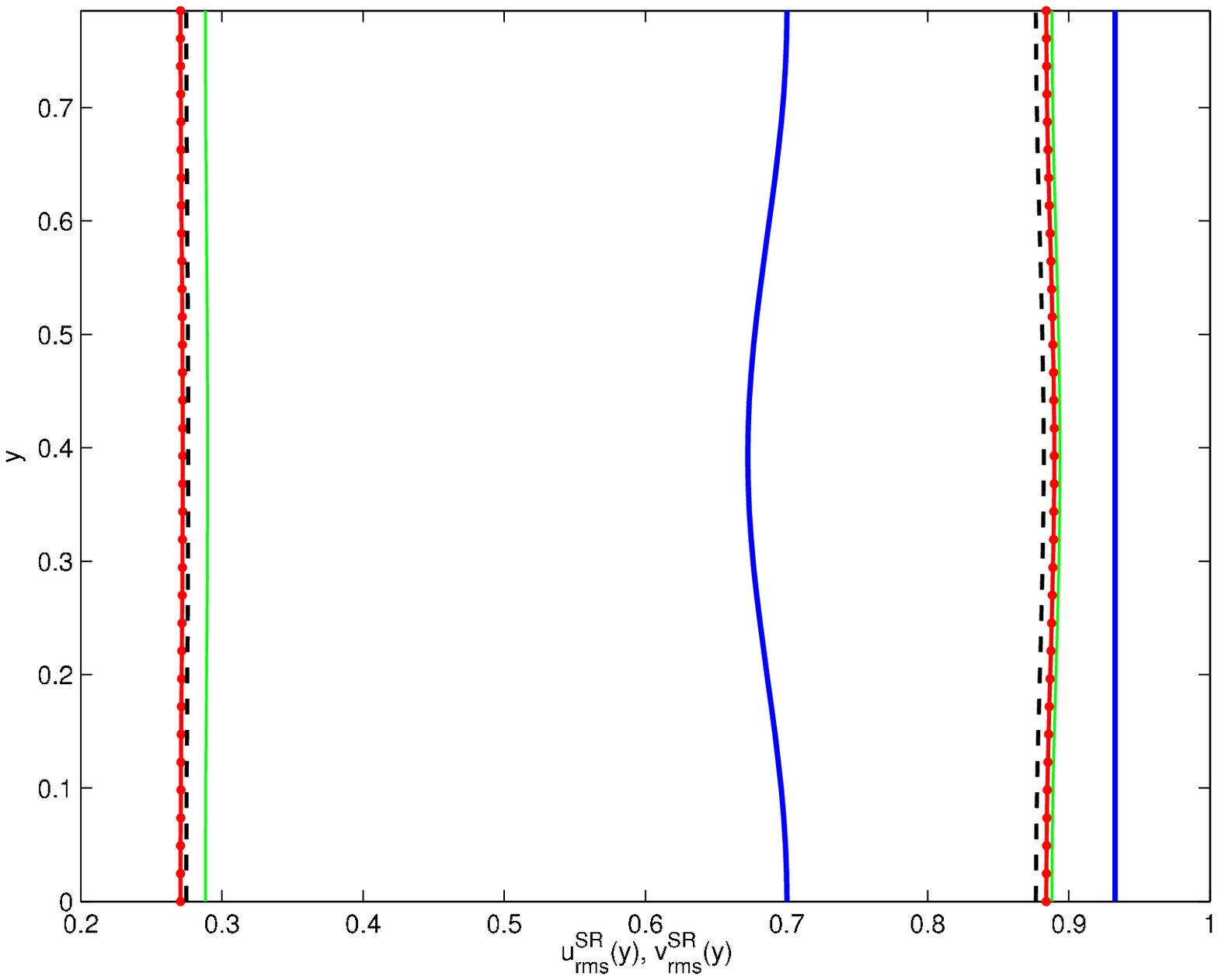,width=6.5cm,height=7cm,clip=true}}
\end{picture}
\end{center}
\caption{The symmetrised mean flow $U^{SR}(y)$ (left) and symmetrised
  $u^{RS}_{rms}(y)$ (right plot left part) and $v^{RS}_{rms}(y)$
  (right part of right plot) from DNS (blue thick line) and
  predictions using weighting protocol 1 (red, thick line with dots),
  2 (green, thin line) and 3 (black thick dashed) at $Re=40$. }
\label{pred_mean_40}
\end{figure}

The pdfs of $E(t)/E_{lam}$ and $D(t)/D_{lam}$ at $Re=40$ are shown in
figure \ref{pdf_40} along with the predictions using protocols 1-3,
that is, the individual pdfs of each recurrent flow are weighted
together appropriately to produce an overall pdf as in
(\ref{expansion}). In the case of $E(t)/E_{lam}$ all three predictions
are very good at the central peak of the pdf but each fails to capture
the clear shoulder at higher energies in the DNS albeit at a level of
the pdf a factor of $\approx 30$ lower. There is a similar story for
the $D(t)/D_{lam}$ pdf although in this case, the performance of the
stability-motivated protocols 1 and 2 seem noticeably more effective
in capturing the DNS pdf. Again, the extremes of the DNS pdf are not
captured reflecting the fact, as commented earlier, that perhaps not
enough recurrent flows with large or small dissipation excursions have
been found. Given the reduced value of the pdf there, these are
infrequently visited and thus require very long runs to realistically
have a chance to extract them. As already mentioned, what initially
looked like long runs of $10^5$ time units were actually not long
enough. As further independent evidence of this, plots of the mean
profiles $\bar{u}(y)$ from each `long' run at the same $Re$ (see Table
1) showed noticeable differences between each other and to the
expected asymptotic state which respects all the symmetries of the
system. In particular, the obvious symmetry that the mean profile
should be invariant under $\pi/2$ shifts in $y$ was clearly
violated. To ameliorate this, we decided to
`symmetrise' the DNS mean profile by extracting that part
($U^{SR}(y)$) from the signal ($\bar{u}$) which does satisfy all the
symmetries listed in \S \ref{symmetries}. Explicitly
\beq
U^{SR}(y):= \frac{1}{2n}\sum^{2n-1}_{m=0} {\cal S}^{-m}U^{R}({\cal S}^m y) \qquad
{\rm with} \quad U^{R}:=\frac{1}{2}[\,\bar{u}(y)+{\cal R}^{-1}\bar{u}({\cal R}y)\,]
\eeq
(recall $n=4$) and similarly for
$u^{SR}_{rms}$ and $v^{SR}_{rms}$. This process picks out the
following Fourier coefficients
\beq U^{SR}(y):=
\sum_{m=0} a_m \sin(4(2m+1)y) \qquad (\,u^{SR}_{rms},v^{SR}_{rms}\,)=
\sum_{m=0} (\,b_m,c_m\,) \cos(8my).  
\label{symmetrised}
\eeq 
from the complete Fourier series for $\bar{u}$, $u_{rms}$ and
$v_{rms}$. In particular, the symmetrised mean profile $U^{SR}(y)$ has
the leading Fourier modal form of $\sin 4y$, which mimicks the
forcing, and a leading correction of $\sin 12y$.  Such a profile needs
only be plotted over $y \in [0,\pi/4]$ which is done in figure
\ref{pred_mean_40} along with the predictions from protocols 1-3. This
comparison looks impressive with $a_0=0.232$ (cf expression
(\ref{symmetrised})) in the DNS, compared to $0.214$ (protocol 1),
$0.215$ (2) and $0.219$ (3). For all, $a_1=O(10^{-6})$ and $a_2=O(10^{-8})$
(as way of comparison, the `raw' mean flow $\bar{u}$ has a leading
non-symmetrised part given by $0.0280 \cos y+0.0268 \sin y$,
i.e. roughly 10\% smaller than the symmetrised part). The explanation
for why the (symmetrised) mean profile matches the forcing profile so well is
currently unclear to us and it is tempting to speculate that actually  $a_n
\rightarrow0$ ($n \geq 1$) with the period of averaging. Sarris et
al. (2007) study  the statistics of 3D Kolmogorov flow for various
computational domains and use two measures to signal whether their
statistics have converged sufficiently over a period of time
integration. One, 
\beq
\gamma_2:= \frac{(\langle I\rangle_t-\langle D\rangle_t)^2}{\langle
  D \rangle_t^2}
\eeq
(equation (22), Sarris et al. 2007) assesses the extent to which the
energy input into the flow matches the the energy dissipated and is
easily calculated from our output data: $\gamma_2$ is at most
$O(10^{-8})$ for all our $10^5$ time unit runs. Even with this small
value, we find evidence that the mean profile is far from converged to
what is expected (i.e. satisfies all the symmetries of the problem)
which emphasizes how easy it is to unwittingly collect unconverged
statistics.

Figure \ref{pred_mean_40} also shows the equivalent plot for $u_{rms}$
and $v_{rms}$. For $u_{rms}$, $b_0=0.687$ for the DNS which clearly
differs from all the predictions - $0.272$ (protocol 1), $0.289$ (2)
and $0.276$ (3). The comparison for $v_{rms}$, however, is much
better: $c_0=0.933$ for the DNS verses $0.887$ (protocol 1), $0.891$
(2) and $0.879$ (3). One possible reason why the $u_{rms}$ comparison
is poor is that $u_{rms}$ is calculated in the DNS `on the fly' by
subtracting the current best estimate of the mean $\bar{u}$ from the
current streamwise velocity (see \ref{fluctuation}) rather than using
the final mean profile to {\em a posteriori} calculate the streamwise
fluctuation field. Given our realisation now that the mean profile
takes a long time to converge to its (symmetric) asymptotic state,
there is likely to be a significant error (henceforth we consider only
$v_{rms}$ for $Re=60,80$ and $100$).

\begin{figure}
\begin{center}\setlength{\unitlength}{1cm}
\begin{picture}(13.5,15)
\put(0,7.5){\epsfig{figure=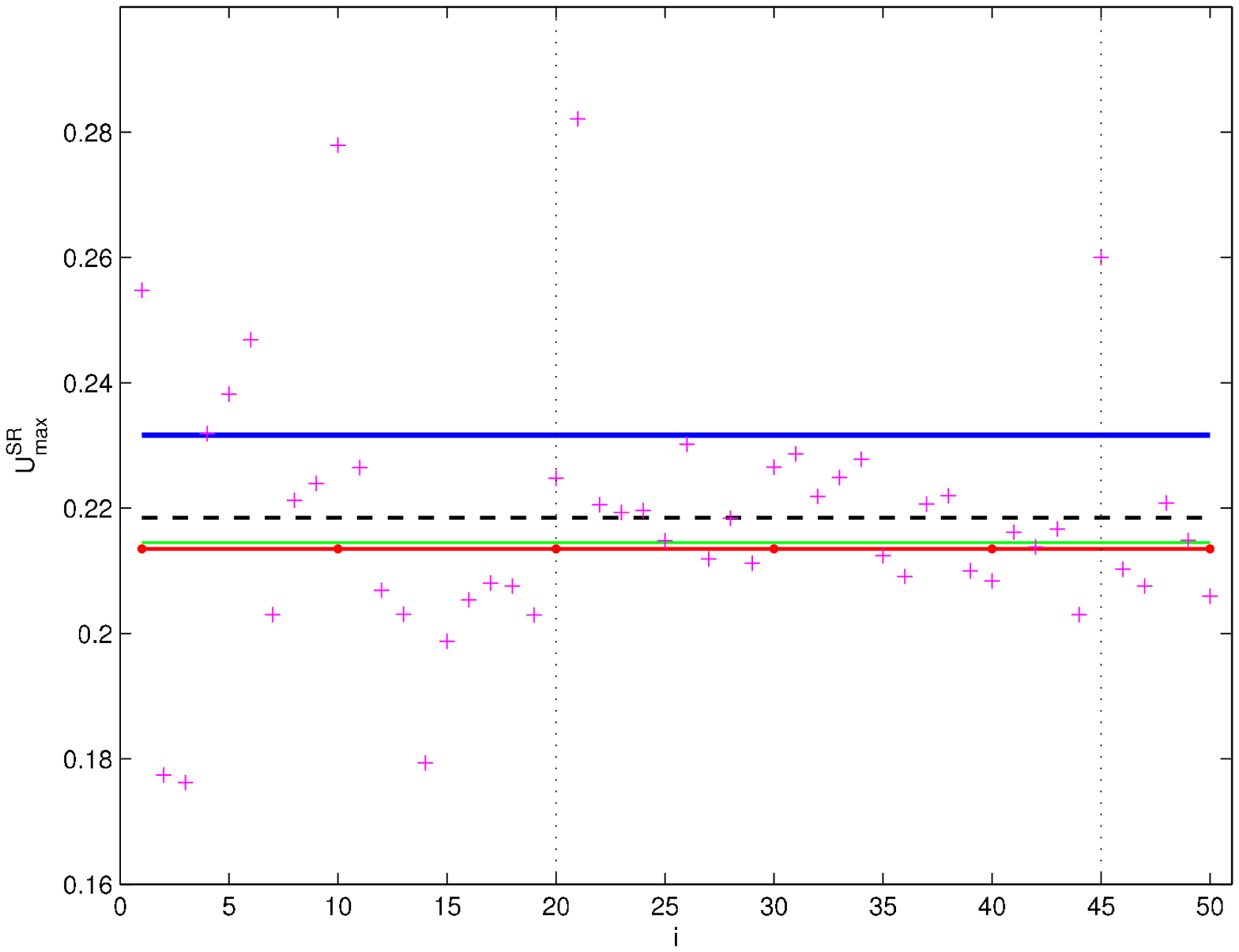,width=12cm,height=7cm,clip=true}}
\put(0,0){\epsfig{figure=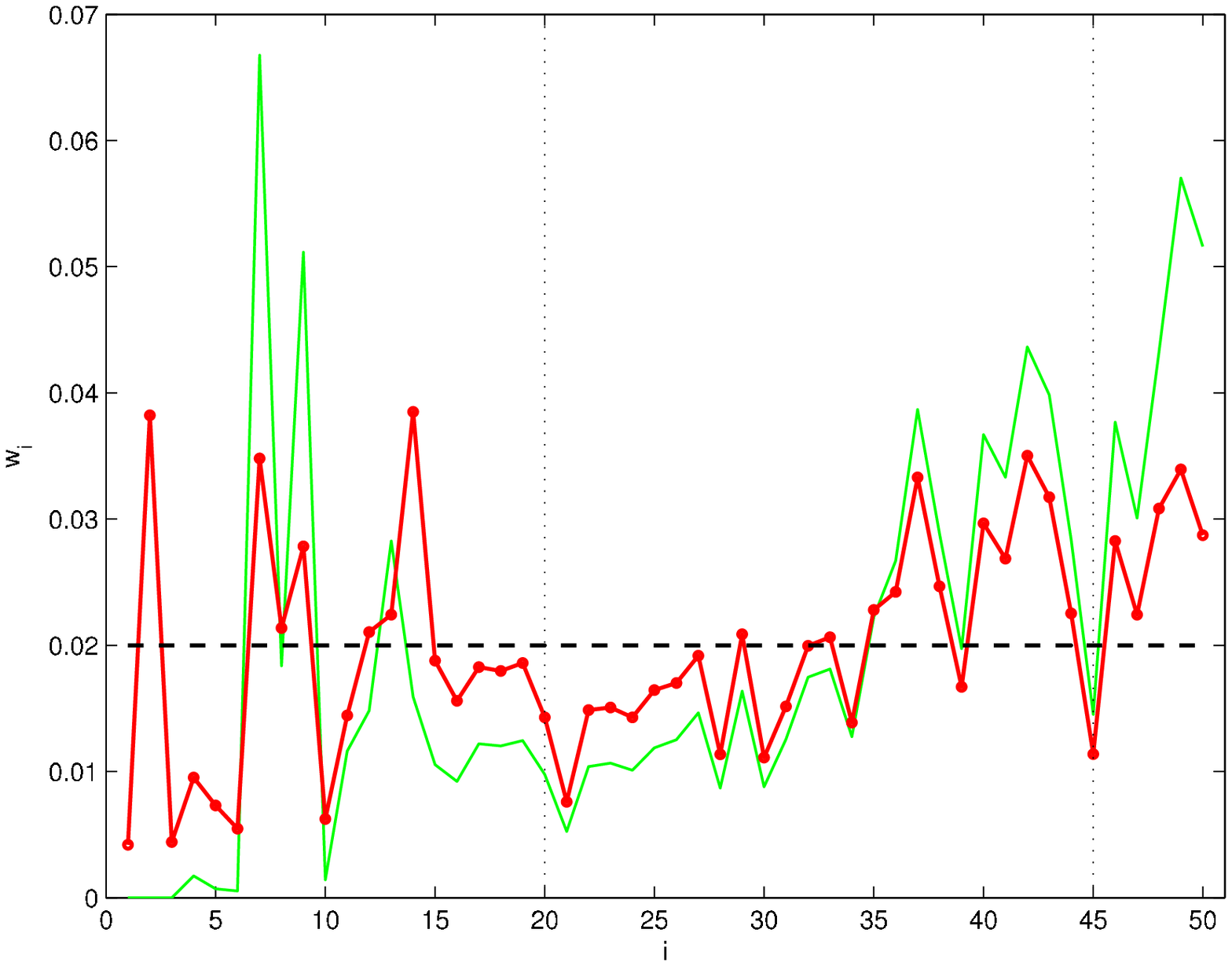,width=12cm,height=7cm,clip=true}}
\end{picture}
\end{center}
\caption{Top: The peak symmetrised mean $U^{RS}(\pi/8)$ for the DNS (thick
  blue horizontal line), prediction using protocol 1 (red dotted line),
  prediction using protocol 2 (green line), the control
  prediction (dashed black line) and for each recurrent flow listed in the
  order given in Table 1 at $Re=40$ (so $i=20$ is $R25$ and $i=45$ is
  $R50$ marked with black dotted lines). Bottom: the weights for the
  three protocols (1-red line with dots, 2-green line, 3-black horizontal
  line) plotted for all the recurrent structures at $Re=40$. Again the
  vertical (black) dotted lines indicate $R25$ and $R50$.}
\label{U_weight_40}
\end{figure}

An inescapable conclusion from these comparisons so far is that the
`control' protocol 3 of actually `no weighting' performs almost as
well as the other stability-motivated protocols 1 and 2. It's worthwhile at
this point to clarify why. The upper plot in figure \ref{U_weight_40}
shows how the peak symmetrised mean value, $U^{SR}(\pi/8)$, varies for
each recurrent flow and compares these values with the DNS and
predictions from the 3 protocols. From this it is clear that most of
the recurrent flows are good predictors individually and so when they
are mixed together the result is still reasonably good. The lower plot
in figure \ref{U_weight_40} shows how the weights vary across the
recurrent flows in the 3 different protocols.  Again, there is not
that much variation over the majority (although notice that $w_i=0$
for $i=1,2,3$ in protocol 2 since $T^{(i)}=0$) which presumably
reflects the fact that the stability characteristics of the recurrent
flows are all pretty similar.

%

\subsection{$Re=60, 80$ \& $100$}

\begin{figure}
\begin{center}\setlength{\unitlength}{1cm}
\begin{picture}(13.5,7.25)
\put(0,0){\epsfig{figure=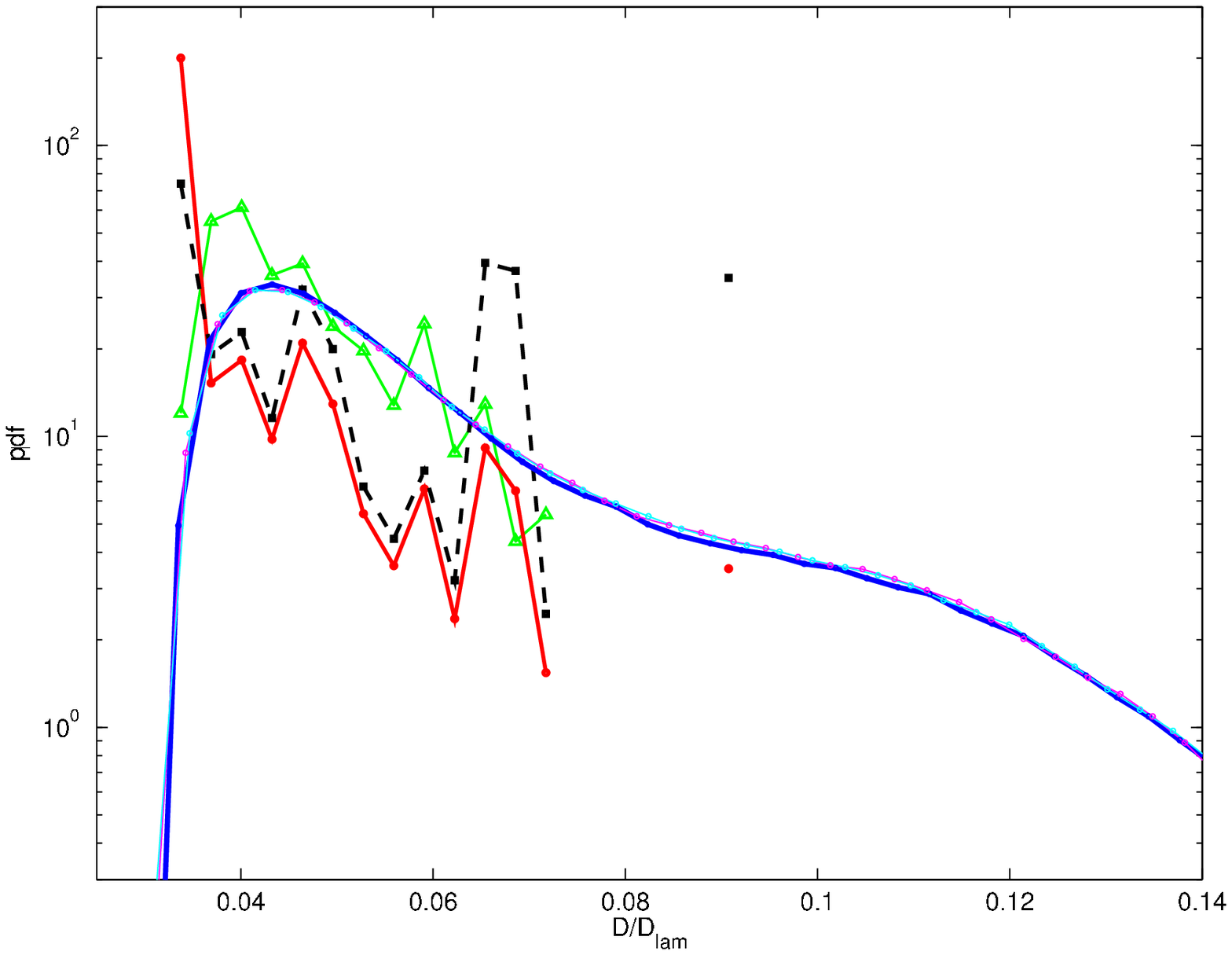,width=6.5cm,height=6cm,clip=true}}
\put(6.75,0){\epsfig{figure=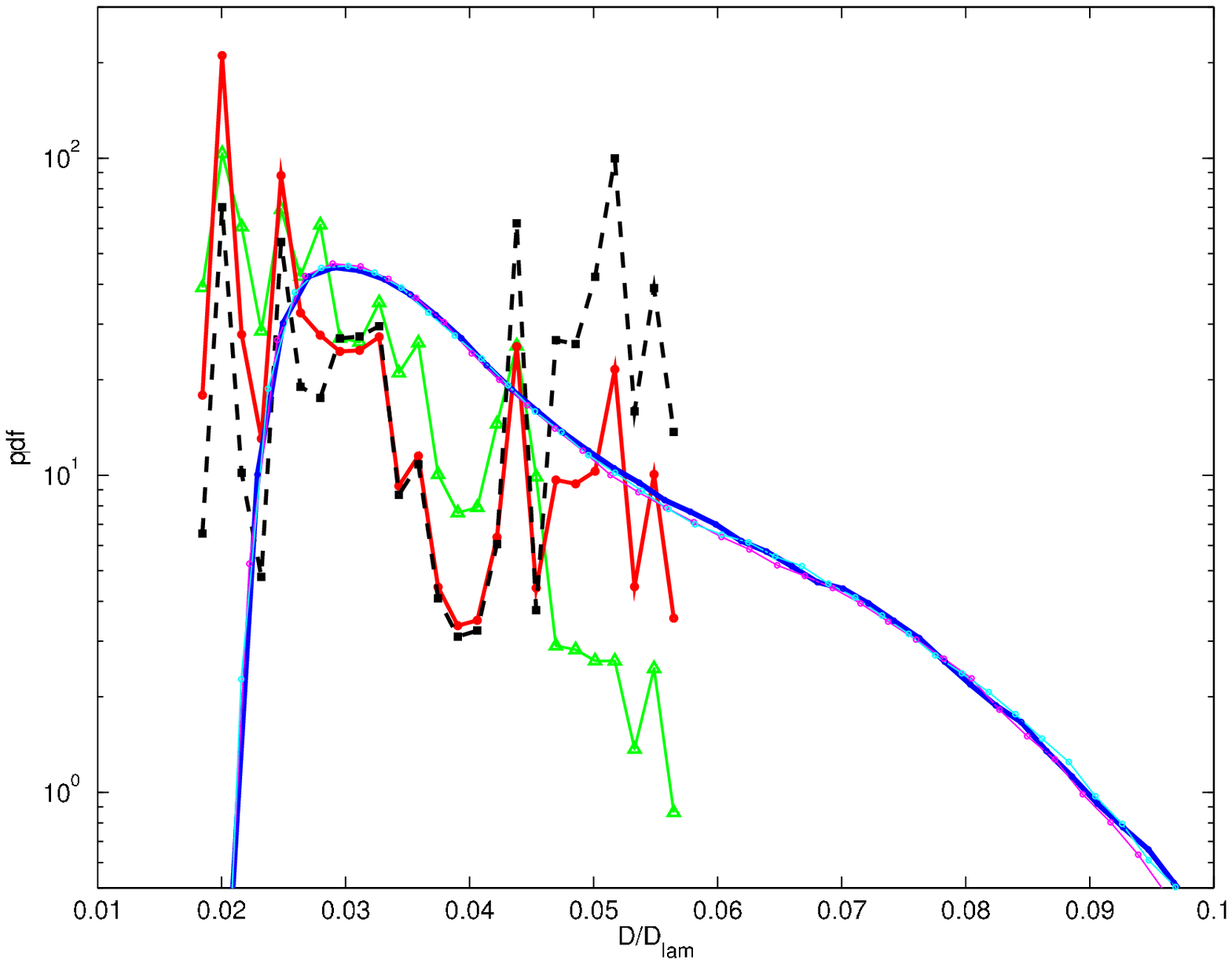,width=6.5cm,height=6cm,clip=true}}
\end{picture}
\end{center}
\caption{The probability density functions for $D(t)/D_{lam}$ from DNS
  (blue thick line, cyan and magenta lines) and predictions using
  weighting protocol 1 (red, thick line with dots), 2 (green, thin
  line with triangles), and 3 (black thick dashed with squares) at
  $Re=60$ (left) and $Re=80$ (right). 60 bins were used to calculate
  the pdfs for the recurrent flows and 100 bins for the DNS due to its
  greater range. These choices gave the best balance of resolution
  with the data available. }
\label{pdf_6080}
\end{figure}

The smaller number of recurrent flows extracted for $Re > 40$ means
that it is harder to generate reasonably smooth predictions for the
pdfs of the energy and dissipation. While the same number of 100 bins
as at $Re=40$ can be used to generate a smooth DNS pdf, only 60 bins
could be used to sum the pdfs of the recurrent flows. This number
produced the best compromise of granularity across the range while
ensuring that there is enough data in each bin for (reasonable)
smoothness at least for $Re=60$ and $Re=80$ (the sparse coverage of
the dissipation range by the recurrent flows at $Re=100$ - see figure
\ref{DvsI_100} - prevented any useful plot to be generated).  Figure
\ref{pdf_6080} shows the result of this procedure for the dissipation
pdf at $Re=60$ and $80$ (plots not shown for the energy are
similar). Here, protocol 2 offers the best partial fit both at $Re=60$
and $Re=80$ although it is clear that much of the turbulent
dissipation range extends higher than any of the recurrent flows found
at both $Re$ (also clear from \ref{DvsI_60} and \ref{DvsI_80}) and that the
`prediction' is of limited quality. It's also worth remarking that the
predictions are now more  distinquishable at these higher $Re$ which
can be traced back to the separating  stability characteristics
of the recurrent flows. For example, $E1$ has $\sum \Re e
(\lambda_j)=5.053$ whereas this is just $0.139$ for $T1$.

\begin{figure}
\begin{center}\setlength{\unitlength}{1cm}
\begin{picture}(13.5,7.25)
\put(0,0){\epsfig{figure=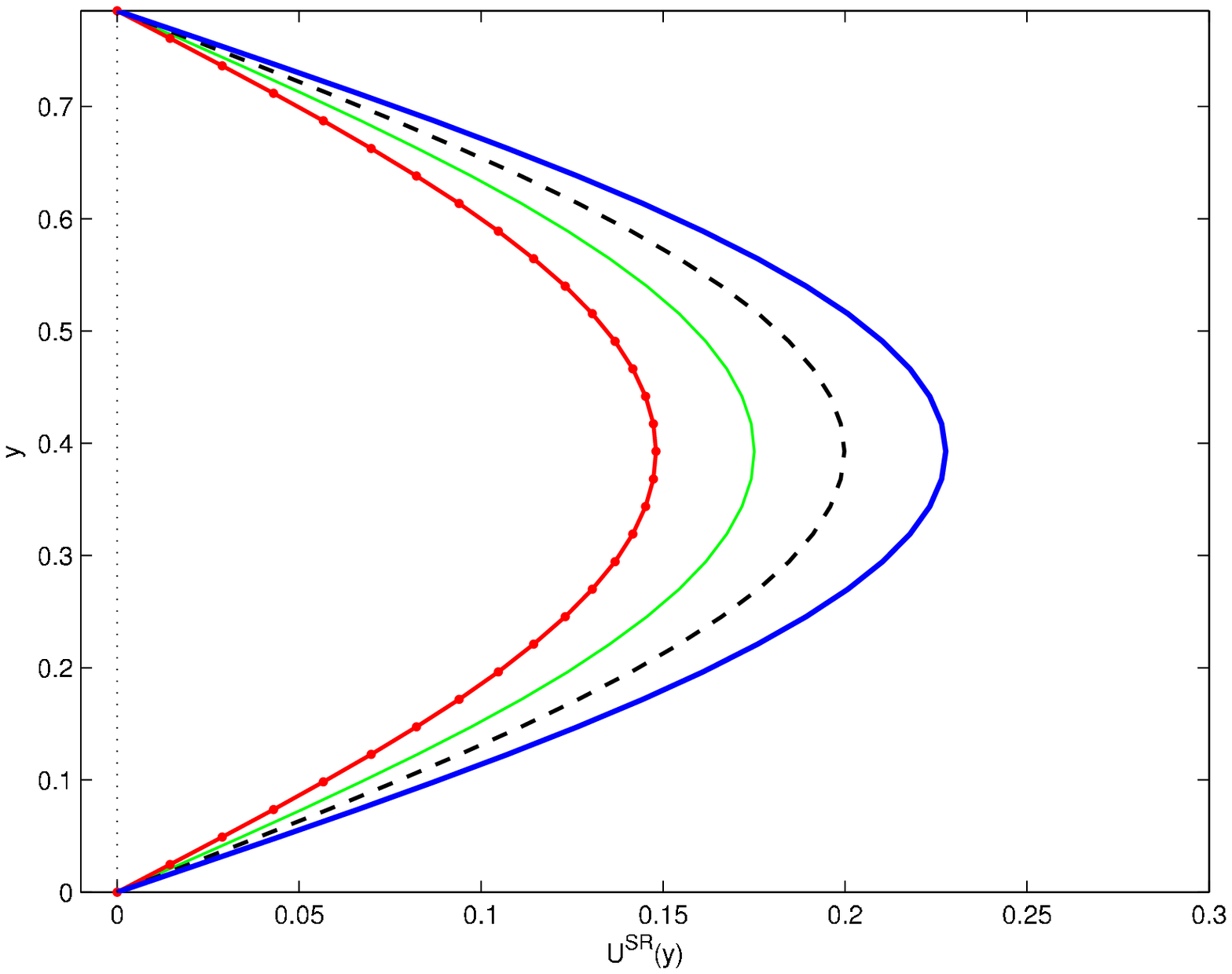,width=6.5cm,height=7cm,clip=true}}
\put(6.75,0){\epsfig{figure=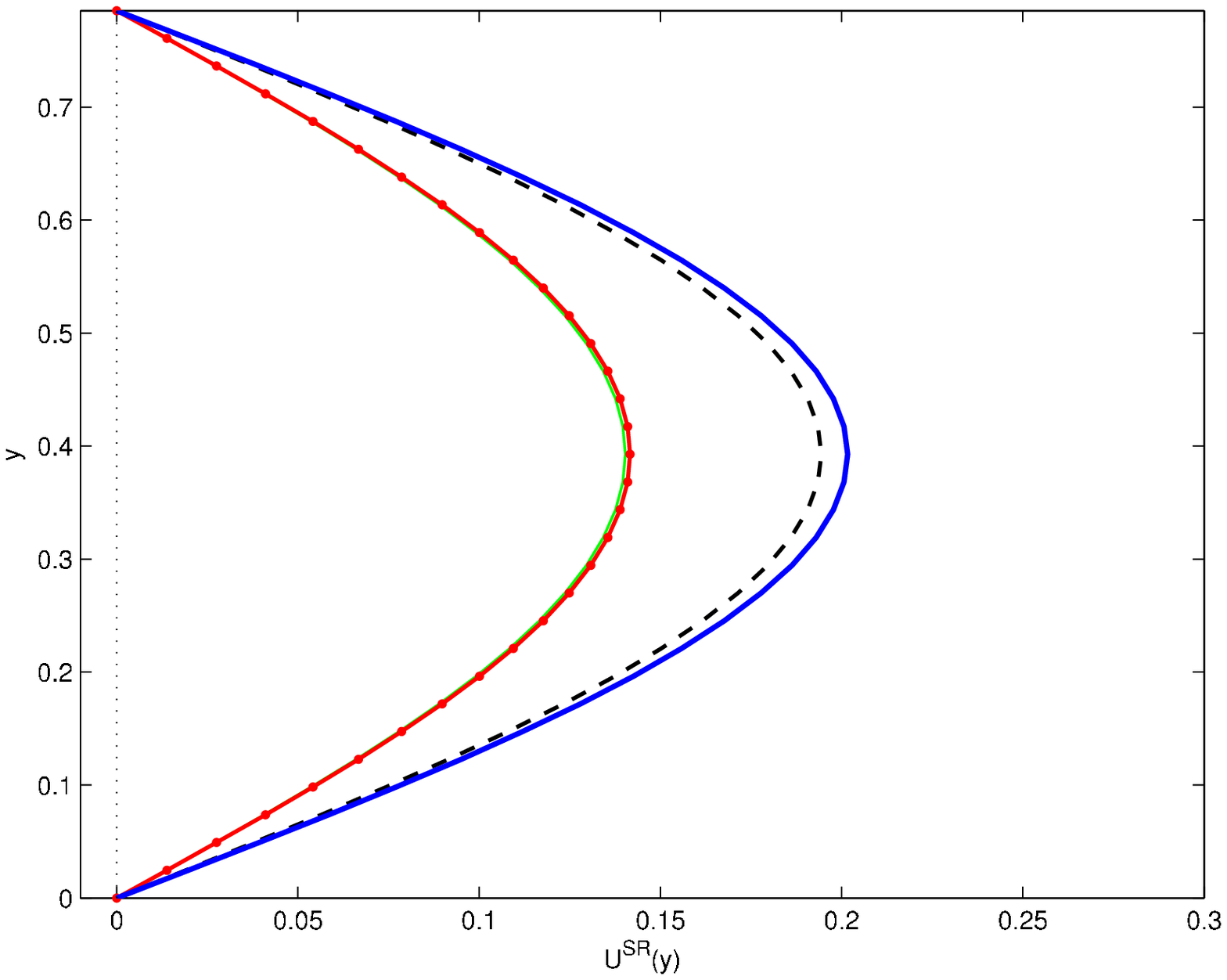,width=6.5cm,height=7cm,clip=true}}
\end{picture}
\end{center}
\caption{The symmetrised mean flow $U^{SR}(y)$ from DNS (blue thick
  line) and predictions using weighting protocol 1 (red, thick line
  with dots), 2 (green, thin line) and 3 (black thick dashed) for
  $Re=60$ (left) and $Re=80$ (right). }
\label{pred_mean_6080}
\end{figure}

Figure \ref{pred_mean_6080} shows the symmetrised mean profile as
calculated from the DNS and predicted by the 3 protocols at $Re=60$
and $Re=80$. Again, somewhat paradoxically, the `control' protocol
does the best job in both cases. At $Re=60$ $a_0=0.2277$, $0.2298$ and
$0.2296$ (cf expression (\ref{symmetrised})) across the three series A
DNS runs listed in Table 1 (with again $\approx 10\%$ non-symmetrised
part in all cases). In comparison, $a_0=0.148$ (protocol 1), $0.175$
(2) and $0.200$ (3). For $v_{rms}$, $c_0=1.099$ in the DNS run to be
compared with the predictions of $1.064$ (protocol 1), $1.057$ (2) and
$1.055$ (3). At $Re=80$, $a_0=0.2009$, $0.2016$ and $0.1998$ across
the three series A DNS runs listed in Table 1. In comparison,
$a_0=0.1416$ (protocol 1), $0.1403$ (2) and $0.1944$ (3).

Finally, we note that at $Re=100$, $a_0=0.1777$, $0.1782$ and $0.1783$
across the three series A DNS runs
listed in Table 1 (with now $\approx 20\%$ non-symmetrised part in all
cases).





 \section{Discussion}


In this study, we have considered 2D turbulence on the torus
$[0,2\pi]^2$ forced monochromatically in one direction (Kolmogorov
flow). By looking for near recurrences of the flow in long direct
numerical simulation runs, sets of exactly closed flow solutions
`embedded' in this turbulence have been extracted at different forcing
amplitudes ($Re$). We have then tried to use these sets of recurrent
flows to reconstruct key statistics of the turbulence motivated by
Periodic Orbit Theory in low-dimensional chaos. The approach has been
reasonably successful at $Re=40$ - see figures \ref{pdf_40} and
\ref{pred_mean_40} - where 50 recurrent flows were found with the
majority buried in the part of phase space most populated by the
turbulence. In contrast, at $Re=60$, $80$ and $100$, the limited size
of the recurrent flow sets found has made the approach largely
impotent. Even at $Re=40$, the success achieved seems more reliant on
just extracting lots of similar-looking recurrent flows buried in the
most popular part of phase space for the turbulence than on any
sophisticated choice of weighting coefficients. Indeed, one is
reminded of Kawahara \& Kida's (2001) conclusion that one judiciously
chosen periodic orbit is `enough' to be a valuable proxy of the
turbulence. We can sympathise with this viewpoint but only if the
comparison with the turbulence statistics is not too demanding.  The
key issue, of course, plaguing this investigation is the paucity of
recurrent flows found from the finite DNS data generated. This is
perhaps the main message to come out of this work: Periodic Orbit
theory for fluid turbulence is a promising approach but only if enough
- O(100)? - recurrent flows are gathered which requires very long
turbulence data sequences. A time sequence of $10^5$ time units seems
marginally adequate for $Re=40$ but is maybe two orders of magnitude
too short for $Re=60$ and beyond.  Unfortunately, without these large
sets, it has been impossible to discern between different weighting
protocols.


Operationally, the work described here has been time-consuming both
computationally in generating near-recurrence episodes and attempting
to converge them, as well as `manually' because of all the careful
processing (e.g. calculating their stability) and archiving of the
recurrent flows needed (e.g. does a new convergence from a DNS guess
represent a new recurrent flow or a repeat of a previously extracted
flow?). Fortunately, there is no reason why this process could not be
automated with the objective being to `automatically' generate a basis
set of recurrent flows for each $Re$.  Indeed, one could hope that
such a set at given $Re=Re_1$ could be used to predict the turbulent
statistics at another $Re=Re_2$. This would require each recurrent
flow at $Re_1$ being continued to $Re_2$ and the fresh weights for an
expansion being generated from the (new) stability information for
each recurrent flow - again painstaking work but readily
automated. One fly in the ointment is the possibility of
bifurcations in the interval $[Re_1,Re_2]$, particularly saddle node
bifurcations where two recurrent flows at $Re_1$ merge and annihilate
before $Re$ reaches $Re_2$. Working with large enough recurrent flow
sets would presumably smooth over this effect somewhat but will not
eliminate it entirely. 

Leaving aside these issues for a moment, it's worth re-emphasizing
that {\em any} recurrent flow extracted from DNS data is a simple
invariant solution `buried' in the turbulence.  As such, each
represents a sustained sequence of dynamical processes which
contributes to, if not underpins, the turbulence itself. Since they
are closed in time, they can be analysed relatively easily in whatever
detail is required to understand key dynamical relationships in the
flow. This seems a very promising byproduct of the analysis whether
one believes a Periodic Theory-type expansion of turbulence is
possible or not (pursuing this has not been the focus here due to the
2-dimensionality of the flow).


Finally, the ever-improving computational resources available now have only
recently made this type of study possible. Even with these, we have
underestimated the demands of data collection in 2D turbulence over
the small torus $[0,2\pi]^2$.  Major challenges ahead include treating
large aspect ratio domains - can we find localised recurrent flows? -
and handling fully 3 dimensional flows - with automated machinery,
will the approach be practical? There is plenty to explore.


\vspace{1cm} Acknowledgements: We both would like to thank Peter
Bartello for generously sharing his DNS code. GJC would like to thank
Iain Waugh for guidance on arc-length continuation and RRK is grateful
to Predrag Cvitanovic for always being willing to talk.

\section*{Appendix A}

The Newton-GMRES-Hook-step algorithm described in the main text is
easily extended to continue solutions over parameter space such as
$Re$ or the domain geometry (e.g. $\alpha$). We briefly describe this
extension for solution branch continuation in $Re$ which was used to
generate figure \ref{Contin_D}. A simple strategy is to use the
solution ${\bf x}(Re)$ as an initial guess in the
Newton-GMRES-Hook-step algorithm with the hope of converging to ${\bf
  x}(Re+\delta Re)$. This should work provided that $\delta Re$ is
`small enough' but is ill-equipped to negotiate turning points in the
solution branch. A standard, more sophisticated approach is arc-length
continuation which uses the branch arc-length as a
natural, monotonically-increasing, parametrisation of the solution
branch. The key idea is to take small controllable steps in the
arc-length rather than $Re$. As a result the state vector needs to be
extended as follows
\beq
\bx=\left[\begin{array}{c} \bOmega\\ s\\ T \\ Re \end{array} \right]
\eeq
and an extra equation 
\beq
\frac{\partial {\bf x}}{\partial r} \cdot
\frac{\partial {\bf x}}{\partial r}=1
\label{continuation}
\eeq
added to determine $Re$. Previous converged solutions $\bx(r_{-1})$
and $\bx(r_0)$ indicate a reasonable step size in $r$, $\delta
r=r_0-r_{-1}=\sqrt{(\bx(r_0)-\bx(r_{-1}) )^2}$ and allow a prediction
to be made for the next solution 
\beq
\bx(r_1) \approx \bx(r_0)+\delta r \frac{\partial \bx}{\partial
  r}\biggl|_{r_0} \biggr.
\eeq
Given $\bx(r_0)$ and  $\delta r$, the extra constraint for the Newton
method comes from approximating (\ref{continuation}) as follows
\beq
{\cal N}(\bx^n):= \frac{\partial \bx}{\partial               
  r}\biggl|_{r_0} \biggr.  (\bx^n-\bx(r_0)) -\delta r  \approx 0        
\eeq
for the $n$th iterate to estimate $\bx(r_1)$. Writing $\delta
\bx^n:=\bx^{n+1}-\bx^n$, then setting
\beq {\cal N}(\bx^{n+1})=\delta x^n. \frac{\partial \bx}{\partial
  r}\left|_{r_0} \right.  +{\cal N}(\bx^n) = 0
\eeq
puts the required extra constraint on the iterative improvement  $\delta \bx^n$. The Newton problem 
(\ref{NR}) then becomes

\beq                                                                                           
\left[\begin{array}{c|c|c|c}
\ddots \hspace*{2cm} & \vspace{0.0cm} \vdots & \vspace{0.0cm} \vdots&
\vdots\\       
\hspace*{1cm} {\displaystyle \frac{\partial {\bf
      \hat{\Omega}_s}}{\partial \bOmega_0}}-{\bf I} 
\hspace*{1cm} & \T_x {\bf \hat{\Omega}_s} &                                                    
{\displaystyle \frac{\partial {\bf \hat{\Omega}_s}}{\partial T}} &
{\displaystyle \frac{\partial {\bf \hat{\Omega}_s}}{\partial
    Re}}\\                          
\hspace*{2cm}\ddots& \vdots & \vdots & \vdots\\\hline \cdots \quad
(\T_x \bOmega_0)^T \quad \cdots & 0 & 0 & 0 \\\hline \cdots \quad
        {\displaystyle \frac{\partial \bOmega_0}{\partial
            t}^T}\quad\cdots & 0 & 0 & 0\\\hline
        \multicolumn{4}{c}{\cdots \quad {\displaystyle \frac{\partial
              \bx_0}{\partial r}^T} \quad \cdots}\\
\end{array}                                                                                    
\right] \left[
\begin{array}{c}                                                                               
\vdots\\   
\\                                                                                            
\vspace{0.25cm}\delta \bOmega\\ \vdots\\ \\ \delta s\\ \\ \\ \delta T
\\ \\ \\ \delta Re \\
\end{array}                                                                                    
\right] =-\left[ \begin{array}{c} 
\vdots\\ 
\\
\vspace{0.25cm}{\bf F}(   {\bf \Omega}_0,s_0,T_0;m)\\                                          
\vdots\\                                                                                       
\\                                                                                             
0\\                                                                                            
\\                                                                                             
\\                                                                                             
0 \\
\\
\\
{\cal N}(\bx_0)\\                                                                              
\end{array} \right]                                                                            
\label{NRextended}    
\eeq         

Depending on how easily convergence is obtained, $\delta r$ can be
increased or decreased if the algorithm shows signs of divergence.  A
second order approach to estimating $\partial \bx/\partial r$ was
actually adopted for the predictive step but the first order estimate
proved sufficient for the constraint present in (\ref{NRextended}).

\end{document}